\documentclass[11pt]{article}
\pdfoutput=1 
\usepackage{caption} 
\usepackage{nicefrac} 
\usepackage{graphicx}  
\usepackage{jheppubg} 
\usepackage{orcidlink}
\usepackage{accents} 
\usepackage[cmyk]{xcolor} 
\usepackage{tikz} 
\pagecolor{white} 

\definecolor{lyell}{cmyk}{0,0,1,.05}
\definecolor{dyell}{cmyk}{0,0,1,.2}
\definecolor{lgree}{cmyk}{.8, 0, .8, 0.08}
\definecolor{lblue}{cmyk}{.8, .4, 0, .1}
\definecolor{myblack}{cmyk}{0,0,0,1}
\definecolor{mywhite}{cmyk}{0,0,0,.01}
\definecolor{lgray}{cmyk}{0,0,0,.15}
\definecolor{mgray}{cmyk}{0,0,0,.5}
\definecolor{myell}{cmyk}{0,0,1,.07}
\definecolor{mrora}{cmyk}{0,.5,.9,.02}
\definecolor{mred}{cmyk}{0,.7,.7,0.15}
\definecolor{mygre}{cmyk}{.3,0,.75,.035}
\definecolor{mgree}{cmyk}{.9,0,.9,0.4}
\definecolor{mbvio}{cmyk}{.3,.4,0,0.03}
\definecolor{mblue}{cmyk}{.7,.35,0,0.2}

\def \gsqus#1#2{\tikz\draw[scale={#2}, color=mgray, fill={#1}, line width=0.1ex, rounded corners=0.05ex, line cap=round] (0,0) -- (.9ex,0) -- (.9ex,.9ex) -- (0,.9ex) -- (0,0);}
\def \gdias#1#2{\tikz[baseline=.12ex]\draw[scale={#2}, color=mgray, fill={#1}, line width=0.1ex, rounded corners=0.05ex, line cap=round] (0,.6ex) -- (.6ex,1.2ex) -- (1.2ex,.6ex) -- (.6ex,0) -- (0,.6ex);}
\def \gcirs#1#2{\tikz\draw[scale={#2}, color=mgray, fill={#1}, line width=0.1ex, rounded corners=0.05ex, line cap=round] (0,0) circle (0.5ex);}

\def \gplus#1#2{\tikz\draw[scale=#2, color=mgray, fill={#1}, line width=0.1ex, rounded corners=0.05ex, line cap=round] (0,0) -- (0.25ex,0.43ex) -- (0.25ex,0.87ex) -- (0.5ex,0.87ex) -- (1ex,0) -- (0,0);}
\def \gplds#1#2{\tikz\draw[scale=#2, color=mgray, fill={#1}, line width=0.1ex, rounded corners=0.05ex, line cap=round] (0,0) -- (0,0.43ex) -- (0.25ex,0.43ex) -- (0.5ex,0.87ex) -- (1ex,0) -- (0,0);}
\def \gprus#1#2{\tikz\draw[scale=#2, color=mgray, fill={#1}, line width=0.1ex, rounded corners=0.05ex, line cap=round] (1ex,0) -- (0.75ex,0.43ex) -- (0.75ex,0.87ex) -- (0.5ex,0.87ex) -- (0,0) -- (1ex,0);}
\def \gprds#1#2{\tikz\draw[scale=#2,color=mgray, fill={#1}, line width=0.1ex, rounded corners=0.05ex, line cap=round] (1ex,0) -- (1ex,0.43ex) -- (0.75ex,0.43ex) -- (0.5ex,0.87ex) -- (0,0) -- (1ex,0);}
\def \galds#1#2{\tikz\draw[scale=#2, color=mgray, fill={#1}, line width=0.1ex, rounded corners=0.05ex, line cap=round] (0,0.58ex) -- (0.25ex,0.14ex) -- (0.25ex,-0.29ex) -- (0.5ex,-0.29ex) -- (1ex,0.58ex) -- (0,0.58ex);}
\def \galus#1#2{\tikz\draw[scale=#2, color=mgray, fill={#1}, line width=0.1ex, rounded corners=0.05ex, line cap=round] (0,0.58ex) -- (0,0.14ex) -- (0.25ex,0.14ex) -- (0.5ex,-0.29ex) -- (1ex,0.58ex) -- (0,0.58ex);}
\def \gards#1#2{\tikz\draw[scale=#2, color=mgray, fill={#1}, line width=0.1ex, rounded corners=0.05ex, line cap=round] (1ex,0.58ex) -- (0.75ex,0.14ex) -- (0.75ex,-0.29ex) -- (0.5ex,-0.29ex) -- (0,0.58ex) -- (1ex,0.58ex);}
\def \garus#1#2{\tikz\draw[scale=#2, color=mgray, fill={#1}, line width=0.1ex, rounded corners=0.05ex, line cap=round] (1ex,0.58ex) -- (1ex,0.14ex) -- (0.75ex,0.14ex) -- (0.5ex,-0.29ex) -- (0,0.58ex) -- (1ex,0.58ex);}
\def \gpas#1#2{\tikz\draw[scale=#2, color=mgray, fill={#1}, line width=0.1ex, rounded corners=0.05ex, line cap=round] (0,0) -- (0.5ex,0.87ex) -- (1ex,0) -- (0,0);}
\def \gaps#1#2{\tikz\draw[scale=#2, color=mgray, fill={#1}, line width=0.1ex, rounded corners=0.05ex, line cap=round] (0,0.58ex) -- (0.5ex,-0.29ex) -- (1ex,0.58ex) -- (0,0.58ex);}

\def \al{\alpha}
\def \be{\beta}
\def \ga{\gamma}
\def \de{\delta}
\def \ep{\epsilon}

\def \et{\eta}
\def \th{\theta}

\def \si{\sigma}
\def \ta{\tau}

\def \ph{\phi}
\def \ch{\chi}
\def \ps{\psi}
\def \om{\omega}
\def \Ga{\Gamma}

\def \Th{\Theta}

\def \Ph{\Phi}
\def \Ps{\Psi}

\def \beq{\begin{equation}}
\def \eeq{\end{equation}}
\def \ba{\begin{array}}
\def \ea{\end{array}}

\def \lb{\left[}
\def \rb{\right]}
\def \lp{\left(}
\def \rp{\right)}

\def \q{\;\;\;\;}
\def \s{\;\;\;\;\;\;}
\def \os#1{{\tilde{#1}}}

\def \fr#1#2{{\textstyle \frac{#1}{#2}}}
\def \ha{\fr{1}{2}}
\def \nfr#1#2{\nicefrac{#1}{#2}}
\def \nha{\nfr{1}{2}}

\def\DynkinNodeSize{3.5mm}
\def\DynkinArrowLength{3mm}
\tikzset{
  dnode/.style={
    circle,
    inner sep=0pt,
    minimum size=\DynkinNodeSize,
    fill=white,
    draw},
  middlearrow/.style={
    decoration={markings,
      mark=at position 0.6 with
      {\draw (0:0mm) -- +(+135:\DynkinArrowLength); \draw (0:0mm) -- +(-135:\DynkinArrowLength);}
    },
    postaction={decorate}
  },
  sedge/.style={},
  dedge/.style={
    middlearrow,
    double distance=0.5mm
  },
  tedge/.style={
    middlearrow,
    double distance=1.0mm+\pgflinewidth,
    postaction={draw}
  }
}

\subheader{$ $}

\title{Division Algebras, Triality, and Exceptional Magic}

\author{A. Garrett Lisi\,\orcidlink{0000-0002-4378-244X}}

\affiliation{Pacific Science Institute, Makawao, HI, USA}

\emailAdd{Gar@Li.si}

\abstract{
We describe in explicit detail the rich relationship between division algebras, split composition algebras, triality, Clifford algebras, spinors, triality Lie algebras, generalized reflections, exceptional magic square Lie algebras, triality eigenspaces, 3-symmetric spaces, Vinberg theta algebras, and Exceptional Unification in particle physics.
}


\begin{document}

\maketitle


\newpage

\section{Introduction}

A growing community of researchers has become interested in the application of division algebras and the corresponding split composition algebras to a structural description of the Standard Model of particle physics \cite{Boy19, Che23, Che20, Dix94, Dra15, Dra10, Dub18, Fur22, Fur24, Gil19, Kra21, Lis07, Lis10, Man22, Per21, Ram03, Vai21, Wil24}. The use of triality in this context, relating three generations of fermions \cite{Lis24}, has gathered increasing interest. The existence of the quaternion group, $Q_8$, within the $CPT$ Group generated by charge, parity, and time conjugation symmetries, and its extension to the $CPTt$ Group --- acting on three generations of fermions related by triality --- strongly indicates division algebras are intricately woven into the fabric of reality. Despite this indication of usefulness, explicit mathematical descriptions of exactly how triality can be used in model building are sparse. It is the purpose of this work to remedy this deficit --- to provide a detailed description of the mathematical scaffolding relating division algebras, triality, Clifford algebras, and Lie algebras related to the Standard Model and gravity. This paper largely follows and complements Baez's excellent paper, ``The Octonions'' \cite{Bae01}, but in more painful detail, and with an intended application in physics.

We begin by introducing division algebras: the complex numbers, $\mathbb{C}$, quaternions, $\mathbb{H}$, and octonions, $\mathbb{O}$, and the related split-signature composition algebras, $\mathbb{C}'$, $\mathbb{H}'$, and $\mathbb{O}'$, and use these to construct Clifford algebras. The Clifford bivectors generate $Spin$ groups, which act on spinors with positive and negative-chiral halves. A structural isomorphism (or ``confusion'') then exists, between $2$, $4$, or $8$-dimensional vectors, $v$, negative-chiral spinors, $\ps$, positive-chiral spinors, $\ch$, and sets of three division algebra elements. (We casually use ``division algebra'', $\mathbb{D}$, to also encompass the corresponding split-signature composition algebras, and sometimes not the reals.)

A real, cyclic, trilinear triality function is defined by the division algebra product (or vice versa), and is invariant under the triality group of symmetries on its arguments. These symmetries relate to generalized reflections, producing duality automorphisms related to twistor incidence relations \cite{Woi21}, as well as triality automorphisms, which transform and cycle the arguments. Each division algebra has a Lie algebra, its triality algebra, corresponding to this triality group. Each triality algebra is a subalgebra of the triality Lie algebra formed by the joining of a triality algebra with the three elements corresponding to a vector, negative-chiral spinor, and positive-chiral spinor. Using these elements, these triality Lie algebras can be expressed heuristically as $su(3,\mathbb{D})$ \cite{Dra15, Wil22}. Triality inner-automorphisms act within these Lie algebras, and can be displayed graphically in their root systems.

Two division algebras can be combined to construct a compound division algebra representation of a Clifford algebra. The corresponding two triality Lie algebras combine to give Lie algebras in the exceptional magic square \cite{BaSu03, Eva09, Fre64, Tit66}. These magic square Lie algebras, and their triality automorphisms, are formulated explicitly. Triality automorphisms partition these Lie algebras into three eigenspaces, with the triality invariant subalgebra acting on the positive eigenspace forming a Vinberg $\Th$-algebra \cite{Vin76}. It is proposed that three generations of fermions can mix within the positive and negative eigenspaces of the $\Th$-algebra decomposition. Explicit, detailed descriptions are provided for Lie algebras $e_6$, $e_7$, and $e_8$. The algebra of the $SO(10)$ Grand Unified Theory, including one generation of fermions, embeds in real compact $e_6$. Three generations of Dixon algebra fermions, related to $\mathbb{C}\otimes\mathbb{H}\otimes\mathbb{O}$, embed in complex $e_7$, or one real generation of fermions, with spin, embeds in compact real $e_7$. And three generations of fermions, with spin, related by triality, embed in real forms of $e_8$, along with Standard Model gauge, Higgs, and gravitational fields. The paper concludes with a brief description of division algebra automorphisms, which brings in the remaining exceptional Lie algebra, $g_2$.


\section{Division Algebra Representation of Clifford Algebras}

A $n$-dimensional division algebra, $\mathbb{D}$, or split-signature composition algebra, $\mathbb{D}'$, is spanned by its basis elements, $e_a$, which have a conjugation,
$$
\os{e}_0 = e_{\os{0}} = e_0 = 1 \s\s \os{e}_1 = e_{\os{1}} = -e_1 \s\s ... \s\s \os{e}_{n-1} = e_{\widetilde{n-1}} = -e_{n-1} 
$$ 
and a multiplication table, $e_a e_b = M_{ab}{}^c e_c$, allowing the definition of its metric,
$$
(e_a, e_b) = \ha ( \os{e}_a e_b + \os{e}_b e_a ) = n_{ab}
$$
with $n_{ab} = \de_{ab}$ for the usual division algebras, and $n_{ab}$ having split signature, $\{+,-\}$, for the split-algebras. Under conjugation, division algebra multiplication satisfies
$$
\widetilde{(e_a e_b)} = e_{\os{b}} e_{\os{a}}  \s \s \s \s \s  M_{ab}{}^{\os{c}} = M_{\os{b} \os{a}}{}^{c}
$$
Standard multiplication tables, $M_{ab}{}^c e_c$, for the division algebras and their split-algebras are:
\beq
\ba{l}
\mathbb{C} \; : \;\,  \small{\lb \ba{cc} e_0 & e_1 \\ e_1 & -e_0  \ea \rb} \\[20pt]
\mathbb{H} \; : \;  \small{\lb \ba{cccc}
e_0 & e_1 & e_2 & e_3 \\
e_1 & -e_0 & e_3 & -e_2 \\
e_2 & -e_3 & -e_0 & e_1 \\
e_3 & e_2 & -e_1 & -e_0
\ea \rb} \\[36pt]
\mathbb{O} \; : \;  \small{ \lb \ba{cccccccc}
e_0 &  e_1 &  e_2 &  e_3 & e_4 & e_5 &  e_6 & e_7 \\
e_1 & -e_0 & e_4 & e_7 & -e_2 & e_6 & -e_5 & -e_3 \\
e_2 & -e_4 & -e_0 & e_5 & e_1 & -e_3 & e_7 & -e_6 \\
e_3 & -e_7 & -e_5 & -e_0 & e_6 & e_2 & -e_4 & e_1  \\
e_4 & e_2 & -e_1 & -e_6 & -e_0 & e_7 & e_3 & -e_5  \\
e_5 & -e_6 & e_3 & -e_2 & -e_7 & -e_0 & e_1 & e_4 \\
e_6 & e_5 & -e_7 & e_4 & -e_3 & -e_1 & -e_0 & e_2 \\
e_7 & e_3 & e_6 & -e_1 & e_5 & -e_4 & -e_2 & -e_0 
\ea \rb} \\[50pt]
\ea
\ba{r}
\!\!\!\!
\mathbb{O}' \; : \;  \small{ \lb \ba{cccccccc}
e_0 &  e_1 &  e_2 &  e_3 & e_4 & e_5 &  e_6 & e_7 \\
e_1 & -e_0 & e_3 & -e_2 & -e_5 & e_4 & -e_7 & e_6 \\
e_2 & -e_3 & -e_0 & e_1 & -e_6 & e_7 & e_4 & -e_5 \\
e_3 & e_2 & -e_1 & -e_0 & -e_7 & -e_6 & e_5 & e_4  \\
e_4 & e_5 & e_6 & e_7 &  e_0 & e_1 & e_2 & e_3  \\
e_5 & -e_4 & -e_7 & e_6 & -e_1 & e_0 & e_3 & -e_2 \\
e_6 & e_7 & -e_4 & -e_5 & -e_2 & -e_3 & e_0 & e_1 \\
e_7 & -e_6 & e_5 & -e_4 & -e_3 & e_2 & -e_1 & e_0 
\ea \rb} \\[60pt]
\mathbb{H}' \; : \; \small{ \lb \ba{cccc}
e_0 & e_1 & e_2 & e_3 \\
e_1 & e_0 & e_3 & e_2 \\
e_2 & -e_3 & -e_0 & e_1 \\
e_3 & -e_2 & -e_1 & e_0
\ea \rb} \\[36pt]
\mathbb{C}' \; : \;  \small{\lb \ba{cc} e_0 & e_1 \\ e_1 & e_0  \ea \rb} \, \\[20pt]
\ea
\label{M}
\vspace{10pt}
\eeq
Division algebra multiplication allows the construction of chiral Clifford basis elements of $Cl(n)$, $Cl(0,n)$, or $Cl(\fr{n}{2},\fr{n}{2})$, which act on chiral \emph{division algebra spinors},
\beq
{\ba{rcl}
\ga_c &=&
\lb \begin{array}{cc}
0 & \pm \os{e}_c \\
e_c & 0
\end{array} \rb
\;\sim\;
\lb \begin{array}{cc}
0 &  \pm (\bar{\Ga}_c)^a{}_b \\
(\Ga_c)^b{}_a & 0
\end{array} \rb
\; = \;
\lb \begin{array}{cc}
0 &  \pm M_{\os{c}\os{b}}{}^a \\
M_{ca}{}^{\os{b}}  & 0
\end{array} \rb
\\[20pt]
\Ps &=& \lb \ba{c} \ps \\ \os{\ch} \ea \rb
\;=\; \lb \ba{c} \ps^a e_a \\ \ch^b \os{e}_b \ea \rb 
\;\sim\; \lb \ba{c} \ps^a Q_a^- \\ \ch^b Q_b^+ \ea \rb \\[-12pt]
\ea}
\label{ClD}
\vspace{12pt}
\eeq
with the definition of a signature-adjusted matrix transpose, $(\bar{\Ga}_c) = n_{cc} (\Ga_c)^T$, and multiplication understood to be to the right by division algebra elements (accounting for non-associativity of octonions), or represented equivalently as $2n \times 2n$ real matrices built from the multiplication table coefficients,
$$
(\Ga_c)^b{}_a = M_{ca}{}^{\os{b}} \s \s \s (\bar{\Ga}_c)^a{}_b = M_{\os{c}\os{b}}{}^a = M_{bc}{}^{\os{a}} = M_c{}^a{}_{\os{b}} = (\Ga_c)_b{}^a = n_{cc} (\Ga_c)^b{}_a
$$
It is worth emphasizing the ``trick'' here, first brought to the author's attention by Mia Hughes, because it is at the heart of this paper. In our cases of interest a correct set of real, chiral, Clifford basis vector representative matrices, $(\Ga_c)^b{}_a$, can be numerically identified with a division algebra multiplication table, $M_{ca}{}^{\os{b}}$, with a suitable shuffling of indices and a conjugation. We subsequently use $M$ and $\Ga$ matrices interchangeably, depending on whether the division algebra or Clifford algebra context is more appropriate.

The representative Clifford basis vector elements, (\ref{ClD}), satisfy the fundamental Clifford identity,
\beq
\ga_a \cdot \ga_b = \ha \lp \ga_a \ga_b + \ga_b \ga_a  \rp = 
\pm \ha
\lb \begin{array}{cc}
 \os{e}_a e_b + \os{e}_b e_a & 0 \\
0 &  e_a \os{e}_b + e_b \os{e}_a
\end{array} \rb
= \pm n_{ab} = \et_{ab}
\label{FCI}
\eeq
with the ``$\pm$'' signature in (\ref{ClD}) and above usually chosen to be ``$-$'' for our purposes. Division algebra multiplication coefficients, and the corresponding Clifford matrix elements, satisfy a cyclic identity,
\beq
\bar{\Ga}_{abc} = \bar{\Ga}_{bca}
= \Ga_{acb} = \Ga_{cba}  
= M_{ca\os{b}} = M_{ab\os{c}}
= M_{\os{a}\os{c}b} = M_{\os{c}\os{b}a}
\label{cyclic}
\eeq
in which $n_{ab}$ is used to lower indices. For the quaternionic and octonionic algebras, including their split forms, the Clifford pseudoscalar is $\ga = \ga_0 ... \ga_{n-1} = \pm \scalebox{.5}{$\lb \ba{cc} I_n & \\[-3pt] & -I_n \\[-1pt] \ea \rb$}$.

The $\ha n (n-1)$ representative Clifford bivector basis elements, for $c < d$, are:
\beq
\ba{rcl}
\ga_{cd} \,=\, \lb \! \ba{cc} \pm \os{e}_c e_d  &  \\   & \pm e_c \os{e}_d \ea \! \rb
&\sim& \lb \! \ba{cc} \pm (\bar{\Ga}_c)^a{}_b (\Ga_d)^b{}_e  &  \\   & \pm (\Ga_c)^b{}_a (\bar{\Ga}_d)^a{}_f \ea \! \rb \\[20pt]
&=& \lb \! \ba{cc} \pm M_{\os{c}\os{b}}{}^a M_{de}{}^\os{b} &  \\   & \pm M_{ca}{}^\os{b} M_{\os{d}\os{f}}{}^a \ea \! \rb \\[4pt]
\ea
\label{bbe}
\eeq
with it understood that, for example, $e_d$, multiplies to the right before $\os{e}_c$ multiplies the result. Since $e_c \os{e}_d = - e_d \os{e}_c$ for $c \ne d$, we also have the reverse-indexed bivectors, $\ga_{d c} = - \ga_{c d}$, with $\ga_{cc}=0$. Bivector sums are unrestricted, with $\ha$ factors avoiding double counting of antisymmetric coefficients, as in $B = \ha B^{cd} \ga_{cd}$. These bi-product division algebra operators are the chiral basis elements of the corresponding spin Lie algebra, which act on division algebra spinors. Spinors are the fundamental representation space of spin groups, which have spin Lie algebras spanned by Clifford algebra bivectors represented by matrices that act on the spinors. For the complex numbers and quaternions, multiplication is associative, so these bi-product basis elements are themselves purely imaginary complex numbers or quaternions, $\os{e}_c e_d \in \mathbb{B} = \mathrm{Im}(\mathbb{D})$, and the corresponding spin Lie algebras, $\mathbb{B}_2 = so(2)=u(1)$ and $\mathbb{B}_4 = so(4) = su(2) + su(2)$, are $1$ and $6$-dimensional. For the octonions, multiplication is not associative, and these octonionic bi-products span $28$-dimensional $\mathbb{B}_8=so(8) \ne \mathrm{Im}(\mathbb{O})$.

To add clarity for the reader, it is worth presenting the simplest example of this construction: $Cl(0,2)$ constructed from complex algebra. From the complex multiplication table, (\ref{M}), and using $(\Ga_c)^b{}_a=M_{ca}{}^{\os{b}}$, we obtain a real chiral matrix representation of Clifford basis vectors, $\ga_c$,
$$
\ba{c}
M_{00}{}^\os{0} = 1  \s \s   M_{01}{}^\os{1} = M_{10}{}^\os{1} = -1 \s \s   M_{11}{}^\os{0} = -1 \\[10pt]
\Ga_0 = \lb  \ba{cc} \, 1 & \\ &-1\ea  \rb \s \s
\Ga_1 = \lb  \ba{cc} &-1 \\-1 &\ea  \rb \s \s 
\ga_c = \lb  \ba{cc}0 & -\bar{\Ga}_c\\ \, \Ga_c&0\ea  \rb
\ea
$$
To be painfully explicit, the resulting representative unit basis vectors and unit basis spinors of $Cl(0,2)$ are:
$$
\ba{c}
\ga_0 = \scalebox{.9}{$ \lb \ba{cccc} & & -1 \; & \\[-2pt] & & & 1 \\[-2pt] 1 & & & \\[-2pt] & -1 & & \\[-2pt] \ea  \rb $} \s \s
\ga_1 = \scalebox{.9}{$ \lb \ba{cccc} & &  & 1 \\[-2pt] & & 1 \; &  \\[-2pt] & -1 \; & & \\[-2pt] -1 & & & \\[-2pt] \ea  \rb $}
\\[30pt]
Q^-_0 = \scalebox{.9}{$ \lb  \ba{c} 1 \\[-2pt]  \\[-2pt]  \\[-2pt]  \\[-2pt] \ea  \rb $} \s \s
Q^-_1 = \scalebox{.9}{$ \lb  \ba{c} \\[-2pt] 1\\[-2pt]   \\[-2pt]  \\[-2pt] \ea  \rb $} \s \s
Q^+_0 = \scalebox{.9}{$ \lb  \ba{c} \\[-2pt]  \\[-2pt]  1 \\[-2pt]  \\[-2pt] \ea  \rb $} \s \s
Q^+_1 = \scalebox{.9}{$ \lb  \ba{c} \\[-2pt]  \\[-2pt]  \\[-2pt]  1 \\[-2pt]   \ea  \rb $}
\ea
$$
Physicists usually think of spinors, such as the unit spinors, $Q^\pm_c$, as matrix columns. Depending on context, they may also be thought of as ideal Clifford algebra elements.

From the division algebra representation of Clifford algebras, we have a natural \emph{confusion} between sets of three $n$-dimensional division algebra elements related by division algebra multiplication, and corresponding sets of Clifford algebra vectors, negative real chiral spinors, and positive real chiral spinors, related by Clifford algebra multiplication,
$$
\ba{lcl}
v = v^c e_c & \s \sim \s & v = v^c \ga_c \\[2pt]
\ps = \ps^a e_a &  \sim  & \ps=\ps^a Q^-_a \\[2pt]
\os{\ch} = \ch^b \os{e}_b &  \sim  & \ch=\ch^b Q^+_b \\[4pt]
\os{\ch} = v \, \ps &  \sim  & \ch = v \, \ps
\ea
$$
It is the chiral division algebra representative matrices of Clifford algebras, $\Ga_c{}^b{}_a = M_{ca}{}^\os{b}$, that allow this direct identification between a set of vector, negative, and positive-chiral spinors, $(v,\ps,\ch)$, and a triplet of division algebra elements with the same coefficients, and their equivalent relationship under division algebra and Clifford multiplication,
$$
\ch^b e_\os{b} = \os{\ch} = v \, \ps = v^c \ps^a M_{ca}{}^\os{b} e_\os{b}
\s \sim \s
\ch^b = v^c \ps^a (\Ga_c)^b{}_a
$$
This confusion of vectors and spinors with division algebra elements, and the division algebra construction of Clifford algebras, leads to the explicit construction and understanding of the structure of many Lie algebras and their automorphisms.

\section{Generalized Reflections and Triality}

Division algebras (and their split versions) have a cubic form --- a real, cyclic, trilinear \emph{triality function}, $T(v, \ps, \ch)$, of three elements, or, equivalently, of vectors and chiral spinors,
\beq
\ch^b v^c \ps^a \Ga_{cba} = \bar{\ch} v \, \ps = T(v, \ps, \ch) = \lp \os{\ch}, v \, \ps \rp = \ch^b v^c \ps^a M_{ca\os{b}} 
\label{trif}
\eeq
in which we again are using a signature-adjusted transpose, $\bar{Q}^+_b = n_{bb} Q^{+T}_b$, so the index in $Q^{+T}_b (\Ga_c) Q^-_a = \Ga_c{}^b{}_a$ will be lowered. The triality function is cyclic, $T(v, \ps, \ch) = T(\ps, \ch, v)$, by virtue of the cyclic nature of division algebra multiplication, (\ref{cyclic}). Although one usually considers the triality function as built from the division algebra product, it is possible, alternatively, to use the existence of a triality function, as a cyclic cubic form on a vector space, to define the division algebra product. The triality function is invariant under the \emph{triality group}, ${\rm Tri}(\mathbb{D})$, with elements $r \in {\rm Tri}(\mathbb{D})$ satisfying:
$$
r \; : \; (v, \ps, \ch) \; \mapsto \; (v', \ps', \ch') 
\s \ni \s
T(v', \ps', \ch') = T(v, \ps, \ch)
$$
The triality group acts linearly on $(v,\ps,\ch)$, preserving their quadratic norms up to permutation of the three arguments.

Consider reflections, $R^u_v$, through a unit-length Clifford vector or division algebra element,
$$
- \, u \cdot u = u^a u^b n_{ab} = s_u = \os{u} u = \pm 1  
$$
in which the signature, $s_u$, is space-like, $+1$, for division algebra elements or $Cl(0,n)$ Clifford vectors, or can be time-like, $-1$, for some split-composition algebra elements or the corresponding time-like $Cl(\fr{n}{2},\fr{n}{2})$ Clifford vectors. Note that we have chosen the ``$-$'' sign in (\ref{ClD}, \ref{FCI}), to later match Lie algebra elements. For quaternionic and octonionic algebras, including their split forms, reflections, $R_v^u$, of a Clifford vector and spinor through $u$ can be written as
$$
v' = R^u_v v  = - u v u^-
\s \s \s
\Ps' = R^u_v \Ps  = \pm \sqrt{s_u} u \ga \Ps
$$
with the ``$+$'' for division algebras and ``$-$'' for split, and unit real or imaginary $\sqrt{s_u}$ introduced so that $R^u_v R^u_v = 1$. Although this expression matches a known description of reflections \cite{Lou01, Por95}, the use of the Clifford pseudoscalar allows the reflection of any Clifford algebra element to be described as a Clifford adjoint, $R^u A = (u \ga) A (u \ga)^-$, with $(u \ga)$ an element of the pin group \cite{Lis24}.\footnotemark[1]\footnotetext[1]{This mathematical construction is likely original to the author.} These reflections can also be expressed using division algebra elements, and, since triality cycles vectors and spinors, we also have \emph{generalized reflections}, $R^u_m$ and $R^u_p$, acting as reflections through negative and positive chiral spinors, using division algebra multiplication,
\beq
\ba{rclcrclcrcl}
  & & R^u_v  & & \s \s & & R^u_m & & \s \s  & & R^u_p  \\[6pt]
v' &=& R^u_v v  = - s_u u \os{v} u              &  & v' &=& R^u_m \ch  =  \sqrt{s_u} \os{\ch} \os{u}  &  & v' &=& R^u_p \ps  = \sqrt{s_u} \os{u} \os{\ps} \\[6pt]
\ps' &=& R^u_v \ch = \sqrt{s_u} \os{u} \os{\ch} &  & \ps' &=& R^u_m \ps  = - s_u u \os{\ps} u  &  & \ps' &=& R^u_p v  =  \sqrt{s_u} \os{v} \os{u} \\[6pt]
\ch' &=& R^u_v \ps = \sqrt{s_u} \os{\ps} \os{u} &  & \ch' &=&  R^u_m v  =  \sqrt{s_u} \os{u} \os{v} &  & \ch' &=& R^u_p \ch  = - s_u u \os{\ch} u
\ea
\eeq
These generalized reflections, through a space-like or time-like unit element, $u$, can be equivalently expressed as operations on Clifford basis vector and chiral spinor elements, and written out explicitly using indices,\footnotemark[1]
\beq\s
\ba{rclcrclcrcl}
  & & R^u_v  &  & \s \q   & & R^u_m &  & \s \q   & & R^u_p  \\[6pt]
\ga'_c \!&=&\! (\de^a_c - 2 s_u u^a u_c) \ga_a &  & \ga'_c \!&=&\! \sqrt{s_u} u^a (\bar{\Ga}_a)^b{}_c Q^+_b &  & \ga'_c \!&=&\! \sqrt{s_u} u^b (\Ga_b)^a{}_c Q^-_a \\[6pt]
\! {Q^-_a}' \!\!&=&\! \sqrt{s_u} u^c (\Ga_c)^b{}_a Q^+_b  & & {Q^-_a}' \!\!&=&\! (\de^b_a \!-\! 2 s_u u^b u_a) Q^-_b  & & {Q^-_a}' \!\!&=&\! \sqrt{s_u} u^c (\bar{\Ga}_c)^b{}_a \ga_b  \\[6pt]
\! {Q^+_b}' \!\!&=&\! \sqrt{s_u} u^c (\bar{\Ga}_c)^a{}_b Q^-_a &  & {Q^+_b}' \!\!&=&\!  \sqrt{s_u} u^c (\Ga_c)^a{}_b \ga_a & & {Q^+_b}' \!\!&=&\! (\de^a_b \!-\! 2 s_u u^a u_b) Q^+_a
\ea
\label{genref}
\eeq
In these expressions, $\ga_a$ are the basis vectors, $Q^+_a$ are the basis positive chiral spinors, and $Q^-_a$ are the basis negative chiral spinors. $R^u_v$ is a reflection through the $u$ unit vector, mapping vectors to vectors, positive chiral spinors to negative chiral spinors, and negative chiral spinors to positive chiral spinors. $R^u_m$ is a generalized reflection through the $u$ unit negative chiral spinor, mapping positive chiral spinors to vectors, negative chiral spinors to negative chiral spinors, and vectors to positive chiral spinors. $R^u_p$ is a generalized reflection through the $u$ unit positive chiral spinor, mapping negative chiral spinors to vectors, vectors to negative chiral spinors, and positive chiral spinors to positive chiral spinors. The triality function is anti-invariant under generalized reflections, such as
$$
\ba{rcl}
T(v',\ps',\ch') &=& T(R^u_v v, R^u_v \ch, R^u_v \ps) = T(-s_u u \os{v} u, \sqrt{s_u} \os{u} \os{\ch}, \sqrt{s_u} \os{\ps} \os{u}) \\[6pt]
 &=& \lp u \ps, ( - u \os{v} u) \os{u} \os{\ch} \rp = - T(v, \ps, \ch)
\ea
$$

If we impose a constraint that $T(v,\ps,\ch) = 1$ for a matched set of $\{ v, \ps, \ch \}$, then we can obtain expressions for either the vector, negative spinor, or positive spinor from the two others such that the constraint is satisfied,
$$
v = \frac{1}{|\ps \ch|^2} \widetilde{\ps \ch}
\s \s \s
\ps = \frac{1}{|\ch v|^2} \widetilde{\ch v}
\s \s \s
\ch = \frac{1}{|v \ps|^2} \widetilde{v \ps}
$$
This is called dualizing the triality function. More generally, generalized reflections through a non-unit-length division algebra element, $v$, give \emph{duality functions},\footnotemark[1] such as $\ps = \widetilde{\ch v}$. We can use division-Clifford algebra confusion to get the equivalent equation for a negative chiral spinor from the Clifford multiplication of a vector and a positive chiral spinor:
$$
\ps = \os{v} \os{\ch} \q \leftrightarrow \q \ps^a = v^c \ch^b M_{\os{c} \os{b}}{}^a \q \leftrightarrow \q \ps^a = (v^c \bar{\Ga}_c{}^a{}_b) \ch^b
$$ 
For the quaternionic case, this is the ``incidence relation'' for a Euclidean twistor.\cite{Woi21} The main idea of twistor theory is that if you know a $\ps$ and $\ch$, the incidence relation lets you solve for $v$, so you can do fun things with a ``twistor'', $(\ps,\ch)$, that translates to things having to do with a vector, $v$. This relates triality to twistor theory, but is not the focus of this work.

Combining two generalized reflections of the same type gives a generalized rotation --- an element of the triality group. Combining two generalized reflections of different types gives an element of the triality group that isn't a rotation. Combining four generalized reflections through two unit-length elements, $u$ and $w$, gives a \emph{triality automorphism}, such as\footnotemark[1]
$$
t^{uw} = R^w_p R^u_v R^\os{u}_m R^\os{u}_p
$$
an element of the triality group that takes vectors to positive spinors, positive spinors to negative spinors, and negative spinors to vectors,
$$
t^{uw}  :  (v, \ps, \ch)  \mapsto  (v', \ps', \ch') = \Big( \sqrt{s_u}\sqrt{s_w} \os{w} (u \ps), \, \sqrt{s_u}\sqrt{s_w} (\ch u) \os{w}, \, s_u s_w w ( \os{u} v \os{u}) w \big)
$$
with $T(v', \ps', \ch') = T(v, \ps, \ch)$. Via Clifford algebra confusion, pairs of unit elements, $u$ and $w$, produce general triality automorphisms of sets of three division algebra elements or of the corresponding Clifford vector and spinors,
\beq
{\ba{rcl}
v' &=& \sqrt{s_u}\sqrt{s_w} \os{w} (u \ps) = \sqrt{s_u}\sqrt{s_w} w^d u^c \ps^a M_{\os{d}\os{b}}{}^f M_{ca}{}^\os{b} e_f \\[8pt]
& & \sim \; v' = \sqrt{s_u}\sqrt{s_w} w^d u^c \ps^a (\bar{\Ga}_d)^f{}_b (\Ga_c)^b{}_a Q^-_f  = - \sqrt{s_u}\sqrt{s_w} w u \ps \\[10pt]
\ps' &=& \sqrt{s_u}\sqrt{s_w} (\ch u) \os{w} = \sqrt{s_u}\sqrt{s_w} \ch^b u^c w^d M_{bc}{}^\os{a} M_{\os{a}\os{d}}{}^f e_f \\[8pt] 
& &  \sim \; \ps' =  \sqrt{s_u}\sqrt{s_w} \ch^b u^c w^d (\Ga_d)^f{}_a (\bar{\Ga}_c)^a{}_b Q^+_f = - \sqrt{s_u}\sqrt{s_w} w u \ch \\[10pt]
\ch' &=& s_u s_w w (\os{u} v \os{u}) w = s_u s_w w^a u^b v^c u^d w^e M_{af}{}^g M_{\os{b}c}{}^j M_{j\os{d}}{}^f M_{ge}{}^h e_h \\[8pt]
& &  \sim \; \ch' =  v^b (\de^c_a - 2 s_w w^c w_a)(\de^a_b - 2 s_u u^a u_b) \ga_c  = s_u s_w w u v u w 
\ea}
\label{gentri}
\eeq
Choosing $u=1$ ($\sim \, u=\ga_0$) and $w=1$ ($\sim \, w=\ga_0$) gives a \emph{canonical triality automorphism},
$$
t \; : \; (v, \ps, \ch) \; \mapsto \; (v', \ps', \ch') = (\ps, \ch, v)
$$
consistent with the invariance of triality under cyclic permutation of its arguments. Even numbers of generalized reflections --- elements of the triality group --- can be factored into canonical triality automorphisms and rotations, such as
$$
t^{uw} = R^w_p R^u_v R^\os{u}_m R^\os{u}_p = R^w_p R^u_p \, t = R^{uw}_p \, t = t \, R^{uw}_v
$$

A simple rotation, $R^{uw}_v = R^w_v R^u_v$, can be described as two reflections through vectors, $u$ and $w$, spanning the plane of rotation, resulting in a transformation of vectors to vectors, negative chiral spinors to negative chiral spinors, and positive chiral spinors to positive chiral spinors. Alternatively, a rotation, $R^B_v$, can be described using a Clifford algebra rotor, $U = e^{\nfr{B}{2}}$, acting on vectors and spinors,
$$
v' = R^B_v v = U v U^- \simeq v + \ha B v - \ha v B \s \s \s
\Ps' = R^B_v \Ps = U \Ps \simeq \Ps + \ha B \Ps
$$
in which
$$
B = \ha B^{cd} \ga_{cd} = \lb \ba{cc}  B_m & \\ &  B_p  \ea \rb
$$
is a Clifford bivector, (\ref{bbe}). Generalized rotations, $R^B_m \!=\! t^2 R^B_v t$ and $R^B_p \!=\! t R^B_v t^2$, can be described by combining rotations with a canonical triality automorphism.

The triality group of a division algebra, ${\rm Tri}(\mathbb{D})$, includes rotations and triality automorphisms. The Lie algebra of this triality group is the \emph{triality algebra} of the division algebra, ${\rm tri}(\mathbb{D})$; the Lie algebra of generalized rotations. Its elements, $R \in {\rm tri}(\mathbb{D})$, satisfy:
$$
R \, : \, (v, \ps, \ch) \, \mapsto \, (v', \ps', \ch') 
\s \s
T(v', \ps, \ch) + T(v, \ps', \ch) + T(v, \ps, \ch') = 0
$$
A bivector rotation generator, $B$, acts on chiral vectors and spinors as
$$
v' = \lb \ba{cc} &  v'_p  \\  v'_m  & \ea \rb = \ha \! \lb \ba{cc} & \! B_m v_p \!-\! v_p B_p  \\  B_p v_m \!-\! v_m B_m \! & \ea \rb
 \s \s \ps' = \ha B_m \ps \s \s \ch' = \ha B_p \ch 
$$
Which, using (\ref{trif}), we can see satisfies
$$
\ba{rcl}
0 &=& T(v', \ps, \ch) + T(v, \ps', \ch) + T(v, \ps, \ch') \\[6pt]
 &=& \bar{\ch} \ha (B_p v_m - v_m B_m) \ps + \bar{\ch} v_m \ha B_m \ps + \bar{\ch} \ha \bar{B}_p v_m \ps
 \ea
$$
using $\bar{B}_p = - B_p$. We can similarly calculate the action of generalized rotation generators on vectors and spinors by using a canonical triality automorphism. In this way, the rotation symmetry algebras of the complex numbers, $so(2)=u(1)$, quaternions, $so(4) = su(2) + su(2)$, octonions, $so(8)$, and split algebras are enlarged by triality. The resulting triality algebras \cite{BaSu03, Eva09} are:
$$
\ba{rclcrcl}
{\rm tri}(\mathbb{C}) &=& u(1) + u(1) & \s \s & {\rm tri}(\mathbb{C}') &=& gl(1) + gl(1) \\[6pt]
{\rm tri}(\mathbb{H}) &=& su(2) + su(2) + su(2) & & {\rm tri}(\mathbb{H}') &=& sl(2) + sl(2) + sl(2) \\[6pt]
{\rm tri}(\mathbb{O}) &=& so(8) &  & {\rm tri}(\mathbb{O}') &=& so(4,4)
\ea
$$
These triality algebras act on the triplets, $(v,\ps,\ch)$, as a vector space symmetrically under triality. If we combine them and close the Lie algebra brackets consistently, we obtain the \emph{triality Lie algebras}.

\section{Triality Lie Algebras}

A good way to understand precisely how triality algebras act on the triplets, $( v,\ps, \ch )$, is via their embedding in the corresponding triality Lie algebras \cite{BaSu03, Eva09},
$$
\ba{rcl}
su(3) &=& {\rm tri}(\mathbb{C}) + \mathbb{C} + \mathbb{C} + \mathbb{C}
= u(1) + u(1) + (1+\bar{1})_v + (1+\bar{1})_m + (1+\bar{1})_p \\[4pt]
sp(3) &=& {\rm tri}(\mathbb{H}) + \mathbb{H} + \mathbb{H} + \mathbb{H}
=  su(2) + su(2) + su(2) + (2,2,1)_v + (2,1,2)_m + (1,2,2)_p \\[4pt]
f_{4(-52)} &=& {\rm tri}(\mathbb{O}) + \mathbb{O} + \mathbb{O} + \mathbb{O}
=  so(8) + 8_v + 8_{s-} + 8_{s+} \\[8pt]
sl(3) &=& {\rm tri}(\mathbb{C}') + \mathbb{C}' + \mathbb{C}' + \mathbb{C}'
= gl(1) + gl(1) + (1+\bar{1})_v + (1+\bar{1})_m + (1+\bar{1})_p \\[4pt]
sp(6,\mathbb{R}) &=& {\rm tri}(\mathbb{H}') + \mathbb{H}' + \mathbb{H}' + \mathbb{H}'
=  sl(2) + sl(2) + sl(2) + (2,2,1)_v + (2,1,2)_m + (1,2,2)_p \\[4pt]
f_{4(4)} &=& {\rm tri}(\mathbb{O}') + \mathbb{O}' + \mathbb{O}' + \mathbb{O}'
=  so(4,4) + 8_v + 8_{s-} + 8_{s+}
\ea
$$
Generalized reflections, generalized rotations, and triality automorphisms are real automorphisms of the compact triality Lie algebras. For split composition algebras, timelike generalized reflections as defined above involve $\sqrt{s_u}=i$ and must be interpreted on the complexified Lie algebra; only transformations preserving the real form give real automorphisms. The structures of these Lie algebras fully elucidate these symmetries, and are worth examining in each case.

\subsection{\texorpdfstring{$su(3)$}{su(3)}}

Lie algebra elements, $A$, corresponding to the special unitary group, $SU(3)$, may be represented by $3 \times 3$ traceless, anti-Hermitian matrices of complex numbers, related to the eight basis Gell-Mann matrices,
\beq
\ba{rcl}
\!\! A(B^1, B^2, v, \ps, \ch) \, &=&
\lb\ba{ccc}
i \, B^1 + \fr{i}{\sqrt{3}} B^2 & -v^0 + i \, v^1 & \ps^0 + i \, \ps^1 \\ 
v^0 + i \, v^1 & - i \, B^1 + \fr{i}{\sqrt{3}} B^2 & -\ch^0 + i \, \ch^1 \\
-\ps^0 + i \, \ps^1 & \ch^0 + i \, \ch^1 & -\fr{2i}{\sqrt{3}} B^2
\ea\rb \\[24pt]
&=&
\lb\ba{ccc}
V-M & -v^* & \ps \\
v & P-V & -\ch^* \\
-\ps^* & \ch & M-P 
\ea\rb \\ [24pt]
\phantom{\!\! A(B^1, B^2, v, \ps, \ch)}
&=& B^1 T_1 + B^2 T_2 + v^a \ga_a + \ps^a Q^-_a + \ch^a Q^+_a  \\[6pt]
&=& -i \lp V \, H_v + M \, H_m + P \, H_p \rp \\[3pt]
 & & + (v \, E^-_v - v^* E^+_v) + (\ps \, E^-_m - \ps^* E^+_m) + (\ch \, E^-_p - \ch^* E^+_p) \\[4pt]
 & \in & su(3)
\ea
\label{su3}
\eeq
with $\{ v, \ps, \ch \}$ complex parameters, $v = v^0 + i \, v^1 = v^0 e_0 + v^1 e_1$, $\{ B^1, B^2 \}$ real parameters, related $\{ V, M, P \}$ pure imaginary parameters, $\{ T_1, T_2, \ga_a, Q^-_a,$ $Q^+_a \}$ Lie algebra basis generators related to the Gell-Mann matrices, $\{ H_v, H_m,$ $H_p \}$ Cartan generators, and $\{ E^\pm_v, E^\pm_m, E^\pm_p  \}$ root vector generators. Note that since $su(3)$ elements are traceless, $V$, $M$, and $P$ correspond to only two degrees of freedom, $B^1$ and $B^2$ --- the same $su(3)$ element is specified if $V$, $M$, and $P$ are all shifted by a constant. These $V$, $M$, and $P$ parameters are motivated by the existence of three overlapping $su(2)$ subalgebras, spanned by $\{H_v, \ga_a \}$, $\{H_m, Q^-_a\}$, and $\{H_p, Q^+_a\}$. As a triality Lie algebra, $su(3)$ relates to the complex division algebra representation of $Cl(0,2)$, with basis vectors $\ga_0 = \scalebox{.5}{$\lb\ba{cc}0 & -1\\1&0\ea\rb$}$ and $\ga_1 = \scalebox{.5}{$\lb\ba{cc}0 & i\\i&0\ea\rb$}$. This $\mathbb{C}$ division algebra representation of $Cl(0,2)$ is not the representation from the complex multiplication table, (\ref{M}), which is instead, from $(\Ga_c)^b{}_a=M_{ca}{}^{\os{b}}$,
$$
\ba{c}
M_{00}{}^\os{0} = 1  \s \s  M_{01}{}^\os{1} = M_{10}{}^\os{1} = -1 \s \s  M_{11}{}^\os{0} = -1 \\[10pt]
\Ga_0 = \lb \! \ba{cc}1 & \\ &-1\ea \! \rb \s \s
\Ga_1 = \lb \! \ba{cc} &-1 \\-1 &\ea \! \rb \s \s
\ga_c = \lb \! \ba{cc}0 & -\bar{\Ga}_c\\ \Ga_c&0\ea \! \rb
\ea
$$

In general, a triality Lie algebra can either be described directly as $\sim \! su(3,\mathbb{D})$, or equivalently by constructing the corresponding Clifford algebra and its bivector and vector matrix representatives (identified with the upper-left $2 \times 2$ block in $su(3,\mathbb{D})$) which act on negative and positive-chiral spinors, then closing the algebra via the Lie brackets between spinors. The representation of the Clifford algebra may be by division algebra elements, by their matrix representatives, or from the equivalent division algebra multiplication table coefficients.

We can compute the $su(3)$ Lie brackets directly from the commutator of its representative matrices, (\ref{su3}),
$$
\ba{c}
\lb A\lp B^1_1, B^2_1, v_1, \ps_1, \ch_1 \rp, A\lp B^1_2, B^2_2, v_2, \ps_2, \ch_2 \rp \rb = A\lp B^1_3, B^2_3, v_3, \ps_3, \ch_3 \rp \\[14pt]
\ba{rcl}
2i \, B^1_3 &=&  - 2  ( v^*_1 v_2 - v^*_2 v_1) - ( \ps_1 \ps^*_2 - \ps_2  \ps^*_1)  - (\ch_1 \ch^*_2 - \ch_2 \ch^*_1) \\[4pt]
\fr{2 i}{\sqrt{3}} B^2_3  &=&  (\ch_1 \ch^*_2 - \ch_2 \ch^*_1) + (\ps^*_1 \ps_2 - \ps^*_2 \ps_1) \\[4pt]
v_3 &=& - 2 i (B^1_1 v_2 - B^1_2 v_1)  + (\ch^*_1 \ps^*_2 - \ch^*_2 \ps^*_1) \\[4pt]
\ps_3 &=&  i (B^1_1 + \sqrt{3} B^2_1) \ps_2 - i (B^1_2 + \sqrt{3} B^2_2) \ps_1 + (v^*_1 \ch^*_2 - v^*_2 \ch^*_1) \\[4pt]
\ch_3 &=&   i (B^1_1 - \sqrt{3} B^2_1) \ch_2 - i (B^1_2 - \sqrt{3} B^2_2) \ch_1+ (\ps^*_1 v^*_2 - \ps^*_2 v^*_1)
\ea
\ea
$$
Alternatively, using our basis elements, $\{T_1, T_2, \ga_a, Q^-_a, Q^+_a \}$, the non-zero $su(3)$ brackets between them are, explicitly,\footnotemark[2]\footnotetext[2]{These formula were constructed and checked with the assistance of symbolic manipulation software, Sage and/or Mathematica.}
\beq
\ba{rclcrclcrcl}
\lb T_1, \ga_0 \rb &=& -2 \, \ga_1 & \s &
   \lb T_1, Q^-_0 \rb &=& +Q^-_1 & \s &
 \lb T_1, Q^+_0 \rb &=& +Q^+_1   \\[6pt]

\lb T_1, \ga_1 \rb &=& +2 \, \ga_0 &  &
   \lb T_1, Q^-_1 \rb &=& -Q^-_0 &  &
 \lb T_1, Q^+_1 \rb &=& -Q^+_0  \\[6pt]

 & &  &  &
   \lb T_2, Q^-_0 \rb &=& +\sqrt{3} \, Q^-_1 &  &
 \lb T_2, Q^+_0 \rb &=& -\sqrt{3} \, Q^+_1  \\[6pt]
 
  & &  &  &
   \lb T_2, Q^-_1 \rb &=& -\sqrt{3} \, Q^-_0 &  &
 \lb T_2, Q^+_1 \rb &=& +\sqrt{3} \, Q^+_0  \\[6pt]
 
 \lb \ga_0, \ga_1 \rb &=& -2 \, T_1 &  &
   \lb Q^-_0, Q^-_1 \rb &=& T_1 + \sqrt{3} \, T_2  &  &
 \lb Q^+_0, Q^+_1 \rb &=& T_1 - \sqrt{3} \, T_2  \\[6pt]
 
  \lb \ga_a, Q^-_b \rb &=& - M_{\os{a} \os{b}}{}^c \, Q^+_c &  &
   \lb \ga_a, Q^+_b \rb &=& M_{\os{a} \os{b}}{}^c \, Q^-_c  &  &
  \lb Q^-_a, Q^+_b \rb &=& - M_{\os{a} \os{b}}{}^c \, \ga_c  \\
\ea
\label{su3L}
\eeq
and their anti-symmeterized partners.

With orthogonal Cartan subalgebra basis generators, $\{ T_1, T_2 \}$, or non-orthogonal Cartan basis generators, $\{  H_v = T_1, H_m = (- \ha T_1 - \fr{\sqrt{3}}{2} T_2), H_p = (- \ha T_1 + \fr{\sqrt{3}}{2} T_2) \}$, the root vectors and their Lie brackets are:\footnotemark[2]
$$
\ba{rclcrclcrcl}
E^+_v  &=& \ha ( -\ga_0 - i \, \ga_1 )   & \s &
 \lb T_1, E^\pm_\al \rb &=& \pm i \, g^1_\al \, E^\pm_\al  & \s &
   \lb E^\pm_v, E^\pm_m \rb &=& \mp E^\mp_p \\[5pt]
E^-_v  &=& \ha ( +\ga_0 - i \, \ga_1 )     & \!\!\!\! &
 \lb T_2, E^\pm_\al \rb &=& \pm i \, g^2_\al \, E^\pm_\al                 & \!\!\!\! &
   \lb E^\pm_m, E^\pm_p \rb &=& \mp E^\mp_v \\[5pt]
E^+_m  &=& \ha ( -Q^-_0 - i \, Q^-_1 )  & &                                          & &                                                         & &
 \lb E^\pm_p, E^\pm_v \rb &=& \mp E^\mp_m \\[5pt]        
E^-_m  &=& \ha ( +Q^-_0 - i \, Q^-_1 )   & \!\!\!\! &
 \lb H_v, E^\pm_\al \rb &=& \pm  i \, v_\al \, E^\pm_\al     &\!\!\!\! &
   \lb E^+_v, E^-_v \rb &=& -i \, H_v \\[5pt]
E^+_p  &=& \ha ( -Q^+_0 - i \, Q^+_1 )  & \!\!\!\! &
  \lb H_m, E^\pm_\al \rb &=& \pm  i \, m_\al \, E^\pm_\al  & \!\!\!\! &
     \lb E^+_m, E^-_m \rb &=& -i \, H_m \\[5pt] 
E^-_p  &=& \ha ( +Q^+_0 - i \, Q^+_1 )    & \!\!\!\! &
 \lb H_p, E^\pm_\al \rb &=& \pm  i \, p_\al \, E^\pm_\al      & \!\!\!\! &
  \lb E^+_p, E^-_p \rb &=& -i \, H_p 
\ea
$$
with the $\{ g^1_\al, g^2_\al, v_\al, m_\al, p_\al \}$ roots shown in Table~\ref{tab:su3roots}. This structure of $su(3)$ is consistent with its triality decomposition, in which each of the three triples, $\{ H_{v/m/p}, E^+_{v/m/p}, E^-_{v/m/p} \} \sim \{ H , E^+ , E^- \}$, corresponds to a different $su(2)$, related to each other by triality, with disjoint root vectors but overlapping Cartan generators. The relevant triality function is,
$$
T(v,\ps,\ch)  = v^c \ps^a \ch^b M_{ab\os{c}} = v^0 \ps^0 \ch^0 - v^0 \ps^1 \ch^1 - v^1 \ps^1 \ch^0 - v^1 \ps^0 \ch^1
$$
and a canonical inner triality automorphism of $su(3)$ is:
$$
t \; : \; A \; \mapsto \; A' = g_t \, A \, g_t^-
\s \s \s
g_t = \lb  \ba{ccc} & & \! 1 \\[-6pt] 1 & &  \\[-6pt] & 1  & \\[-2pt]  \ea  \rb \; \in \; SU(3)
$$
in which $g_t$ is an element of the $3 \times 3$ representation of the $SU(3)$ Lie group, and $t$ transforms the generators, root vectors, and Cartan subalgebra elements as:
$$
t  \q : \q  \ga_a \mapsto Q^+_a \mapsto Q^-_a \mapsto \ga_a
\s \q
 E^\pm_v \mapsto E^\pm_p \mapsto E^\pm_m \mapsto E^\pm_v
\s \q
H_v \mapsto H_p \mapsto H_m \mapsto H_v
$$
On root space coordinates, $(g^1_\al, g^2_\al)$, measured against the fixed Cartan basis, this triality automorphism acts by the \emph{triality matrix}, $t$. On a column listing the transformed Cartan basis elements, $T_j'=g_t T_j g_t^-$, the corresponding formula uses $t^T=t^-$,
$$
\lb \ba{c} g_\al^1{}' \\[4pt] g_\al^2{}' \ea \rb = \lb \! \ba{cc} -\ha & -\fr{\sqrt{3}}{2} \\[3pt] \fr{\sqrt{3}}{2} & -\ha  \ea  \rb \lb  \ba{c} g_\al^1 \\[4pt] g_\al^2 \ea  \rb
\s \s
\lb \ba{c} T_1{}' \\[4pt] T_2{}' \ea  \rb = \lb \! \ba{cc} -\ha &  \fr{\sqrt{3}}{2} \\[3pt] -\fr{\sqrt{3}}{2} &  -\ha  \ea  \rb \lb  \ba{c} T_1 \\[4pt] T_2 \ea  \rb
\s \s
t = \lb \! \ba{cc} -\ha &  -\fr{\sqrt{3}}{2} \\[3pt] \fr{\sqrt{3}}{2} &  -\ha  \ea  \rb
$$
Within the triality algebra of $su(3)$, which is also its Cartan subalgebra, the two basis generators, $\{ T_1, T_2 \}$, are each rotated between three directions by a canonical triality automorphism. Specifically, $T_1^I = H_v = T_1$, $T_1^{II} = H_p = - \ha T_1 + \fr{\sqrt{3}}{2} T_2$, and $T_1^{III} = H_m = - \ha T_1 - \fr{\sqrt{3}}{2} T_2$.

\begin{table}[h!t]
$\q${\centerline{\parbox{.45\textwidth}{\centerline{
\scalebox{0.85}{
\renewcommand{\arraystretch}{1.4} 
\begin{tabular}
{@{\vrule width1.0pt}c@{\vrule width0.2pt}c@{\vrule width1.0pt}c@{\vrule width0.0pt}c@{\vrule width1.0pt}c@{\vrule width0.0pt}c@{\vrule width0.0pt}c@{\vrule width1.0pt}}
\noalign{\hrule height 1.0pt}
\multicolumn{2}{@{\vrule width1.0pt}c@{\vrule width1.0pt}}{$\;\; su(3) \;\;$} & $\q g^1 \q$  & {\vrule width0.2pt} $\;\;\; g^2 \q$ & $\q v \;\;\;$ {\vrule width0.2pt} &  $\;\;\; m \;\;$ {\vrule width0.2pt} & $\q p \q$ \\
\noalign{\hrule height 1.0pt}
$\,$ \gpas{lyell}{2} $\,$ & $\;\; v^+ \;\;$ & $ +2 \;$ & $ 0 \;$ & $ +2 \;$ & $ -1 \;$ & $ -1 \;$ \\
\noalign{\hrule height 0.2pt}
$\,$ \raisebox{-1pt}{\gaps{lyell}{2}} $\,$ & $\;\; v^- \;\;$ & $ -2 \;$ & $ 0 \;$ & $ -2 \;$ & $ +1 \;$ & $ +1 \;$ \\
\noalign{\hrule height .8pt}
$\,$ \gpas{lyell}{1.7} $\,$ & $\;\; m^+ \;\;$ & $ -1 \;$ & $\! -\sqrt{3} \;$ & $ -1 \;$ & $ +2 \;$ & $ -1 \;$ \\
\noalign{\hrule height 0.2pt}
$\,$ \gaps{lyell}{1.7} $\,$ & $\;\; m^- \;\;$ & $ +1 \;$ & $\! +\sqrt{3} \;$ & $ +1 \;$ & $ -2 \;$ & $ +1 \;$ \\
\noalign{\hrule height .8pt}
$\,$ \gpas{lyell}{1.5} $\,$ & $\;\; p^+ \;\;$ & $ -1 \;$ & $\! +\sqrt{3} \;$ & $ -1 \;$ & $ -1 \;$ & $ +2 \;$ \\
\noalign{\hrule height 0.2pt}
$\,$ \gaps{lyell}{1.5} $\,$ & $\;\; p^- \;\;$ & $ +1 \;$ & $\! -\sqrt{3} \;$ & $ +1 \;$ &$ +1 \;$ & $ -2 \;$ \\
\noalign{\hrule height 1.0pt}
\end{tabular}
}}}$\s$
\parbox{.45\textwidth}{\centerline{
\includegraphics[height=1.8in]{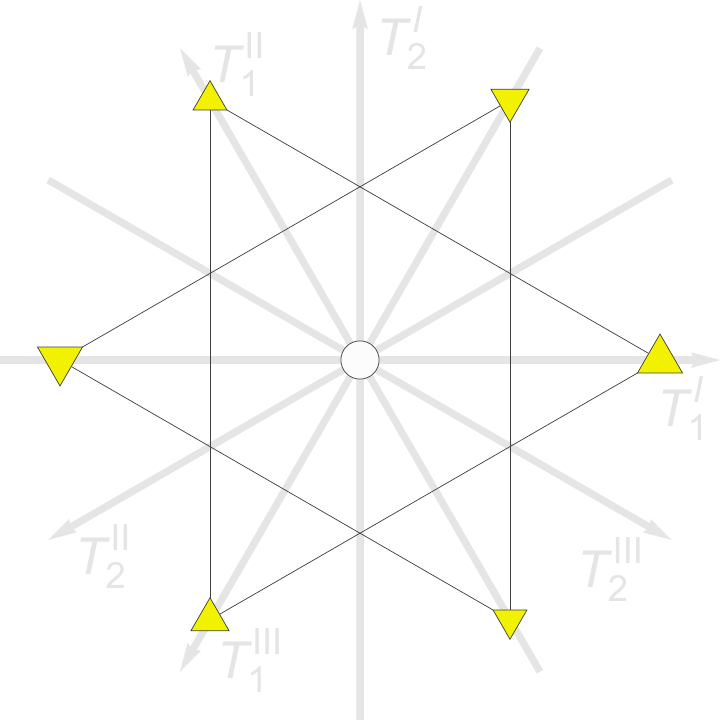}
}}}}
\vspace{8pt}
\caption{Roots of $su(3)$ with respect to the orthogonal Cartan subalgebra basis generators, $\{T_1,T_2\}$, or non-orthogonal Cartan basis generators, $\{H_v,H_m,H_p\}$, and their canonical triality automorphism. The ``particle'' roots related by triality are shown as yellow triangles, \protect\gpas{lyell}{1.7}, of different sizes, while their ``antiparticle'' roots related by triality are inverted triangles, \protect\gaps{lyell}{1.7}.}
\vspace{-8pt}
\label{tab:su3roots}
\end{table}

Another way of looking at the decomposition of $su(3)$ is by grouping any of the three $u(1)$ subalgebras generated by $H_v$, $H_m$, or $H_p$ within the $u(1) \!+\! u(1)$ triality subalgebra with the corresponding $E^\pm$ root vector pair to make a $su(2)$ subalgebra, resulting in the decomposition:
$$
\ba{rcl}
su(3)  &=&  u(1) + \lp u(1) + (1+\bar{1})_v \rp + (1+\bar{1})_m + (1+\bar{1})_p \\[6pt]
  &=&  u(1) + su(2) + 2_{+\sqrt{3}} + 2_{-\sqrt{3}}
\ea
$$
in which the $\pm \sqrt{3}$ are the $u(1)$ charges, equal to the $g^2$ charges visible in the figure and Table \ref{tab:su3roots}. This $su(2)$ subalgebra of $su(3)$ corresponds to the upper left $2 \times 2$ block of the $3 \times 3$ matrix representation, (\ref{su3}), spanned by the $T_1$ and $\ga_a$ generators. A canonical triality automorphism of $su(3)$ with this decomposition produces three different $su(2)$ subalgebras.

The split-complex numbers, a composition algebra, are represented by $\{e'_0 = 1, e'_1 = I\}$, with $M'_{11}{}^0 = I^2 = 1$. Repeating our Lie algebra construction, using real Gell-Mann matrices, we get the Lie algebra $sl(3)$, with non-vanishing brackets:\footnotemark[2]
\beq
\scalebox{.9}{$
\ba{rclcrclcrcl}
\lb T'_1, \ga'_0 \rb &=& -2 \, \ga'_1 & \s  &
   \lb T'_1, Q'^-_0 \rb &=& +Q'^-_1 & \s &
 \lb T'_1, Q'^+_0 \rb &=& +Q'^+_1   \\[6pt]

\lb T'_1, \ga'_1 \rb &=& -2 \, \ga'_0 &  &
   \lb T'_1, Q'^-_1 \rb &=& +Q'^-_0 &  &
 \lb T'_1, Q'^+_1 \rb &=& +Q'^+_0  \\[6pt]

 & &  & \!\!\!\!\!\!\!\! &
   \lb T'_2, Q'^-_0 \rb &=& +\sqrt{3} \, Q'^-_1 &  &
 \lb T'_2, Q'^+_0 \rb &=& -\sqrt{3} \, Q'^+_1  \\[6pt]
 
  & &  & \!\!\!\!\!\!\!\! &
   \lb T'_2, Q'^-_1 \rb &=& +\sqrt{3} \, Q'^-_0 &  &
 \lb T'_2, Q'^+_1 \rb &=& -\sqrt{3} \, Q'^+_0  \\[6pt]
 
 \lb \ga'_0, \ga'_1 \rb &=& -2 \, T'_1 & &
   \lb Q'^-_0, Q'^-_1 \rb &=& T'_1 + \sqrt{3} \, T'_2  &  &
 \lb Q'^+_0, Q'^+_1 \rb &=& T'_1 - \sqrt{3} \, T'_2  \\[6pt]
 
  \lb \ga'_a, Q'^-_b \rb &=& - M'_{\os{a} \os{b}}{}^c \, Q'^+_c &  &
   \lb \ga'_a, Q'^+_b \rb &=& M'_{\os{a} \os{b}}{}^c \, Q'^-_c  &  &
  \lb Q'^-_a, Q'^+_b \rb &=& -M'_{\os{a} \os{b}}{}^c \, \ga'_c  \\
\ea
$}
\label{su3Lp}
\eeq
The triality automorphism structure for $sl(3)$ is the same as for $su(3)$.

\subsection{\texorpdfstring{$sp(3)$}{sp3}}

The structure of the $21$-dimensional symplectic Lie algebra, $sp(3)$ --- the triality Lie algebra of the quaternions --- is similar to that of $su(3)$. Instead of $3 \times 3$ traceless matrices of complex numbers, elements of $sp(3)$ can be represented by matrices of quaternions,
\beq
A(M,P,V,v,\ps,\ch) \,  =
\lb \ba{ccc}
M & -\os{v} & \ps \\
v & P & -\os{\ch} \\
-\os{\ps} & \ch & V 
\ea \rb
\;\; \in \; sp(3) = su(3,\mathbb{H})
\label{sp3}
\eeq
with $\{ v, \ps, \ch \}$ quaternions and $\{ M, P, V \}$ purely imaginary quaternions. The Lie brackets are thus:
$$
\lb A \! \lp M_1, P_1, V_1, v_1, \ps_1, \ch_1 \rp \!, A \! \lp M_2, P_2, V_2, v_2, \ps_2, \ch_2 \rp \rb \! = \! A \! \lp M_3, P_3, V_3, v_3, \ps_3, \ch_3 \rp
$$
$$
\ba{rcl}
M_3 &=& M_1 M_2 - M_2 M_1 - (\os{v}_1 v_2 - \os{v}_2 v_1) - (\ps_1 \os{\ps}_2 - \ps_2 \os{\ps}_1) \\[4pt]
P_3 &=& P_1 P_2 - P_2 P_1 - (\os{\ch}_1 \ch_2 - \os{\ch}_2 \ch_1) - (v_1 \os{v}_2 - v_2 \os{v}_1) \\[4pt]
V_3 &=& V_1 V_2 - V_2 V_1 - (\os{\ps}_1 \ps_2 - \os{\ps}_2 \ps_1) - (\ch_1 \os{\ch}_2 - \ch_2 \os{\ch}_1) \\[4pt]
v_3 &=& (P_1 v_2 - P_2 v_1) + (v_1 M_2 - v_2 M_1) + (\os{\ch}_1 \os{\ps}_2 - \os{\ch}_2 \os{\ps}_1) \\[4pt]
\ps_3 &=& (M_1 \ps_2 - M_2 \ps_1) + (\ps_1 V_2 - \ps_2 V_1) + (\os{v}_1 \os{\ch}_2 - \os{v}_2 \os{\ch}_1) \\[4pt]
\ch_3 &=& (V_1 \ch_2 - V_2 \ch_1) + (\ch_1 P_2 - \ch_2 P_1) + (\os{\ps}_1 \os{v}_2 - \os{\ps}_2 \os{v}_1)
\ea
$$
Each of the three diagonal matrix elements is a $su(2)$ subalgebra, which act on two out of the three off-diagonal elements. The Cartan subalgebra basis consists of one element from each diagonal element, $\{ T^M_3, T^P_3, T^V_3 \}$.

The $sp(3)$ Lie algebra, and its brackets, relate to the Clifford algebra matrix representation of $Cl(0,4)$ from the quaternion multiplication table, as in (\ref{ClD}). The quaternion matrix (\ref{sp3}) has a $12 \times 12$ real representation. Or, using the usual Pauli matrix representation of quaternions,
$$
e_0 = \si_0 \s \s e_1 = -i \, \si_1 \s \s e_2 = -i \, \si_2 \s \s e_3 = -i \, \si_3
$$
(\ref{sp3}) results in a $6 \times 6$ complex representation. Using the quaternion multiplication table, (\ref{M}), the $Cl(0,4)$ chiral representative matrices (\ref{ClD}) are $(\Ga_c)^b{}_a = M_{ca}{}^\os{b}$ and $-(\bar{\Ga}_c)^a{}_b = - M_{\os{c}\os{b}}{}^a$, so
$$
v \sim v^c \Ga_c =\!
\scalebox{.9}{$ \lb \! \ba{cccc}
v^0 \!& -v^1 \!& -v^2 \!& -v^3 \\
-v^1 \!& -v^0 \!& v^3 \!& -v^2 \\
-v^2 \!& -v^3 \!& -v^0 \!& v^1 \\
-v^3 \!& v^2 \!& -v^1 \!& -v^0
\ea \! \rb$}
\s \s
- \os{v} \sim - v^c \bar{\Ga}_c =\!
\scalebox{.9}{$ \lb \! \ba{cccc}
-v^0 \!& v^1 \!& v^2 \!& v^3 \\
v^1 \!& v^0 \!& v^3 \!& -v^2 \\
v^2 \!& -v^3 \!& v^0 \!& v^1 \\
v^3 \!& v^2 \!& -v^1 \!& v^0
\ea \! \rb$}
$$
The negative and positive $Cl(0,4)$ chiral bivector matrices separate into independent degrees of freedom, corresponding to $so(4) = su(2)_M + su(2)_P$,
$$
M \sim B_M = - \ha B^{ab} \bar{\Ga}_a \Ga_b = B_M^A \, \bar{\Ga}_0 \Ga_A
\s \s \s
P \sim B_P = - \ha B^{ab} \Ga_a \bar{\Ga}_b = B_P^A  \, \Ga_A \bar{\Ga}_0
$$
in which $B_{M/P}^A = \mp B^{0 A} + \ha \ep^{B C A} B^{B C}$, and the capital indices, $\{A,B,C\}$, are bivector indices, ranging over $\{1,2,3\}$.

The $sp(3)$ Lie algebra elements can be written in an orthoganal, Killing-normalized basis as
$$
\ba{rcl}
A(M,P,V,v,\ps,\ch) \, &=& M^A T^M_A + P^A T^P_A + V^A T^V_A + v^a \ga_a + \ps^a Q^-_a + \ch^a Q^+_a \\[8pt]
&=&
\lb \ba{ccc}
M^A e_A & - v^a \fr{1}{\sqrt{2}} \os{e}_a & \ps^a \fr{1}{\sqrt{2}} e_a \\[6pt]
v^a \fr{1}{\sqrt{2}} e_a & P^A e_A & - \ch^a \fr{1}{\sqrt{2}} \os{e}_a \\[6pt]
-\ps^a \fr{1}{\sqrt{2}} \os{e}_a & \ch^a \fr{1}{\sqrt{2}} e_a & V^A e_A 
\ea \rb
\ea
$$
with the $sp(3)$ Lie brackets between these basis generators computed explicitly:\footnotemark[2]
\beq
\ba{rclcrcl}
\lb T^M_A, T^M_B \rb &=& T^M_C (2 M_{[AB]}{}^C) & \s \s &
   \lb \ga_a, Q^-_b \rb &=& Q^+_c  \fr{1}{\sqrt{2}} (- M_{\os{b}\os{a}}{}^c)  \\[6pt]
   
\lb T^P_A, T^P_B \rb &=& T^P_C (2 M_{[AB]}{}^C) & &
 \lb \ga_a, Q^+_b \rb &=& Q^-_c  \fr{1}{\sqrt{2}} (M_{\os{a}\os{b}}{}^c)  \\[6pt]
 
\lb T^V_A, T^V_B \rb &=& T^V_C (2 M_{[AB]}{}^C) & &
  \lb Q^-_a, Q^+_b \rb &=& \ga_c \fr{1}{\sqrt{2}} (- M_{\os{b}\os{a}}{}^c)  \\[6pt]
  
 \lb T^M_A, \ga_b \rb &=& \ga_c \, (- M_{bA}{}^c) & &
  \lb T^M_A, Q^-_b \rb &=& Q^-_c \, (M_{Ab}{}^c) \\[6pt]
  
 \lb T^P_A, \ga_b \rb &=& \ga_c \, (M_{Ab}{}^c)  & &
  \lb T^P_A, Q^+_b \rb &=& Q^+_c \,(- M_{bA}{}^c)  \\[6pt]
  
   \lb T^V_A, Q^-_b \rb &=& Q^-_c \,(- M_{bA}{}^c) & &
 \lb T^V_A, Q^+_b \rb &=& Q^+_c \,( M_{Ab}{}^c)    \\[6pt]

 \lb \ga_a, \ga_b \rb &=& T^M_C \ha (M_{\os{b}a}{}^C - M_{\os{a}b}{}^C) + T^P_C \ha (M_{b\os{a}}{}^C - M_{a\os{b}}{}^C) 
 \!\!\!\! \!\!\!\! \!\!\!\! \!\!\!\! \!\!\!\! \!\!\!\! \!\!\!\! \!\!\!\! \!\!\!\! \!\!\!\! \!\!\!\! \!\!\!\! \!\!\!\! \!\!\!\! \!\!\!\! \!\!\!\! \!\!\!\! \!\!\!\!
 & &
 & &   \\[6pt]

\lb Q^-_a, Q^-_b \rb &=& T^V_C \ha (M_{\os{b}a}{}^C - M_{\os{a}b}{}^C) + T^M_C \ha (M_{b\os{a}}{}^C - M_{a\os{b}}{}^C) 
\!\!\!\! \!\!\!\! \!\!\!\! \!\!\!\! \!\!\!\! \!\!\!\! \!\!\!\! \!\!\!\! \!\!\!\! \!\!\!\! \!\!\!\! \!\!\!\! \!\!\!\! \!\!\!\! \!\!\!\! \!\!\!\! \!\!\!\! \!\!\!\!
& &
 & &   \\[6pt]
 
\lb Q^+_a, Q^+_b \rb &=& T^P_C \ha (M_{\os{b}a}{}^C - M_{\os{a}b}{}^C) + T^V_C \ha (M_{b\os{a}}{}^C - M_{a\os{b}}{}^C) 
\!\!\!\! \!\!\!\! \!\!\!\! \!\!\!\! \!\!\!\! \!\!\!\! \!\!\!\! \!\!\!\! \!\!\!\! \!\!\!\! \!\!\!\! \!\!\!\! \!\!\!\! \!\!\!\! \!\!\!\! \!\!\!\! \!\!\!\! \!\!\!\!
& &
 & & 
\ea
\label{sp3L}
\eeq

\begin{table}[h!t]
$\q${\centerline{\parbox{.45\textwidth}{\centerline{
\renewcommand{\arraystretch}{1.2}
\scalebox{.92}{
\begin{tabular}
{@{\vrule width1.0pt}c@{\vrule width0.2pt}c@{\vrule width1.0pt}c@{\vrule width0.0pt}c@{\vrule width0.0pt}c@{\vrule width1.0pt}}
\noalign{\hrule height 1.0pt}
\multicolumn{2}{@{\vrule width1.0pt}c@{\vrule width1.0pt}}{$\;\; sp(3) \;\;$} & $\;\;\; M \;\,$ {\vrule width0.2pt} & $\;\;\;\, P \;\;$ {\vrule width0.2pt} & $\;\;\; V \;\;\;$ \\
\noalign{\hrule height 1.0pt}
$\,$ \raisebox{-1pt}{\gcirs{lyell}{1.7}} $\,$ & $\;\; M^\pm \;\;$ & $ \pm 2 \;$ & $ 0 \;$ & $ 0 \;$ \\
\noalign{\hrule height 0.2pt}
$\,$ \raisebox{-1pt}{\gcirs{mywhite}{1.7}} $\,$ & $\;\; P^\pm \;\;$ & $ 0 \;$ & $ \pm 2 \;$ & $ 0 \;$ \\
\noalign{\hrule height 0.2pt}
$\,$ \raisebox{-1pt}{\gcirs{lgray}{1.7}} $\,$ & $\;\; V^\pm \;\;$ & $ 0 \;$ & $ 0 \;$ & $ \pm 2 \;$ \\
\noalign{\hrule height .8pt}
$\,$ \raisebox{-1.5pt}{\gplus{lgray}{2} \gards{lgray}{2}} $\,$ & $\;\; \ps^\pm_e \;\;$ & $ \pm 1 \;$ & $ 0 \;$ & $ \pm 1 \;$ \\
\noalign{\hrule height 0.2pt}
$\,$ \raisebox{-1.5pt}{\gplus{lyell}{2} \gards{lyell}{2}} $\,$ & $\;\; \ps^\pm_o \;\;$ & $ \pm 1 \;$ & $ 0 \;$ & $ \mp 1 \;$ \\
\noalign{\hrule height .8pt}
$\,$ \raisebox{-1pt}{\gplus{lgray}{1.7} \gards{lgray}{1.7}} $\,$ & $\;\; \ch^\pm_e \;\;$ & $ 0 \;$ & $ \pm 1 \;$ & $ \pm 1 \;$ \\
\noalign{\hrule height 0.2pt}
$\,$ \raisebox{-1pt}{\gplus{lyell}{1.7} \gards{lyell}{1.7}} $\,$ & $\;\; \ch^\pm_o \;\;$ & $ 0 \;$ & $ \mp 1 \;$ & $ \pm 1 \;$ \\
\noalign{\hrule height .8pt}
$\,$ \raisebox{-.5pt}{\gplus{lgray}{1.5} \gards{lgray}{1.5}} $\,$ & $\;\; v^\pm_e \;\;$ & $ \pm 1 \;$ & $ \pm 1 \;$ & $ 0 \;$ \\
\noalign{\hrule height 0.2pt}
$\,$ \raisebox{-.5pt}{\gplus{lyell}{1.5} \gards{lyell}{1.5}} $\,$ & $\;\; v^\pm_o \;\;$ & $ \mp 1 \;$ & $ \pm 1 \;$ & $ 0 \;$ \\
\noalign{\hrule height 1.0pt}
\end{tabular}
}
		}}$\q$
		\parbox{.5\textwidth}{\centerline{
		\includegraphics[height=2.4in]{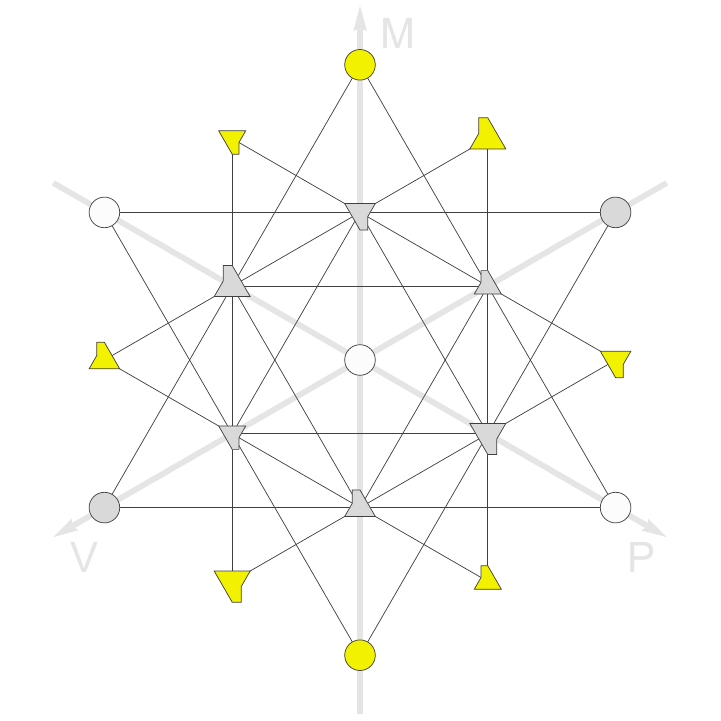}		
			}}}}
\vspace{8pt}
\caption{Roots of $sp(3)$ with respect to the orthogonal Cartan subalgebra basis generators, $\{ T^M_3, T^P_3, T^V_3 \}$, and their canonical triality automorphism. Roots are labeled with circles, \protect\gcirs{lyell}{1.5} \protect\gcirs{mywhite}{1.5} \protect\gcirs{lgray}{1.5}, of different colors for each triality-related $su(2)$, with upper-left triangles of different colors, \protect\gplus{lgray}{1.7} \protect\gplus{lyell}{1.7}, for even or odd charge combinations, and lower-right inverted triangles for their antiparticles, \protect\gards{lgray}{1.7} \protect\gards{lyell}{1.7}, with triangle glyphs of different sizes related by triality.}
\vspace{-8pt}
\label{tab:sp3roots}
\end{table}

Generalized quaternionic reflections give automorphisms of $sp(3)$. For example, the generalized reflection along a unit positive spinor, $R^0_p$, gives
\beq
R^0_p \; : \; A(M, P, V, v, \ps, \ch) \; \mapsto \; A(M', P', V', v', \ps', \ch') \; = \; A(M, V, P, \os{\ps}, \os{v}, - \os{\ch})
\label{R1p}
\eeq
which is an inner automorphism,
$$
R^0_p \;\; : \;\; A \; \mapsto \; A' = g_R \, A \, g_R^-
\s\s\s
g_R = \scalebox{.9}{$ \lb \! \ba{ccc} -1 & & \\[-6pt] &  & 1 \\[-3pt] &  \!\!1 & \\[-2pt]  \ea \! \rb$} \;\; \in \; SP(3) = SU(3,\mathbb{H})
$$
This corresponds to a reflection in root space coordinates, $(\al_M, \al_P, \al_V)$, by the $R$ matrix, and a transformation of the Cartan subalgebra basis elements, $\{ M, P, V \}$, by $R$,
$$
\scalebox{.9}{$
\lb \! \ba{c} \al_M' \\[1pt] \al_P' \\[1pt] \al_V' \\[1pt] \ea \! \rb = \lb \ba{ccc} 1 & & \\[1pt] & & \;1 \\[1pt] & \;1  & \\[1pt]  \ea  \rb \lb \! \ba{c} \al_M \\[1pt] \al_P \\[1pt] \al_V \\[1pt] \ea \! \rb = \lb \! \ba{c} \al_M \\[1pt] \al_V \\[1pt] \al_P \\[1pt] \ea \! \rb
\s \s
\lb \! \ba{c} M' \\[1pt] P' \\[1pt] V' \\[1pt] \ea \! \rb = \lb \ba{ccc} 1 & & \\[1pt] & & \;1 \\[1pt] & \;1  & \\[1pt]  \ea  \rb \lb \! \ba{c} M \\[1pt] P \\[1pt] V \\[1pt] \ea \! \rb = \lb \! \ba{c} M \\[1pt] V \\[1pt] P \\[1pt] \ea \! \rb 
\s \s
R =  \lb  \ba{ccc} 1 & & \\[1pt] & & \;1 \\[1pt] & \;1  & \\[1pt]  \ea  \rb
$}
$$
This reflection visually reflects the $sp(3)$ root diagram in Table \ref{tab:sp3roots} across the $M$ axis.

These generalized quaternion reflections and their compositions form a subgroup of the inner automorphism group of $sp(3)$ preserving the triality decomposition. A canonical triality automorphism of $sp(3)$ is:
$$
t \q : \q  \ga_a \mapsto Q^+_a  \mapsto Q^-_a  \mapsto \ga_a \s T^M_A \mapsto T^P_A  \mapsto  T^V_A \mapsto T^M_A 
$$
which, as a composition of reflections, $t = R^0_p R^0_v R^0_m R^0_p$, is an inner automorphism,
$$
t \;\; : \;\; A \; \mapsto \; A' = g_t \, A \, g_t^-
\s \s
g_t = \lb \ba{ccc} & & \!\!1 \\[-6pt] 1 & &  \\[-6pt] & 1  & \\[-2pt]  \ea  \rb \;\; \in \; SP(3) = SU(3,\mathbb{H})
$$
This corresponds to a rotation of root space coordinates, $(\al_M, \al_P, \al_V)$, measured against the fixed Cartan basis, by the triality matrix, $t$. The diagonal quaternion parameters, $(M,P,V)$, in (\ref{sp3}) transform by the same matrix,
$$
\scalebox{.9}{$
\lb \! \ba{c} \al_M' \\[1pt] \al_P' \\[1pt] \al_V' \\[1pt] \ea \! \rb = \lb \ba{ccc} & & 1 \\[1pt] 1 & & \\[1pt] & \;1 & \\[1pt] \ea \rb \lb \! \ba{c} \al_M \\[1pt] \al_P \\[1pt] \al_V \\[1pt] \ea \! \rb = \lb \! \ba{c} \al_V \\[1pt] \al_M \\[1pt] \al_P \\[1pt] \ea \! \rb
\s \s
\lb \! \ba{c} M' \\[1pt] P' \\[-2pt] V' \\[1pt] \ea \! \rb = \lb  \ba{ccc} & & 1 \\[1pt] 1 & &  \\[1pt] & \;1  & \\[1pt]  \ea  \rb \lb \! \ba{c} M \\[1pt] P \\[1pt] V \\[1pt] \ea \! \rb = \lb \! \ba{c} V \\[1pt] M \\[1pt] P \\[1pt] \ea \! \rb 
\s \s
t  =   \lb \ba{ccc} & & 1 \\[1pt] 1 & &  \\[1pt] & \;1  & \\[1pt]  \ea  \rb
$}
$$
Although one can describe these sorts of Lie algebra automorphisms as reflections or rotations of roots in root space, which correspond to reflections and rotations within the Cartan subalgebra and maps between root vectors, these descriptions do not specify the phases of the maps between root vectors. To obtain a consistent set of phases for such maps, it is usually easiest to describe the automorphisms directly, as transformations of the Lie algebra generators, and then transform to the Cartan-Weyl basis to get a complete description of the maps between root vectors, including phases.

Another way of looking at the decomposition of $sp(3)$ is by grouping any two of the three $su(2)$ factors of the $su(2)\!+\!su(2)\!+\!su(2)$ triality subalgebra with the corresponding root vectors to make a $so(5)\!=\!sp(2)$ subalgebra, resulting in the decomposition:
$$
\ba{rcl}
sp(3)  &=&  \lp su(2)_M + su(2)_P + (2,2,1)_v \rp + su(2)_V + (2,1,2)_m + (1,2,2)_p \\[6pt]
 &=&  sp(2) + su(2)_V + 4 \otimes 2
\ea
$$
which can be seen in the figure and Table \ref{tab:sp3roots}. This $sp(2)$ subalgebra of $sp(3)$ corresponds to the upper left $2 \times 2$ block of the $3 \times 3$ matrix representation, (\ref{sp3}), spanned by the $T^M_A$, $T^P_A$, and $\ga_a$ generators. A canonical triality automorphism of $sp(3)$ with this decomposition produces three different $sp(2)$ subalgebras.

To describe the split real form, $sp(6,\mathbb{R})$, of the $sp(3)$ Lie algebra, we can use split-quaternions in (\ref{sp3}) instead of quaternions, representing the basis elements as $2 \times 2$ real matrices,
$$
e'_0 = \si_0 \s \s e'_1 = \si_1 \s \s e'_2 = -i \, \si_2 \s \s e'_3 = \si_3
$$
A real Lie algebra can be represented by matrices with complex entries. Entrywise complex conjugation, $K$, changes $i$ to $-i$ and fixes each $e'_a$ above. Within the complexified algebra, its fixed matrices form the usual real realization. After a complex change of basis, the same real Lie algebra is fixed by a different conjugation, called a \emph{real structure}. For the following example we use $\si'=R_v^0K$: first complex-conjugate, then apply the reflection through the unit $e'_0$. The elements fixed by $\si'$ give another realization of $sp(6,\mathbb{R})$, and $SP(6,\mathbb{R})$ below denotes the corresponding real group. The reflection below preserves these fixed elements, so its explicit $i$'s are compatible with a real automorphism.

Automorphisms of this split real Lie algebra are similar to those of compact $sp(3)$; however, an interesting difference can occur. For example, a reflection, $R^3_v$, around the $u=e'_3$ unit split-quaternion vector has $s_u = -1$, and the corresponding reflection element,
$$
g_R = \lb \ba{ccc}  & - i e'_3 & \\[-2pt] i e'_3 &  & \\[-1pt] & & -1 \\[-1pt]  \ea \rb \;\; \in \; SP(6,\mathbb{R}) = SU(3,\mathbb{H}')
$$
produces a real $sp(6,\mathbb{R})$ inner automorphism,
$$
M' = e'_3 P e'_3 \s P' = e'_3 M e'_3 \s V' = V \s v' = e'_3 \os{v} e'_3 \s \ps' = - i e'_3 \os{\ch} \s \ch' = - i \os{\ps} e'_3
$$
which doesn't look like a real automorphism, but is. This will be further discussed in the next section.

\subsection{\texorpdfstring{$f_4$}{f4}}

The triality Lie algebra of the octonions, $f_4$, has a similar structure to that of $sp(3)$:
\beq
\ba{rcl}
A(B,v,\ps,\ch)  &=&
 \ha B^{ab} \ga_{ab} + v^a \ga_a + \ps^a Q^-_a + \ch^a Q^+_a \\[9pt]
  &=&
\lb \!\! \ba{ccc}
B_M & -\os{v} & \ps \\
v & B_P & -\os{\ch} \\
-\os{\ps} & \ch & B_V 
\ea \!\! \rb
\;\; \os{\in} \;\, su(3,\mathbb{O}) \, \sim f_4
\ea
\label{su3O}
\eeq
Here (\ref{su3O}) is a schematic octonionic array, with compositional diagonal entries; its Lie bracket is defined by (\ref{f4L}--\ref{f4oct}). Octonion bi-products act compositionally, with octonions multiplying to their right before they are then multiplied by octonions from their left. The matrix representation of $Cl(0,8)$ vectors, bivectors, and chiral spinors, from compositional octonionic multiplication, is
$$
\scalebox{.95}{$
B = \ha B^{ab} \ga_{ab} = \lb \! \ba{cc} - \ha B^{ab} \bar{\Ga}_a \Ga_b & \\ & \!\!-\ha B^{ab} \Ga_a \bar{\Ga}_b \ea \! \rb
\s \s
v = v^c \ga_c =  \lb \! \ba{cc} & \!\!-v^c \bar{\Ga}_c \\ v^c \Ga_c &  \ea \! \rb
\s \s
\Ps = \lb \! \ba{c} \ps^a Q^-_a \\ \ch^b Q^+_b \ea \! \rb
$}
$$
using the octonion multiplication table, (\ref{M}), and $(\Ga_c)^b{}_a = M_{ca}{}^\os{b}$. The $su(3,\mathbb{O})$-inspired Lie brackets for the $f_4$ basis generators are:\footnotemark[2]
\beq
\ba{rclcrcl}
\lb \ga_{a b}, \ga_{c d} \rb &=& 2 \left\{ n_{a c} \ga_{b d} - n_{a d} \ga_{b c} - n_{b c} \ga_{a d} + n_{b d} \ga_{a c} \right\} \!\!\!\!\!\!\!\!\!\!\!\!\!\!\!\!\!\!\!\!\!\!\!\!\!\!\!\!\!\!\!\!\!\!\!\!\!\!\!\! \!\!\!\! \!\!\!\! \!\!\!\! \!\!\!\! \!\!\!\! \!\!\!\! \!\!\!\! \!\!\!\! \!\!\!\!
& & & & \\[6pt]

\lb \ga_{ab}, \ga_{c} \rb &=& 2 \left\{ - n_{bc} \ga_{a} + n_{ac} \ga_{b} \right\} & \s \s &
\lb \ga_{ a}, \ga_{ b} \rb &=& 2 \, \ga_{ab}  \\[6pt]

\lb \ga_{ab}, Q^-_c \rb &=&  Q^-_d ( - \bar{\Ga}_a \Ga_b )^d{}_c & &
\lb \ga_{c}, Q^-_a \rb &=& Q^+_b ( \Ga_c )^b{}_a \\[6pt]

\lb \ga_{ab}, Q^+_c \rb &=&  Q^+_d ( - \Ga_a \bar{\Ga}_b )^d{}_c & &
\lb \ga_{c}, Q^+_b \rb &=& Q^-_a ( - \bar{\Ga}_c )^a{}_b \\[6pt]

\lb Q^-_a, Q^-_b \rb &=&  \ha \ga_{c d}  ( \pm \bar{\Ga}{}^c \Ga^d )_{ab} & &
\lb Q^-_a, Q^+_b \rb &=&  \ga_{ c}  ( \pm \bar{\Ga}{}^c )_{ab} \\[6pt]

\lb Q^+_a, Q^+_b \rb &=&  \ha \ga_{cd}  ( \pm \Ga^c \bar{\Ga}{}^d )_{ab} & & & &
\ea
\label{f4L}
\eeq
in which $n_{ab} = \de_{ab}$ is used to raise or lower indices, and we have $28$ independent bivector generators, $\ga_{ab}=-\ga_{ba}$, with $a<b$. The choice of ``$+$'' signs in the ``$\pm$'' gives the compact real form, $f_{4(-52)}$, while ``$-$'' gives $f_{4(-20)}$. For the split real form, $f_{4(4)}$, we use the split-octonionic representation of $Cl(4,4)$ and $n'_{ab}=\text{diag}(++++----)$. Here $\ga_0\ga_1\ga_2\ga_3=\text{diag}(n',-n')$; choosing the relative sign of the invariant chiral metrics gives $n'^-=n'^+=n'$. The split brackets then have the same ``$+$'' signs as the compact brackets, with indices raised and lowered using $n'$.
For a canonical triality automorphism to exist, the metrics of the vectors, negative spinors, and positive spinors in $f_{4(-52)}$ or $f_{4(4)}$ must be equal and proportional to the corresponding restrictions of the Killing form.
 
A canonical triality automorphism between vectors and spinors,
$$
t \q : \q  \ga_a \mapsto Q^+_a  \mapsto Q^-_a \mapsto \ga_a
$$
is an automorphism of $f_{4(-52)}$ or $f_{4(4)}$, with corresponding automorphisms of its $so(8)$ or $so(4,4)$ subalgebra,
$$
t \q : \q \ga_{ab} \;\mapsto\; \ga'_{ab} = \ha \ga_{cd} \, t^{cd}{}_{ab}
= \ha \lb \ga'_a, \ga'_b \rb = \ha \lb Q^+_a, Q^+_b \rb
= \ga_{cd} \lp \fr{1}{4} \Ga^c \bar{\Ga}^d \rp_{ab}
$$
and
$$
t^2 \q : \q \ga_{ab} \;\mapsto\; \ga''_{ab} = \ha \ga_{cd} \, t^2{}^{\; cd}{}_{ab} 
= \ha \lb \ga''_a, \ga''_b \rb = \ha \lb Q^-_a, Q^-_b \rb
= \ga_{cd} \lp \fr{1}{4} \bar{\Ga}^c \Ga^d \rp_{ab}
$$
in which $a < b$ labels the independent basis bivectors.

The description of $f_4$ can be made more explicitly octonionic by remembering that $\bar{\Ga}_a \Ga_b$ comes from multiplying to the right by $e_b$ then multiplying to the right by $\os{e}_a$, so each $so(8)$ chiral bivector element corresponds to a sum of compositional octonion multiplications, $B_- \sim - \ha B^{ab} \os{e}_a e_b$. A canonical triality automorphism of $so(8)$ thus corresponds to a transformation of compositional multiplications of octonions,
$$
\big[t(\os{e}_a e_b)\big](x) = \overline{e_a(\os{e}_b\os{x})}
$$
where $x\in\mathbb{O}$ and $t$ acts on multiplication operators. The diagonal entries of (\ref{su3O}) are then:
$$
\ba{rcl}
B_M(x) &=& - \ha B^{ab} \os{e}_a(e_b x) \\[6pt]
B_P(x) &=& [t(B_M)](x) = - \ha B^{ab}\overline{e_a(\os{e}_b\os{x})} \\[6pt]
B_V(x) &=& [t^2(B_M)](x)
\ea
$$
This allows us to write the Lie brackets between $f_{4(-52)}$ elements,
$$
\lb \, A \! \lp B_1, v_1, \ps_1, \ch_1 \rp, A \! \lp B_2, v_2, \ps_2, \ch_2 \rp \, \rb = A \! \lp B_3, v_3, \ps_3, \ch_3 \rp
$$
using compositional octonionic multiplication and triality \cite{Kol18},
\beq
{\ba{rcl}
B_3 &=& B_1 B_2 - B_2 B_1 -  ( \os{v}_1 v_2 - \os{v}_2 v_1 ) -  t^2 ( \os{\ps}_1 \ps_2 - \os{\ps}_2 \ps_1 ) -  t ( \os{\ch}_1 \ch_2 - \os{\ch}_2 \ch_1 ) \\[6pt]
v_3 &=& t^2(B_1) v_2 - t^2(B_2) v_1 - \os{\ch}_1 \os{\ps}_2 + \os{\ch}_2 \os{\ps}_1 \\[6pt]
\ps_3 &=& B_1 \ps_2 - B_2 \ps_1 - \os{v}_1 \os{\ch}_2 + \os{v}_2 \os{\ch}_1 \\[6pt]
\ch_3 &=& t(B_1) \ch_2 - t(B_2) \ch_1 - \os{\ps}_1 \os{v}_2 + \os{\ps}_2 \os{v}_1 \\[0pt]
\ea}
\label{f4oct}
\eeq
in which the bivectors here are pairs of octonions multiplying to the right in order, such as $B_1 \ps_2 = - \ha B_1^{ab} \ps_2^c \, \os{e}_a ( e_b e_c)$.

Generalized reflection symmetries of $f_{4(-52)}$, $f_{4(-20)}$, or $f_{4(4)}$, through a space-like or time-like unit vector, $u$, are real Lie algebra automorphisms when they preserve the chosen real structure, and their combinations belong to the corresponding $F_4$ Lie group. For $f_{4(-20)}$, using (\ref{f4L}), the opposite Killing-form signs of vectors and spinors require extra factors of $i$ in reflections exchanging them. These are obtained from the compact reflection maps $R$ by
$$
\widetilde R = D^{-1} R D \s\s D(B,v,\ps,\ch)=(B,v,i\ps,i\ch)
$$
For the compact and split presentations, $F_4$ group elements corresponding to generalized reflections are represented schematically by
$$
R^u_v : \lb \! \ba{ccc} &  \!\! \sqrt{s_u} \os{u} & \\[0pt] \sqrt{s_u} u & & \\[0pt] & & \!\!-1 \\[0pt] \ea \! \rb \s\s
R^u_m  : \lb \! \ba{ccc} &  & \!\! \sqrt{s_u} u \\[0pt]  & \!\!-1 & \\[0pt] \sqrt{s_u} \os{u} \; & & \\[0pt] \ea \! \rb \s\s
R^u_p  :  \lb \! \ba{ccc} -1 & & \\[0pt]  & & \!\! \sqrt{s_u} \os{u} \\[0pt] & \!\! \sqrt{s_u} u & \\[0pt] \ea \! \rb \\[5pt]
$$
Explicitly, in addition to the action of $R^u_v$, $R^u_m$, and $R^u_p$ on $f_4$ vector and chiral spinor basis generators, (\ref{genref}), these each extend to corresponding actions on the $so(8)$ or $so(4,4)$ subalgebra basis generators,
$$
\ba{rcl}
R^u_v &\;:\;& \ga_{ab}  \;\mapsto\;  \ga'_{ab} = \ha \lb R^u_v \ga_a, R^u_v \ga_b \rb
= -s_u u \ga_{ab} u = (\de^c_a - 2 s_u u^c u_a)(\de^d_b - 2 s_u u^d u_b) \ga_{cd} \\[8pt]
R^u_m  &\;:\;&  \ga_{ab}  \;\mapsto\;  \ga'_{ab} = \ha \lb R^u_m \ga_a, R^u_m \ga_b \rb
= \fr{1}{4} s_u u^e (\bar{\Ga}_e)^c{}_a u^f (\bar{\Ga}_f)^d{}_b ( \Ga^g \bar{\Ga}^h)_{cd} \ga_{gh} \\[8pt]
R^u_p  &\;:\;&  \ga_{ab}  \;\mapsto\;  \ga'_{ab} = \ha \lb R^u_p \ga_a, R^u_p \ga_b \rb
= \fr{1}{4} s_u u^e (\Ga_e)^c{}_a u^f (\Ga_f)^d{}_b ( \bar{\Ga}^g \Ga^h)_{cd} \ga_{gh} \\[0pt]
\ea
$$
producing automorphisms of the $f_4$ Lie algebra described via these maps of basis generators.

For triality automorphisms, from (\ref{gentri}), using unit vectors $u$ and $w$, we have $t^{uw} : A (\ga_c,Q^-_a,Q^+_b,\ga_{ab}) \mapsto A (\ga'_c,{Q^-_a}',{Q^+_b}',\ga'_{ab})$, with
$$
\ba{rcl}
\ga'_c &=& (\de^a_c - 2 s_u u^a u_c)(\de^b_a - 2 s_w w^b w_a) Q^+_b \\[8pt]
{Q^-_a}' &=& \sqrt{s_u}\sqrt{s_w} w^d (\bar{\Ga}_d)^f{}_b u^c (\Ga_c)^b{}_a \ga_f \\[8pt]
{Q^+_b}' &=& \sqrt{s_u}\sqrt{s_w} w^d (\Ga_d)^f{}_a u^c (\bar{\Ga}_c)^a{}_b Q^-_f \\[8pt]
\ga'_{ab} &=& \ha \lb \ga'_a, \ga'_b \rb \\[6pt]
&=& \fr{1}{4} (\de^c_a - 2 s_u u^c u_a)(\de^k_c - 2 s_w w^k w_c) \\[4pt]
& & \q \times (\de^d_b - 2 s_u u^d u_b)(\de^m_d - 2 s_w w^m w_d)(\Ga^g \bar{\Ga}^h)_{km}\ga_{gh}
\ea
$$
In the same compositional notation, this generalized triality element is
$$
t^{uw} = R^w_p R^u_v R^\os{u}_m R^\os{u}_p  \;\; : \;\; 
\lb \ba{ccc} & & 1 \\[-1pt] \sqrt{s_u}\sqrt{s_w} \os{w} u & & \\[-1pt] & \sqrt{s_u}\sqrt{s_w} w \os{u} & \\[0pt] \ea \rb
$$
with it again understood that this compositional multiplication operation is ordered to act right-first --- for example, $w \os{u} \ps = w (\os{u} \ps)$.

Real automorphisms can have explicit $i$'s relative to a different real structure. For example, the usual anti-linear conjugation $\si=K$ fixes $f_{4(4)}$ inside complex $f_4$. The real involution $R_v^0$ gives another conjugation, $\si'=R_v^0K$, whose fixed algebra also has Killing signature $(28,24)$ and is therefore another realization of $f_{4(4)}$. For a unit time-like $u$ orthogonal to $\ga_0$, the reflection $\ph=R_v^u$ contains $\sqrt{s_u}=i$ but satisfies
$$
\ph\,\si'=R_v^uR_v^0K=R_v^0KR_v^u=\si'\,\ph.
$$
Thus $\ph$ is real with respect to $\si'$. This commutation need not hold for general time-like $u$; other generalized reflections and triality automorphisms likewise require a compatible real structure.

A canonical triality automorphism matrix, $t$, for $so(8)$ or $so(4,4)$ is a $28 \times 28$ rotation matrix, satisfying $t t^T =1$, of real coefficients,
$$
t^{cd}{}_{ab} = \lp \ha \Ga^c \bar{\Ga}^d \rp_{ab}
$$
derived from the octonionic multiplication table, with the $28$ basis bivectors, $\ga_\al = \ga_{ab}$, indexed by $1 \le \al = \ha (13 a - a^2) + b \le 28$, with $0 \le a < b \le 7$. (A redundant $64\times64$ array follows by antisymmetry, with zero diagonal-index entries.) The $28$ basis bivectors can be separated into $7$ disjoint sets of $4$ intra-commuting basis bivectors, each set spanning a Cartan subalgebra intra-rotated by triality. The triality automorphism matrix, $t_\al{}^\be$, can thus be re-ordered to be block diagonal, consisting of seven $4 \times 4$ Hadamard matrices. Typical blocks look like:
$$
t_1 = \scalebox{.9}{$ \lb \ba{cccc}
+\nha &\, -\nha \,&\, +\nha &\, +\nha \\
+\nha & -\nha & -\nha & -\nha \\
+\nha & +\nha & +\nha & -\nha \\
+\nha & +\nha & -\nha & +\nha 
\ea  \rb $}
\s \s
t_2 = \scalebox{.9}{$  \lb \ba{cccc}
-\nha &\, +\nha &\, +\nha &\, +\nha \\
-\nha & +\nha & -\nha & -\nha \\
-\nha & -\nha & +\nha & -\nha \\
-\nha & -\nha & -\nha & +\nha 
\ea  \rb $}
$$
but signs may vary, based on the octonionic multiplication table, or if we use a non-canonical triality automorphism.

For $so(4,4)$ triality automorphisms, the four bivectors, $\ga_{ab}$, spanning each Cartan subalgebra rotated by triality can be constructed from Clifford basis vectors with space-space, time-time, or space-time signature. The triality-allowed signature sets for $so(4,4)$ Cartan bivectors are $\{ ss, ss, tt, tt \}$ or $\{ st, st, st, st \}$. If we choose one of these triality-adapted Cartan subalgebras for our $f_4$ Cartan-Weyl decomposition, then $t$ acts on coefficient columns in this Cartan basis and on root-coordinate columns; $t^- = t^2 = t^T$ gives the inverse action.

There are two especially interesting $f_4$ Cartan subalgebra transformations we can do that emphasize the $sp(3)$ and $su(3)$ subalgebras of $f_4$, matching their previously described triality automorphisms:
$$
\ba{c}
c_1 =
\scalebox{.9}{$  \lb  \ba{cccc}
\nfr{1}{\sqrt{2}} & \nfr{-1}{\sqrt{2}} &  &  \\
\nfr{1}{\sqrt{2}} & \nfr{1}{\sqrt{2}} &  &  \\
 &  & \nfr{1}{\sqrt{2}} & \nfr{1}{\sqrt{2}} \\
 &  & \nfr{-1}{\sqrt{2}} & \nfr{1}{\sqrt{2}} 
\ea \rb $}
\s\s
c_2 =
\scalebox{.9}{$  \lb \ba{cccc}
1 & 0 & 0 & 0 \\
0 &\, \nfr{-1}{\sqrt{3}} & \nfr{-1}{\sqrt{3}} &\, \nfr{-1}{\sqrt{3}} \\
0 & \nfr{-1}{\sqrt{2}} & \nfr{1}{\sqrt{2}} & 0 \\
0 & \nfr{-1}{\sqrt{6}} & \nfr{-1}{\sqrt{6}} & \nfr{\sqrt{2}}{\sqrt{3}}
\ea \rb $}
\\[30pt]
t'_1 = c_1 \, t_1 \, c_1^- =
\scalebox{.9}{$ \lb  \ba{cccc}
 &  & 1 &  \\
1  &  & &  \\[-2pt]
 & \;1 &  &  \\[-4pt]
 &  &  & 1 \\[-1pt]
\ea \rb $}
\s\s
t'_2 = c_2 \, t_2 \, c_2^- =
\scalebox{.9}{$ \lb  \ba{cccc}
\nfr{-1}{2} & \nfr{-\sqrt{3}}{2}  &\;  &\;  \\
\nfr{\sqrt{3}}{2} & \nfr{-1}{2}  &  &  \\[-2pt]
 & & 1 &  \\[-2pt]
 &  &  & 1 \\[-0.5pt]
\ea  \rb $}
\ea
$$
From $t'_1$ we see that triality cycles three $su(2)'s$ in $f_4$, while leaving a complementary $su(2)$ invariant, matching $sp(3)$ triality; while from $t'_2$ we see that triality rotates in a single plane in $4$-dimensional root space, matching $su(3)$ triality, while leaving a complementary $su(3)$ invariant. The corresponding decompositions of $f_{4(-52)}$ are:
\beq
\ba{rcl}
 f_{4(-52)} &=& so(8) + 8_v + 8_{s-} + 8_{s+} \\[8pt]
      &=& su(2)_{L} + su(2)_{R} + su(2)_{W'} + su(2)_{W}  + (2,2,2,2) \\[6pt]
      & & + \lp (2,2,1,1) + (1,1,2,2) \rp_v + \lp (2,1,2,1) + (1,2,1,2) \rp_{s-} \\[6pt]
      & & + \lp (1,2,2,1) + (2,1,1,2) \rp_{s+}
\ea
\label{f41}
\eeq
and
\beq
\ba{rcl}
f_{4(-52)} &=& so(8) + 8_v + 8_{s-} + 8_{s+} \\[8pt]
      &=& u(1)_p + u(1)_B + su(3)_g + 3_{I} + \bar{3}_{I} + 3_{II} + \bar{3}_{II} + 3_{III} + \bar{3}_{III}  \\[8pt]
      & & + \lp 1 + \bar{1} + 3 + \bar{3} \rp_v + \lp 1 + \bar{1} + 3 + \bar{3} \rp_{s-} + \lp 1 + \bar{1} + 3 + \bar{3} \rp_{s+}
\ea
\label{f42}
\eeq
The $f_4$ roots matching decompositions (\ref{f41}) and (\ref{f42}) are shown in Table \ref{tab:f41roots} and Table \ref{tab:f42roots}.

\begin{table}[h!t]
\vspace{6pt}
{\centerline{\parbox{.45\textwidth}{\centerline{
\!\!\!\! \!\!\!\! \!\!\!\! \!\!\!\! \!\!\!\! \!\!\!\! \!\!\!\! \!\!\!\! \!\!\!\!
\renewcommand{\arraystretch}{1.2} 
\scalebox{0.70}{
\begin{tabular}
{@{\vrule width1.0pt}c@{\vrule width0.2pt}c@{\vrule width1.0pt}c@{\vrule width0.0pt}c@{\vrule width0.0pt}c@{\vrule width0.0pt}c@{\vrule width1.0pt}}
\noalign{\hrule height 1.0pt}
\multicolumn{2}{@{\vrule width1.0pt}c@{\vrule width1.0pt}}{$\;\; f_4 \;\;$} & $\q\; \om_T \;\;\;$ {\vrule width0.2pt} & $\q \om_s \;\;\;$ {\vrule width0.2pt} & $\q U \;\;\;$ {\vrule width0.2pt} & $\q V \q$ \\
\noalign{\hrule height 1.0pt}

 \raisebox{-1pt}{\gcirs{lgray}{2}}  & $\;\; \om_{L}^{\wedge/\vee} \;\;$ & $\; \mp \;$ & $\; \pm \;$ & $\; 0 \;$ & $\; 0 \;$ \\
\noalign{\hrule height 0.2pt}
 \raisebox{-.5pt}{\gcirs{lgray}{1.7}}  & $\;\; \om_{R}^{\wedge/\vee} \;\;$ & $\; \pm \;$ & $\; \pm \;$ & $\; 0 \;$ & $\; 0 \;$ \\
\noalign{\hrule height 0.2pt}
 \raisebox{0pt}{\gcirs{lgray}{1.5}}  & $\;\; W'{}^{\pm} \;\;$ & $\; 0 \;$ & $\; 0 \;$ & $\; \pm \;$ & $\; \pm \;$ \\
\noalign{\hrule height 0.2pt}
\raisebox{-2pt}{\gcirs{lyell}{2}}  & $\;\; W^{\pm} \;\;$ & $\; 0 \;$ & $\; 0 \;$ & $\; \mp \;$ & $\; \pm \;$ \\
\noalign{\hrule height 0.2pt}

 \raisebox{-1pt}{\gsqus{lgray}{2} \gdias{lgray}{2}} & $\;\; e_T^{\wedge/\vee} \ph_{I}^{/*}  \;\;$ & $\; \pm \;$ & $\; 0 \;$ & $\; \mp \;$ & $\; 0 \;$ \\
\noalign{\hrule height 0.2pt}
\gsqus{lgray}{1.7} \gdias{lgray}{1.7} & $\;\; e_T^{\wedge/\vee} \ph_{II}^{/*}  \;\;$ & $\; 0 \;$ & $\; \pm \;$ & $\; 0 \;$ & $\; \pm \;$ \\
\noalign{\hrule height 0.2pt}
\gsqus{lgray}{1.5} \gdias{lgray}{1.5} & $\;\; e_T^{\wedge/\vee} \ph_{III}^{/*}  \;\;$ & $\; 0 \;$ & $\; \mp \;$ & $\; 0 \;$ & $\; \pm \;$ \\
\noalign{\hrule height 0.2pt}

 \raisebox{-1pt}{\gdias{lgray}{2} \gsqus{lgray}{2}} & $\;\; e_s^{\wedge/\vee} \ph_{I}^{*/}  \;\;$ & $\; 0 \;$ & $\; \pm \;$ & $\; \pm \;$ & $\; 0 \;$ \\
\noalign{\hrule height 0.2pt}
\gdias{lgray}{1.7} \gsqus{lgray}{1.7} & $\;\; e_s^{\wedge/\vee} \ph_{II}^{*/}  \;\;$ & $\; 0 \;$ & $\; \mp \;$ & $\; \pm \;$ & $\; 0 \;$ \\
\noalign{\hrule height 0.2pt}
\gdias{lgray}{1.5} \gsqus{lgray}{1.5} & $\;\; e_s^{\wedge/\vee} \ph_{III}^{*/}  \;\;$ & $\; \pm \;$ & $\; 0 \;$ & $\; 0 \;$ & $\; \mp \;$ \\
\noalign{\hrule height 0.2pt}

\raisebox{-1pt}{\gdias{lgray}{2} \gsqus{lgray}{2}} & $\;\; e_T^{\wedge/\vee} \ph^{*/}  \;\;$ & $\; \pm \;$ & $\; 0 \;$ & $\; \pm \;$ & $\; 0 \;$ \\
\noalign{\hrule height 0.2pt}
\raisebox{-1pt}{\gsqus{lyell}{2} \gdias{lyell}{2}} & $\;\; e_T^{\wedge/\vee} \ph_{\pm}  \;\;$ & $\; \pm \;$ & $\; 0 \;$ & $\; 0 \;$ & $\; \pm \;$ \\
\noalign{\hrule height .8pt}

$\;$ \raisebox{-.5pt}{\gplus{lgray}{2} \gplds{lgray}{2}} $\;$ & $\; \nu_{eL}^{\wedge/\vee} \;$ & $ \mp \nha \;$ & $ \, \pm \nha \;$ & $ - \nha \,\;$ & $ + \nha \,\,$ \\
\noalign{\hrule height 0.2pt}
\raisebox{-.5pt}{\gprus{lgray}{2} \gprds{lgray}{2}}  & $\; \nu_{eR}^{\wedge/\vee} \;$ & $ \pm \nha \;$ & $ \, \pm \nha \;$ & $ + \nha \,\;$ & $ + \nha \,\,$ \\
\noalign{\hrule height 0.2pt}
\raisebox{-.5pt}{\gplus{myell}{2} \gplds{myell}{2}}  & $\; e_L^{\wedge/\vee} \;$ & $ \mp \nha \;$ & $ \, \pm \nha \;$ & $ + \nha \,\;$ & $ - \nha \,\,$ \\
\noalign{\hrule height 0.2pt}
\raisebox{-.5pt}{\gprus{myell}{2} \gprds{myell}{2}}  & $\; e_R^{\wedge/\vee} \;$ & $ \pm \nha \;$ & $ \, \pm \nha \;$ & $ - \nha \,\;$ & $ - \nha \,\,$ \\
\noalign{\hrule height .8pt}

\raisebox{-.5pt}{\gplus{lgray}{1.7} \gplds{lgray}{1.7}}  & $\; \nu_{\mu L}^{\wedge/\vee} \;$ & $ \mp \nha \;$ & $ \, \mp \nha \;$ & $ - \nha \,\;$ & $ + \nha \,\,$ \\
\noalign{\hrule height 0.2pt}
\raisebox{-.5pt}{\gprus{lgray}{1.7} \gprds{lgray}{1.7}}  & $\; \nu_{\mu R}^{\wedge/\vee} \;$ & $ + \nha \;$ & $ \, - \nha \;$ & $ \pm \nha \,\;$ & $ \pm \nha \,\,$ \\
\noalign{\hrule height 0.2pt}
\raisebox{-.5pt}{\gplus{myell}{1.7} \gplds{myell}{1.7}}  & $\; \mu_L^{\wedge/\vee} \;$ & $ \mp \nha \;$ & $ \, \mp \nha \;$ & $ + \nha \,\;$ & $ - \nha \,\,$ \\
\noalign{\hrule height 0.2pt}
\raisebox{-.5pt}{\gprus{myell}{1.7} \gprds{myell}{1.7}}  & $\; \mu_R^{\wedge/\vee} \;$ & $ - \nha \;$ & $ \, + \nha \;$ & $ \pm \nha \,\;$ & $ \pm \nha \,\,$ \\
\noalign{\hrule height .8pt}

\raisebox{0pt}{\gplus{lgray}{1.5} \gplds{lyell}{1.5}}  & $\; \nu_{\tau L}^{\wedge} \;\; \tau_L^{\vee} \;$ & $\; 0 \;$ & $\; 0 \;$ & $\; \mp \;$ & $\; 0 \;$ \\
\noalign{\hrule height 0.2pt}
\raisebox{0pt}{\gplds{lgray}{1.5} \gplus{lyell}{1.5}}  & $\; \nu_{\tau L}^{\vee} \;\; \tau_L^{\wedge} \;$ & $\; 0 \;$ & $\; 0 \;$ & $\; 0 \;$ & $\; \pm \;$ \\
\noalign{\hrule height 0.2pt}
\raisebox{0pt}{\gprus{lgray}{1.5} \gprds{lyell}{1.5}}  & $\; \nu_{\tau R}^{\wedge} \;\; \tau_R^{\vee} \;$ & $\; \pm \;$ & $\; 0 \;$ & $\; 0 \;$ & $\; 0 \;$ \\
\noalign{\hrule height 0.2pt}
\raisebox{0pt}{\gprds{lgray}{1.5} \gprus{lyell}{1.5}}  & $\; \nu_{\tau R}^{\vee} \;\; \tau_R^{\wedge} \;$ & $\; 0 \;$ & $\; \pm \;$ & $\; 0 \;$ & $\; 0 \;$ \\
\noalign{\hrule height 1.0pt}
\end{tabular}
}
		}}$\;\;$
		\parbox{.45\textwidth}{\centerline{
		\includegraphics[height=3.0in]{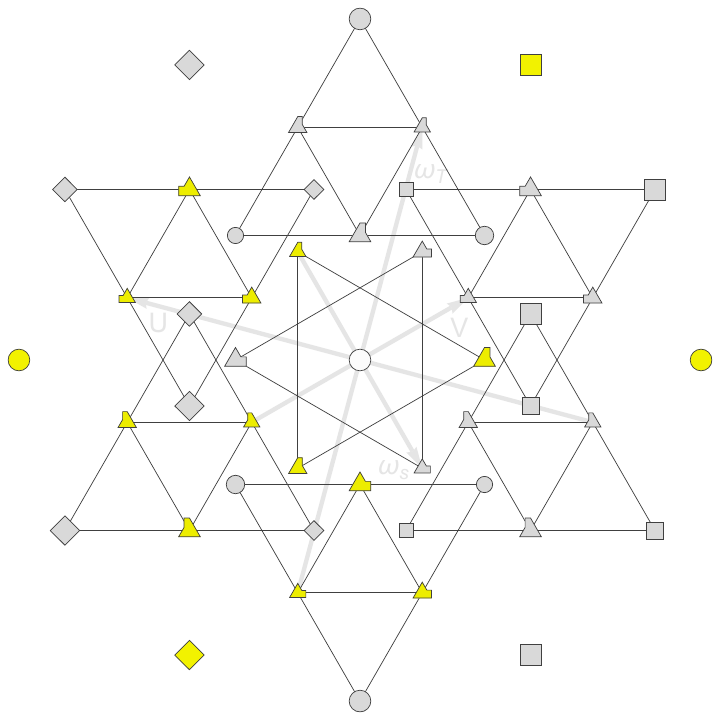}
			}}}}
\vspace{8pt}
\caption{The $48$ roots of $f_4$, labeled suggestively as elementary particles, matching the branching decomposition of (\ref{f41}). In this table $\om_T$ and $\om_s$ are Euclidean boost and spin quantum numbers, while $U$ and $V$ combine to give weak and weaker charge, $W = \ha(-U +V)$ and $W' = \ha(U + V)$ --- see (\ref{charges}). The gravitational spin connection and weaker boson roots, \scalebox{.9}{$\om_{L/R}^{\wedge/\vee}$} and $W'{}^\pm$, correspond to $su(2)$'s related by triality, while the weak boson roots, $W^\pm$, correspond to a triality invariant $su(2)_W$. The \scalebox{.9}{$e^{\wedge/\vee}_{s/T} \ph$} are gravitational frame-Higgs boson roots, related by triality or triality invariant. The first generation lepton roots, \scalebox{.9}{$e^{\wedge/\vee}_{L/R}$} and \scalebox{.9}{$\nu^{\wedge/\vee}_{ eL/R}$}, with spin up or down and left or right chirality, correspond to a $8_{s+}$ of $so(8)$ and relate to the second and third generation leptons by triality. The spin and charges of these second and third generation leptons are not correct when considered independently of their triality relationship to the first.} 
\label{tab:f41roots}
\end{table}

\clearpage
\begin{table}[h!t]
{\centerline{\parbox{.45\textwidth}{\centerline{
\renewcommand{\arraystretch}{1.1}
\!\!\!\! \!\!\!\! \!\!\!\! \!\!\!\! \!\!\!\! \!\!\!\! \!\!\!\! \!\!\!\! \!\!\!\! 
\renewcommand{\arraystretch}{1.2} 
\scalebox{0.70}{
\begin{tabular}
{@{\vrule width1.0pt}c@{\vrule width0.2pt}c@{\vrule width1.0pt}c@{\vrule width0.0pt}c@{\vrule width0.0pt}c@{\vrule width0.0pt}c@{\vrule width1.0pt}}
\noalign{\hrule height 1.0pt}
\multicolumn{2}{@{\vrule width1.0pt}c@{\vrule width1.0pt}}{$\;\; f_4 \;\;$} & $\;\;\;\;\; p \;\;\;\,$ {\vrule width0.2pt} & $\;\;\;\;\; x \;\;\;\,$ {\vrule width0.2pt} & $\;\;\;\;\; y \;\;\;\,$ {\vrule width0.2pt} & $\;\;\;\;\; z \;\;\;\;\;$  \\
\noalign{\hrule height 1.0pt}

 \raisebox{-2pt}{\gcirs{lblue}{2}}  & \raisebox{1pt}{$\;\; g \;\;$} & $ 0 $ & $\! ( +1 \;$ & $ -1 \;$ & $\;\;\; 0 \; ) $ \\
\noalign{\hrule height 0.2pt}
$\;\!$ \raisebox{-2pt}{\gcirs{mrora}{2} \gcirs{mygre}{2} \gcirs{mbvio}{2}} $\;\!$ & $\;\, X_{I}^{rgb} \; \bar{X}_{I}^{rgb} \,\;$ & $ 0 $ & $\! ( \pm 1 \;$ & $ \pm 1 \;$ & $\;\;\; 0 \; ) $ \\
\noalign{\hrule height 0.2pt}
 \raisebox{-1pt}{\gcirs{mrora}{1.7} $\!$ \gcirs{mygre}{1.7} $\!$ \gcirs{mbvio}{1.7}}  & $ X_{II}^{rgb} \; \bar{X}_{II}^{rgb} $ & $ \pm 1 $ & $\! ( \mp 1 \;$ & $ 0 $ & $\;\;\; 0 \; )$ \\
\noalign{\hrule height 0.2pt}
 \gcirs{mrora}{1.5} $\!$ \gcirs{mygre}{1.5} $\!$ \gcirs{mbvio}{1.7}  & $\; X_{III}^{rgb} \; \bar{X}_{III}^{rgb} \;$ & $ \mp 1 $ & $\! ( \mp 1 \;$ & $ 0 $ & $\;\;\; 0 \; )$ \\
\noalign{\hrule height .8pt}

 \raisebox{-1pt}{\gpas{myell}{2}}  & $\; l_I \;$ & $ + \nha \;$ & $\!\! \, + \nha \;$ & $\!\! +\nha \,\;$ & $\!\!\! +\nha \,\,$ \\
\noalign{\hrule height 0.2pt}
 \raisebox{-1pt}{\gaps{myell}{2}}  & $\; \bar{l}_I \;$ & $ -\nha \;$ & $\!\! \, - \nha \;$ & $\!\! -\nha \,\;$ & $\!\!\! -\nha \,\,$ \\
\noalign{\hrule height 0.2pt}
 \raisebox{-1pt}{\gpas{mred}{2} \gpas{mgree}{2} \gpas{mblue}{2}}  & $\; q_I^{(rgb)} \;$ & $ + \nha \;$ & $\!\!\! ( + \nha \;$ & $\!\! -\nha \,\;$ & $\!\! -\nha \, )$ \\
\noalign{\hrule height 0.2pt}
 \raisebox{-1pt}{\gaps{mred}{2} \gaps{mgree}{2} \gaps{mblue}{2}}  & $\; \bar{q}_I^{(rgb)} \;$ & $ - \nha \;$ & $\!\!\! ( - \nha \;$ & $\!\! +\nha \,\;$ & $\!\! +\nha \, )$ \\
\noalign{\hrule height .8pt}

 \gpas{myell}{1.7}  & $\; l_{II} \;$ & $ + \nha \;$ & $\!\! \, - \nha \;$ & $\!\!\! -\nha \,\;$ & $\!\!\! -\nha \,\,$ \\
\noalign{\hrule height 0.2pt}
 \raisebox{-1pt}{\gaps{myell}{1.7}}  & $\; \bar{l}_{II} \;$ & $ -\nha \;$ & $\!\! \, + \nha \;$ & $\!\!\! +\nha \,\;$ & $\!\!\! +\nha \,\,$ \\
\noalign{\hrule height 0.2pt}
 \gpas{mred}{1.7} $\!$ \gpas{mgree}{1.7} $\!$ \gpas{mblue}{1.7}  & $\; q_{II}^{(rgb)} \;$ & $ - \nha \;$ & $\!\!\! ( + \nha \;$ & $\!\! -\nha \,\;$ & $\!\! -\nha \, )$ \\
\noalign{\hrule height 0.2pt}
 \gaps{mred}{1.7} $\!$ \gaps{mgree}{1.7} $\!$ \gaps{mblue}{1.7}  & $\; \bar{q}_{II}^{(rgb)} \;$ & $ + \nha \;$ & $\!\!\! ( - \nha \;$ & $\!\! +\nha \,\;$ & $\!\! +\nha \, )$ \\
\noalign{\hrule height .8pt}

 \gpas{myell}{1.5}  & $\; l_{III} \;$ & $ -1 $ & $ 0 $ & $ 0 $ & $\, 0 $ \\
\noalign{\hrule height 0.2pt}
 \gaps{myell}{1.5} & $\; \bar{l}_{III} \;$ & $ +1 $ & $ 0 $ & $  0 $ & $\, 0 $ \\
\noalign{\hrule height 0.2pt}
 \gpas{mred}{1.5} $\!$ \gpas{mgree}{1.5} $\!$ \gpas{mblue}{1.5}  & $\; q_{III}^{(rgb)} \;$ & $ 0 $ & $\! ( + 1 \;$ & $ 0 $ & $\;\;\; 0 \; )$ \\
\noalign{\hrule height 0.2pt}
 \gaps{mred}{1.5} $\!$ \gaps{mgree}{1.5} $\!$ \gaps{mblue}{1.5}  & $\; \bar{q}_{III}^{(rgb)} \;$ & $ 0 $ & $\! ( -1 \;$ & $ 0 $ & $\;\;\; 0 \; )$ \\
\noalign{\hrule height 1.0pt}
\end{tabular}
}
		}}$\;\;$
		\parbox{.45\textwidth}{\centerline{
		\includegraphics[height=3.0in]{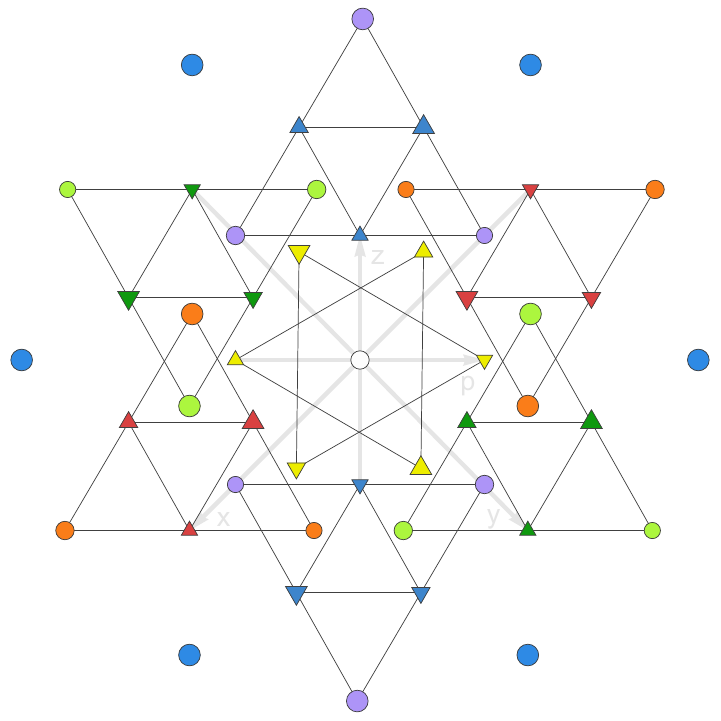} 
			}}}}
\vspace{8pt}
\caption{The $48$ roots of $f_4$, labeled suggestively as elementary particles, showing the triality invariant $su(3)_g$ and triality-related roots, as per (\ref{f42}). Coordinates in parenthesis are permuted over specified columns. The $g$ roots are triality invariant gluons and the $X$ are strongly charged (colored) GUT X-bosons related by triality. The leptons and anti-leptons, $l_I$ and $\bar{l}_I$, and colored quarks and anti-quarks, $q_I$ and $\bar{q}_I$, are in an $8_{s+}$ and relate to two other generations by triality. The $(x, y, z)$ charges correspond to color and $B$ charge, (\ref{charges}), while $p$ indicates particles vs antiparticles.}
\vspace{-8pt}
\label{tab:f42roots}
\end{table}

It is interesting to perform an eigenspace decomposition of $f_{4(-52)}$ with respect to its canonical triality automorphism. This can be done computationally, or via sufficient staring at the figure in Table \ref{tab:f42roots}. After complexification, the $f_4$ decomposition in (\ref{f42}) recombines into three triality eigenspaces,
\beq
\scalebox{.95}{$
\ba{rcl}
\;\, f_{4(-52)} &=& so(8) + 8_v + 8_{s-} + 8_{s+} \\[8pt]
      &=& u(1)_p + u(1)_B + su(3)_g + 3_{I} + \bar{3}_{I} + 3_{II} + \bar{3}_{II} + 3_{III} + \bar{3}_{III}  \\[6pt]
      & & + \lp 1 + \bar{1} + 3 + \bar{3} \rp_v + \lp 1 + \bar{1} + 3 + \bar{3} \rp_{s-} + \lp 1 + \bar{1} + 3 + \bar{3} \rp_{s+} \\[8pt]
      &=& \lb su(3)_g + 3 + \bar{3} + 1 + \bar{1} + 3 + \bar{3}  \rb_{0} + \lb u(1) + 3 + \bar{3} + 1 + \bar{1} + 3 + \bar{3} \rb_{\pm 1} \\[8pt] 
      &=& \Big[ so(7) + u(1)_w \Big]_0 +  \Big[ 7^v_{\pm 1} + 8^s_{\mp \nha} \Big]_{\pm 1}\\[8pt]
      &=& so(9) + 16^s \\
\ea$}
\label{f4t}
\eeq
with eigenvalues $e^{\nfr{2 \pi i k}{3}}$, in which $k \in \{0,+1,-1\}$. Algebraically, the $so(8)$ contains a triality invariant $su(3)$ and the triality invariant average of three $(3 + \bar{3})$'s, which combine into an invariant $g_2$ subalgebra of the $so(8)$. Similarly, the three triality-related $8$'s produce a triality invariant average $8$, which is a $7+1$ under the invariant $g_2$. These combine to give $\lb so(7) + u(1)_w \rb$ as the triality invariant subalgebra of $f_4$. The remaining representation spaces split into $k = \pm 1$ triality eigenspaces, each including a vector, $7^v$, and spinor, $8^s$, of the $so(7)$, with $w$ charges of $\pm 1$ and $\mp \nha$. The invariant $\lb so(7) + u(1)_w \rb$ lives inside a $so(9)$ along with the $7^v_{\pm 1}$, which are intra-rotated by triality. Triality acts on the complex conjugate representation spaces, $7^v_{+1}$ and $7^v_{-1}$, as multiplication by $e^{\nfr{\pm 2 \pi i}{3}}$. This corresponds to a rotation of real Lie algebra generators by $\nfr{\pm 2 \pi }{3}$ in $7$ independent planes spanned by the orthogonal ``real'', $7_v^\mathbb{R}=\ha(7^v_{+1} + 7^v_{-1})$, and ``imaginary'', $7_v^\mathbb{I}=\fr{1}{2 i}(7^v_{+1} - 7^v_{-1})$, generators within $so(9)$ --- mapping between three sets of non-orthogonal $7^v$ vectors. Similarly, triality acts on the complex $8^s_{\mp \nha}$ representation spaces as multiplication by $e^{\nfr{\pm 2 \pi i}{3}}$, corresponding to a rotation of $\nfr{\pm 2 \pi }{3}$ in $8$ independent planes spanned by the orthogonal real, $8_s^\mathbb{R}=\ha(8^s_{-\nha} + 8^s_{+\nha})$, and imaginary, $8_s^\mathbb{I}=\fr{1}{2 i}(8^s_{-\nha} - 8^s_{+\nha})$, generators  --- mapping between three sets of non-orthogonal $8^s$ spinor generators within $16^s$.

For a compact Lie algebra, such as $f_{4(-52)}$, the complex conjugate pairs of root vectors, and conjugate pairs of triality eigenvectors, correspond to sign-conjugate pairs of roots. The $f_{4(-52)}$ roots matching the triality eigenspace decomposition, (\ref{f4t}), are shown in Table \ref{tab:f4troots}. For a non-compact Lie algebra, such as split real $f_{4(4)}$, triality-automorphisms among root vectors can get more complicated.

\begin{table}[h!t]
\vspace{12pt}
{\centerline{\parbox{.45\textwidth}{\centerline{
\renewcommand{\arraystretch}{1.2}
\!\!\!\! \!\!\!\! \!\!\!\! \!\!\!\! \!\!\!\! \!\!\!\! \!\!\!\! \!\!\!\! \!\!\!\! 
\scalebox{0.70}{
\begin{tabular}
{@{\vrule width1.0pt}c@{\vrule width0.2pt}c@{\vrule width1.0pt}c@{\vrule width0.0pt}c@{\vrule width0.0pt}c@{\vrule width0.0pt}c@{\vrule width1.0pt}}
\noalign{\hrule height 1.0pt}
\multicolumn{2}{@{\vrule width1.0pt}c@{\vrule width1.0pt}}{$\;\; f_4 \;\;$} & $\;\;\;\;\; w \;\;\;\,$ {\vrule width0.2pt} & $\;\;\;\;\; x \;\;\;\,$ {\vrule width0.2pt} & $\;\;\;\;\; y \;\;\;\,$ {\vrule width0.2pt} & $\;\;\;\;\; z \;\;\;\;\;$  \\
\noalign{\hrule height 1.0pt}

 \raisebox{-2pt}{\gcirs{lblue}{2}}  & \raisebox{1pt}{$\;\; g \;\;$} & $ 0 $ & $\! ( +1 \;$ & $ -1 \;$ & $\;\;\; 0 \; ) $ \\
\noalign{\hrule height 0.2pt}
$\;\!$ \raisebox{-1pt}{\gcirs{mred}{1.5} \gcirs{mgree}{1.5} \gcirs{mblue}{1.5}} $\;\!$ & $\;\;\; X^{PS} \;\;\;$ & $ 0 $ & $\! ( \pm 1 \;$ & $ \pm 1 \;$ & $\;\;\; 0 \; ) $ \\
\noalign{\hrule height 0.2pt}
 \raisebox{-1pt}{\gcirs{mrora}{1.5}  \gcirs{mygre}{1.5}  \gcirs{mbvio}{1.5}}  & $ X^{GG} $ & $ 0 $ & $\! ( \pm 1 \;$ & $ 0 $ & $\;\;\; 0 \; )$ \\
\noalign{\hrule height 0.8pt}

 \raisebox{-1pt}{\gsqus{mgray}{1.7}}  & $ \!\!\!\!\Ph $ & $ +1 $ & $ 0 $ & $ 0 $ & $ 0 $ \\
 \noalign{\hrule height 0.2pt}
 \raisebox{-1pt}{\gsqus{mrora}{1.7}  \gsqus{mygre}{1.7}  \gsqus{mbvio}{1.7}}  & $ \Ph_+ $ & $ +1 $ & $\! ( + 1 \;$ & $ 0 $ & $\;\;\; 0 \; )$ \\
 \noalign{\hrule height 0.2pt}
 \raisebox{-1pt}{\gdias{mrora}{1.7} $\!\!\!$  \gdias{mygre}{1.7} $\!\!\!$ \gdias{mbvio}{1.7}}  & $ \Ph_- $ & $ +1 $ & $\! ( - 1 \;$ & $ 0 $ & $\;\;\; 0 \; )$ \\
\noalign{\hrule height 0.8pt}

 \raisebox{-.5pt}{\gsqus{mgray}{1.5}}  & $ \Ph^{\,*} $ & $ -1 $ & $ 0 $ & $ 0 $ & $ 0 $ \\
 \noalign{\hrule height 0.2pt}
 \raisebox{-.5pt}{\gdias{mrora}{1.5} $\!\!\!$  \gdias{mygre}{1.5} $\!\!\!$ \gdias{mbvio}{1.5}}  & $ \Ph_+^{\,*} $ & $ -1 $ & $\! ( - 1 \;$ & $ 0 $ & $\;\;\; 0 \; )$ \\
 \noalign{\hrule height 0.2pt}
 \raisebox{-.5pt}{\gsqus{mrora}{1.5}  \gsqus{mygre}{1.5}  \gsqus{mbvio}{1.5}}  & $ \Ph_-^{\,*} $ & $ -1 $ & $\! ( + 1 \;$ & $ 0 $ & $\;\;\; 0 \; )$ \\
\noalign{\hrule height 0.8pt}

 \raisebox{-1pt}{\gpas{myell}{2}}  & $\; l \;$ & $ +\nha \;$ & $\!\! \, + \nha \;$ & $\!\! +\nha \,\;$ & $\!\!\! +\nha \,\,$ \\
\noalign{\hrule height 0.2pt}
 \raisebox{-1pt}{\gaps{myell}{2}}  & $\; \bar{l} \;$ & $ +\nha \;$ & $\!\! \, - \nha \;$ & $\!\! -\nha \,\;$ & $\!\!\! -\nha \,\,$ \\
\noalign{\hrule height 0.2pt}
 \raisebox{-1pt}{\gpas{mred}{2} $\!\!\!$ \gpas{mgree}{2} $\!\!\!$ \gpas{mblue}{2}}  & $\; q \;$ & $ +\nha \;$ & $\!\!\! ( + \nha \;$ & $\!\! -\nha \,\;$ & $\!\! -\nha \, )$ \\
\noalign{\hrule height 0.2pt}
 \raisebox{-1pt}{\gaps{mred}{2} $\!\!\!$ \gaps{mgree}{2} $\!\!\!$ \gaps{mblue}{2}}  & $\; \bar{q} \;$ & $ +\nha \;$ & $\!\!\! ( - \nha \;$ & $\!\! +\nha \,\;$ & $\!\! +\nha \, )$ \\
\noalign{\hrule height .8pt}

 \raisebox{-.5pt}{\gaps{myell}{1.7}}  & $\; l^{\,*} \;$ & $ -\nha \;$ & $\!\! \, -\nha \;$ & $\!\! -\nha \,\;$ & $\!\!\! -\nha \,\,$ \\
\noalign{\hrule height 0.2pt}
 \raisebox{-.5pt}{\gpas{myell}{1.7}}  & $\; \bar{l}^{\,*} \;$ & $ -\nha \;$ & $\!\! \, +\nha \;$ & $\!\! +\nha \,\;$ & $\!\!\! +\nha \,\,$ \\
\noalign{\hrule height 0.2pt}
 \raisebox{-.5pt}{\gaps{mred}{1.7} \gaps{mgree}{1.7} \gaps{mblue}{1.7}}  & $\; q^{\,*} \;$ & $ -\nha \;$ & $\!\!\! ( -\nha \;$ & $\!\! +\nha \,\;$ & $\!\! +\nha \, )$ \\
\noalign{\hrule height 0.2pt}
 \raisebox{-.5pt}{\gpas{mred}{1.7} \gpas{mgree}{1.7} \gpas{mblue}{1.7}}  & $\; \bar{q}^{\,*} \;$ & $ -\nha \;$ & $\!\!\! ( +\nha \;$ & $\!\! -\nha \,\;$ & $\!\! -\nha \, )$ \\
\noalign{\hrule height 1.0pt}
\end{tabular}
}
		}}$\;\;$
		\parbox{.45\textwidth}{\centerline{
		\includegraphics[height=3.0in]{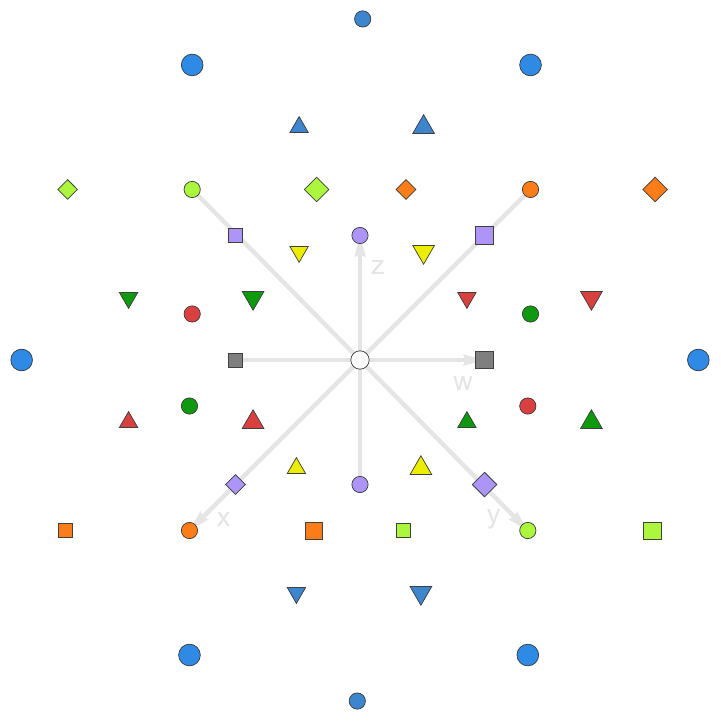} 
			}}}}
\vspace{8pt}
\caption{The $48$ roots of $f_4$, labeled suggestively as elementary particles, sorted into triality eigenspaces, as per (\ref{f4t}). Roots with $w \!\!\in\!\! \{+1,-\nha\}$ and $w \!\!\in\!\! \{-1,+\nha\}$ are triality rotated by $e^{+\nfr{2 \pi i}{3}}$ and $e^{-\nfr{2 \pi i}{3}}$. The $g$ are gluons, $X^{PS}$ Pati-Salam X bosons, $X^{GG}$ Georgi-Glashow X bosons, all invariant under triality. The $\Ph$ are Higgs bosons, and $l$, $\bar{l}$, $q$, $\bar{q}$ are leptons, anti-leptons, quarks, and anti-quarks. Triality multiplies these eigenvectors and their starred conjugates by conjugate phases; complex conjugation exchanges the two eigenspaces.}
\label{tab:f4troots}
\end{table}

\section{Exceptional Magic}

To obtain larger exceptional Lie algebras, we can consider the tensor product of division algebras or their split versions, such as the $32$-dimensional vector space $\mathbb{H} \otimes \mathbb{O}$. From any two division algebras, $\mathbb{D}'$ and $\mathbb{D}$, of dimension $n'$ and $n$ and signature $(p',q')$ and $(p,q)$, we can construct chiral representations of Clifford algebras, $Cl(p'\!+\!p,q'\!+\!q)$ or $Cl(q'\!+\!p,p'\!+\!q)$, in a similar manner to the construction of Clifford division algebra representations. In a \emph{Clifford compound division algebra representation}, $(n'+n)$-dimensional Clifford basis vectors are expressed as:
$$
\vspace{-16pt}
$$
$$
v = v^\al \ga_\al =
\lb \! \ba{cc}
0 & v_- \\
v_+ & 0
\ea \! \rb
\s
v_- =
\lb \! \ba{cc}
1 \!\otimes\! \os{x} & \; \pm \os{y}{}' \!\otimes\! 1 \\[0pt]
y' \!\otimes\! 1 & \; - 1 \!\otimes\! x
\ea \! \rb
\s
v_+ =
\lb \! \ba{cc}
1 \!\otimes\! x & \; \pm \os{y}{}' \!\otimes\! 1 \\[0pt]
y' \!\otimes\! 1 & \; - 1 \!\otimes\! \os{x}
\ea \! \rb
$$
which may be understood as matrices of inter-commuting division algebra elements, $x \in \mathbb{D}$ and $y' \in \mathbb{D}'$, or as $\mathbb{R}(4(n'\!\times\!n))$ matrices via their multiplication coefficients. The result of multiplying two vectors is
$$
\vspace{-18pt}
$$
$$
\ba{c}
u \, v = \lb \! \ba{cc}
u_- v_+ & 0 \\
0 & u_+ v_-
\ea \! \rb
\\[14pt]
u_- v_+ =
\lb \! \ba{cc}
\os{w} & \; \pm \os{z}{}' \\
z' & \; - w
\ea \! \rb
\lb \! \ba{cc}
x & \; \pm \os{y}{}' \\
y' & \; - \os{x}
\ea \! \rb
=
\lb  \ba{cc}
1 \!\otimes\! \os{w} x \pm \os{z}' y' \!\otimes\! 1 
&
\pm \os{y}' \!\otimes\! \os{w} \mp \os{z}' \!\otimes\! \os{x} \\[0pt]
z' \!\otimes\! x -  y' \!\otimes\! w
&
\pm z' \os{y}' \!\otimes\! 1 + 1 \!\otimes\! w \os{x}
\ea  \rb
\ea
$$
from which we see that the result of squaring a Clifford vector is
$$
\vspace{-26pt}
$$
$$
v_- v_+ = v_+ v_- =  \lb \ba{cc}
1 \otimes \os{x} x \pm \os{y}' y' \otimes 1 
&
0 \\
0
&
\pm y' \os{y}' \otimes 1 + 1 \otimes x \os{x}
\ea \rb
$$
and so the represented Clifford algebra has signature $(p'\!+\!p,q'\!+\!q)$ or $(q'\!+\!p,p'\!+\!q)$, depending on the choice of $\pm$. The chiral bivector part of $u v$ is a representative element of a spin Lie algebra, $so(p'\!+\!p,q'\!+\!q)$ or $so(q'\!+\!p,p'\!+\!q)$, which is
$$
\vspace{-26pt}
$$
$$
\ha (u_- v_+ - v_- u_+) \; \in \; 
\lb \ba{cc}
(\mathbb{B}'{}_{\!\! M} \!\otimes\! 1) \oplus (1 \!\otimes\! \mathbb{B}_M) 
& \;
\mathbb{D}' \otimes \mathbb{D} \\[0pt]
\mathbb{D}' \otimes \mathbb{D}
&\;
(\mathbb{B}'{}_{\!\! P} \!\otimes\! 1) \oplus (1 \!\otimes\! \mathbb{B}_P)
\ea \rb
$$
with the direct sum of bi-products on the diagonal, and the tensor product of the division algebras on the off diagonal. Expanding upon the previous description of Lie algebras using $su(3,\mathbb{D})$, we have a family of Lie algebras described heuristically as:
$$
\vspace{-26pt}
$$
$$ 
\lb \ba{ccc}
\mathbb{B}'{}_{\!\! M} \oplus \mathbb{B}_M & \;
\mathbb{D}'{}_{\!\! \os{v}} \otimes \mathbb{D}_\os{v} & \;
\mathbb{D}'{}_{\!\! \ps} \otimes \mathbb{D}_\ps \\
\mathbb{D}'{}_{\!\! v} \otimes \mathbb{D}_v & \;
\mathbb{B}'{}_{\!\! P} \oplus \mathbb{B}_P & \;
\mathbb{D}'{}_{\!\! \os{\ch}} \otimes \mathbb{D}_\os{\ch} \\
\mathbb{D}'{}_{\!\! \os{\ps}} \otimes \mathbb{D}_\os{\ps} & \;
\mathbb{D}'{}_{\!\! \ch} \otimes \mathbb{D}_\ch & \;
\mathbb{B}'{}_{\!\! V} \oplus \mathbb{B}_V \\
\ea \rb
\;\; \sim \;
su(3,\mathbb{D}' \otimes \mathbb{D})
$$
This family is the \emph{magic square of Lie algebras} \cite{Eva09}, shown in Table \ref{table:magic}. Each member of the magic square has a canonical triality automorphism, $t$, constructed from the triality automorphisms of its constituent parts.  Every member also has three $so(p'\!+\!p,q'\!+\!q) \sim su(2,\mathbb{D}' \otimes \mathbb{D})$ subalgebras, related by a canonical triality automorphism. 

\begin{table}[h!t]
\centering
\renewcommand{\arraystretch}{1.0}
\begin{tabular}
{@{\vrule width1.0pt}c@{\vrule width1.0pt}c@{\vrule width0.4pt}c@{\vrule width0.4pt}c@{\vrule width0.4pt}c@{\vrule width0.4pt}c@{\vrule width0.4pt}c@{\vrule width0.4pt}c@{\vrule width1.0pt}}
 \noalign{\hrule height 1.0pt}
\raisebox{2pt}{$ \;\; \mathfrak{g}_{\mathbb{D}' \!\otimes \mathbb{D}} \; $} & $\s \mathbb{R} \s$ & $\s \mathbb{C} \s$ & $\s \mathbb{H} \s$ & $\s \mathbb{O} \s$ & $\s \mathbb{C}' \s$ & $\s \mathbb{H}' \s$ & $\s \mathbb{O}' \s$ \\
\noalign{\hrule height 1.0pt}
$\mathbb{R}$ & $su(2)$ & $su(3)$ & $sp(3)$ & $f_4$ & $sl(3)$ & $sp(6,{\mathbb{R}})$ & $f_{4(4)}$ \\[2pt]
\noalign{\hrule height 0.2pt}
$\mathbb{C}$ & $su(3)$ & $2 \, su(3)$ & $su(6)$ & $e_6$ & $sl(3,{\mathbb{C}})$ & $su(3,3)$ & $e_{6(2)}$ \\[2pt]
\noalign{\hrule height 0.2pt}
$\mathbb{H}$ & $sp(3)$ & $su(6)$ & $so(12)$ & $e_7$ & $sl(3,{\mathbb{H}})$ & $sp(6,{\mathbb{H}})$ & $e_{7(-5)}$ \\[2pt]
\noalign{\hrule height 0.2pt}
$\mathbb{O}$ & $f_4$ & $e_6$ & $e_7$ & $e_8$ & $e_{6(-26)}$ & $e_{7(-25)}$ & $e_{8(-24)}$ \\[2pt]
\noalign{\hrule height 0.2pt}
$\mathbb{C}'$ & $sl(3)$ & $sl(3,{\mathbb{C}})$ & $sl(3,{\mathbb{H}})$ & $e_{6(-26)}$ & $2 sl(3)$ & $sl(6,{\mathbb{R}})$ & $e_{6(6)}$ \\[2pt]
\noalign{\hrule height 0.2pt}
$\mathbb{H}'$ & $sp(6,{\mathbb{R}})$ & $su(3,3)$ & $sp(6,{\mathbb{H}})$ & $e_{7(-25)}$ & $sl(6,{\mathbb{R}})$ & $so(6,6)$ & $e_{7(7)}$ \\[2pt]
\noalign{\hrule height 0.2pt}
$\mathbb{O}'$ & $f_{4(4)}$ & $e_{6(2)}$ & $e_{7(-5)}$ & $e_{8(-24)}$ & $e_{6(6)}$ & $e_{7(7)}$ & $e_{8(8)}$ \\[2pt]
\noalign{\hrule height 1.0pt}
\end{tabular}
\vspace{8pt}
\caption{The magic square Lie algebras, constructed from pairs of division algebras or their split versions.}
\label{table:magic}
\vspace{-8pt}
\end{table}

An easy way to construct the Lie brackets of any member, $\mathfrak{g}_{\mathbb{D}' \!\otimes \mathbb{D}}$, of the magic square is by suitably joining the Lie brackets of its constituent pairing of $su(2)$, $su(3)$, $sp(3)$, $f_4$, $sl(3)$, $sp(6,{\mathbb{R}})$, or $f_{4(4)}$ subalgebras, corresponding to its compound triality decomposition,
$$
\vspace{-22pt}
$$
$$
\mathfrak{g}_{\mathbb{D}' \!\otimes \mathbb{D}} = \mathrm{tri}(\mathbb{D}') \,+\, \mathrm{tri}(\mathbb{D}) \,+\, \mathbb{D}'{}_{\!\! v} \otimes \mathbb{D}_v \,+\, \mathbb{D}'{}_{\!\! m} \otimes \mathbb{D}_m \,+\, \mathbb{D}'{}_{\!\! p} \otimes \mathbb{D}_p 
$$
For the compact real $e_6$, $e_7$, and $e_8$ Lie algebras, their compound triality decompositions are:
$$
\vspace{-20pt}
$$
$$
\ba{rcl}
e_6 &=& u(1) + u(1) + so(8) + (1\!+\!\bar{1})_v \otimes 8_v + (1\!+\!\bar{1})_m \otimes 8_{s-} + (1\!+\!\bar{1})_p \otimes 8_{s+} \\[6pt]
e_7 &=& su(2)_M + su(2)_P + su(2)_V + so(8) + \, (2,2,1)_v \otimes 8_v + (2,1,2)_m \otimes 8_{s-} + (1,2,2)_p \otimes 8_{s+} \\[6pt]
e_8 &=& so(8)' \,+\, so(8) \,+\, 8'_v \otimes 8_v \,+\, 8'_{s-} \otimes 8_{s-} \,+\, 8'_{s+} \otimes 8_{s+} \\[4pt]
\ea
$$

The root system of any magic square Lie algebra may be constructed by suitably joining the roots of the constituent pairing. The triality matrix for its root system is a block diagonal matrix, $t$, constructed from the triality matrices of its constituents, such as
\vspace{-1pt}
\beq
t =
\scalebox{.6}{$
\lb
\ba{cccccccc}
+\nha &\, -\nha &\, +\nha &\, +\nha & 0 & 0 & 0 & 0 \\
+\nha & -\nha & -\nha & -\nha & 0 & 0 & 0 & 0 \\
+\nha & +\nha & +\nha & -\nha & 0 & 0 & 0 & 0 \\
+\nha & +\nha & -\nha & +\nha & 0 & 0 & 0 & 0 \\
0 & 0 & 0 & 0 &\, -\nha &\, +\nha &\, +\nha &\, +\nha \\
0 & 0 & 0 & 0 & -\nha & +\nha & -\nha & -\nha \\
0 & 0 & 0 & 0 & -\nha & -\nha & +\nha & -\nha \\
0 & 0 & 0 & 0 & -\nha & -\nha & -\nha & +\nha \\
\ea
\rb
$}
\vspace{-1pt}
\label{tri}
\eeq
for a real form of $e_8$. Although the roots of $sl(3)$, $sp(6,{\mathbb{R}})$, and $f_{4(4)}$ may have a mixture of imaginary and real components, a canonical triality automorphism only rotates between all compact or between all noncompact Cartan generators, and the triality matrices for these Lie algebras can only inter-mix all real or all imaginary root components. 

The main advantage of having explicit expressions for the structure of a Lie algebra and its triality automorphisms, over the description via roots and a triality matrix, is that the triality matrix alone doesn't determine the phases of the triality maps between root vectors or generators. We could employ tricks to find these phases, but it's usually easier to find them from a direct division algebra description and the corresponding explicit triality automorphism.

\section{Triality Eigenspaces, Vinberg \texorpdfstring{$\Th$}{Theta}-Groups, and Gradings}

For any Lie algebra automorphism, $\Th$, of order $n$, the Lie algebra separates, after complexification, into $n$ eigenspaces (some possibly zero), with eigenvalues $\om^k$, with $\om = e^{\nfr{2 \pi i}{n}}$ and integral $0 \le k \le n-1$. If $n$ is odd, then $\om^k = \om^{k-n}$ and we can index these spaces with $\nfr{-(n-1)}{2} \le k \le \nfr{(n-1)}{2}$. For a triality automorphism, $\Th=t$, the Lie algebra separates into three complex $\mathfrak{g}_{k}$ eigenspaces,
$$
\mathfrak{g} \; = \; \mathfrak{g}_{-1} \; + \; \mathfrak{g}_0 \; + \; \mathfrak{g}_{+1}
$$
with complexification understood here and in the following eigenspace decompositions. Here $\mathfrak{g}_0$ denotes the real triality invariant subalgebra of $\mathfrak{g}$, and the Lie brackets satisfy $\lb \mathfrak{g}_i , \mathfrak{g}_j \rb \subset \mathfrak{g}_{i+j}$, with indices taken modulo $3$. The $\mathfrak{g}_{\pm1}$ eigenspaces are spanned by complex conjugate sets of eigenvectors, $V^{+1}_i \in \mathfrak{g}_{+1}$ and $V^{+1 \, *}_i = V^{-1}_i \in \mathfrak{g}_{-1}$. These conjugate pairs of coset basis generators combine into ``real'' and ``imaginary'' orthogonal pairs of real vectors,
$$
\ba{rcl}
V^\mathbb{R}_i &=& \ha \lp V^{+1}_i + V^{-1}_i \rp \\[2pt]
V^\mathbb{I}_i &=& \fr{1}{2 i} \lp V^{+1}_i - V^{-1}_i \rp \\
\ea
\s\s
\Th  :  \lb \! \ba{c} V{}^\mathbb{R}_i \\[2pt] V{}^\mathbb{I}_i \ea \! \rb \mapsto
\lb \! \ba{c} V'{}^\mathbb{R}_i \\[2pt] V'{}^\mathbb{I}_i \ea \! \rb =
\lb  \ba{cc} -\fr{1}{2} & -\fr{\sqrt{3}}{2} \\[2pt] \fr{\sqrt{3}}{2} & -\fr{1}{2}  \ea  \rb
\lb \! \ba{c} V^\mathbb{R}_i \\[2pt] V^\mathbb{I}_i \ea \! \rb
$$
which are rotated $120^\circ$ in their plane by triality. The real Lie algebra, spanned by $\{V^0_a, V^\mathbb{R}_i, V^\mathbb{I}_i\}$, is invariant under these triality rotations\footnotemark[1] --- the triality automorphism. These orthogonal pairs of basis vectors, $\{V^\mathbb{R}_i, V^\mathbb{I}_i\}$, can also be replaced by non-orthogonal triples,
$$
V^{I}_i = V^\mathbb{R}_i \s\s
V^{II}_i = -\fr{1}{2} V^\mathbb{R}_i - \fr{\sqrt{3}}{2} V^\mathbb{I}_i \s\s
V^{III}_i = -\fr{1}{2} V^\mathbb{R}_i + \fr{\sqrt{3}}{2} V^\mathbb{I}_i
$$
which are permuted by triality, $\Th : V^{I}_i \mapsto V^{II}_i \mapsto V^{III}_i \mapsto V^{I}_i$. It is an intriguing possibility that three such sets of triality-related generators may correspond to the three generations of fermions in the Standard Model of particle physics.

The complement of $\mathfrak{g}_0$ in $\mathfrak{g}$ is spanned by the coset generators of a $3$-symmetric space, $G/G_0$ \cite{WG68},
$$
\mathfrak{g}/\mathfrak{g}_0 \,=\, \mathfrak{g}_{+1} \,+\, \mathfrak{g}_{-1} 
$$
Just as a symmetric space is invariant under inversion of the coset generators, this complex $3$-symmetric space is invariant under $V^k_i \to e^{\nfr{2 \pi i k}{3}} \, V^k_i$, and the corresponding real $3$-symmetric space is invariant under $120^\circ$ rotations in the spanning $\{V^\mathbb{R}_i, V^\mathbb{I}_i\}$ planes.

The action of the complexified fixed group on $\mathfrak{g}_{+1}$ defines the Vinberg $\Th$-group \cite{Vin76}. Its associated semidirect-product Lie algebra, denoted here by $\mathfrak{g}_\Th = \mathfrak{g}_0 + \mathfrak{g}_{+1}$, has $[\mathfrak{g}_0,\mathfrak{g}_0] \subset \mathfrak{g}_0$ and $[\mathfrak{g}_0,\mathfrak{g}_{+1}] \subset \mathfrak{g}_{+1}$ brackets from $\mathfrak{g}$, with $[\mathfrak{g}_{+1},\mathfrak{g}_{+1}] = 0$ imposed. It is not necessarily a subalgebra of $\mathfrak{g}$. For some of the magic square Lie algebras, under triality automorphisms, these associated algebras are shown below, with the real forms of the fixed subalgebras named and their complexifications understood:\footnotemark[2]
$$
\ba{rcl}
su(3)_{\Th} &=& \Big[ u(1) + u(1) \Big]_0 + \Big[ 1_{(-\sqrt{3},1)} + 1_{(0,-2)} + 1_{(\sqrt{3},1)} \Big]_{+1} \\[6pt]
sp(3)_{\Th} &=& \Big[ so(4) + u(1)_w  \Big]_0 + \Big[ 3^{su(2)}_{+1} + 4^v_{-\nha} \Big]_{+1} \\[6pt]
f_{4(-52)\Th} &=& \Big[ so(7) + u(1)_w \Big]_0 + \Big[ 7^v_{+1} + 8^s_{-\nfr{1}{2}} \Big]_{+1} \\ 
\ea
$$
\beq
\ba{rcl}
f_{4(4)\Th} &=& \Big[ so(4,3) + u(1)_w \Big]_0 + \Big[ 7^v_{+1} + 8^s_{-\nfr{1}{2}} \Big]_{+1} \\[6pt]
e_{6(-78)\Th} &=& \Big[ so(8) + u(1) +  u(1) \Big]_0 + \Big[ 8^v + 8^{s+} + 8^{s-} \Big]_{+1} \\[6pt]
e_{7(-133)\Th} &=& \Big[ so(10) + su(2) + u(1)_w \Big]_0 + \Big[ 10^v_{+1} + 2 \otimes 16^{s-}_{-\nfr{1}{2}} \Big]_{+1} \\[6pt]
e_{8(-248)\Th} &=& \Big[ so(14) + u(1)_w \Big]_0 + \Big[ 14^v_{+1} + 64^{s-}_{-\nfr{1}{2}} \Big]_{+1} \\[6pt]
e_{8(8)\Th} &=& \Big[ so(8,6) + u(1)_w \Big]_0 + \Big[ 14^v_{+1} + 64^{s-}_{-\nfr{1}{2}} \Big]_{+1} \\[6pt]
e_{8(-24)\Th} &=& \Big[ so(4,10) + u(1)_w \Big]_0 + \Big[ 14^v_{+1} + 64^{s-}_{-\nfr{1}{2}} \Big]_{+1}
\ea
\label{teig}
\eeq
in which the subscripts are the $u(1)_w$ charge and triality eigenvalue index, $k$, corresponding to $\om^k$. The complex dimension of the associated algebra is $\dim{\mathfrak{g}_\Th} = \dim{\mathfrak{g}_0} + \dim{\mathfrak{g}_{+1}}$. The complex triality eigenspace decomposition of a real Lie algebra might or might not correspond to a graded decomposition of the Lie algebra into a real subalgebra and real representation spaces.

The non-compact real forms of $e_8$ do have triality-related five-gradings and two-gradings,
\beq
\ba{rcl}
e_{8(8)} &=& 14^v_{-1} + 64^{s+}_{+\nfr{1}{2}} + \Big[ so(7,7) + so(1,1)_{w^\mathbb{R}} \Big] + 64^{s-}_{-\nfr{1}{2}} + 14^v_{+1}
=  so(8,8) + 128^{s+} \\[8pt]
\!\!\!\! e_{8(-24)} &=& 14^v_{-1} + 64^{s+}_{+\nfr{1}{2}} + \Big[ so(3,11) + so(1,1)_{w^\mathbb{R}} \Big] + 64^{s-}_{-\nfr{1}{2}}  + 14^v_{+1}
=  so(4,12) + 128^{s+}
\ea
\label{grad}
\eeq
which include subalgebras of the same dimension but different signatures from the corresponding triality invariant subalgebra. Interestingly, the extended graviGUT algebra \cite{Dou14},
\vspace{-8pt}
$$
so(3,11) + 64^{s+}_{+\nfr{1}{2}}  + 14^v_{+1}
$$
is embedded within $e_{8(-24)}$, and can describe the gravitational, gauge, and scalar fields of the Standard Model acting on one generation of fermions, with spin \cite{Lis10}. Note that the intersection of the graviGUT gauge algebra, $so(3,11)$, and the triality invariant gauge algebra, $\lb so(4,10) + u(1) \rb$, within the $so(4,12)$ subalgebra of $e_{8(-24)}$ is $so(3,10)$, which includes precisely the spin and GUT quantum numbers needed to identify a $64$-real-dimensional generation of fermions. The triality-map of $64^{s+}_{+\nfr{1}{2}}$ spinors exists as rotations within the $128^{s+}$ of the two-grading. Explicitly, the $\ps_i \in 64^{s+}_{+\nfr{1}{2}}$ can be identified with a Standard Model generation while the $\ps^m_i \in 64^{s-}_{-\nfr{1}{2}}$ are non-physical ``mirror'' fermions \cite{Dis09}. Since these $w^\mathbb{R} = \pm \nha$ states are not $w = \pm \nha$ triality eigenstates, triality does not act on them as $e^{\pm \nfr{2 \pi i}{3}}$. Rather, a canonical triality automorphism in $e_{8(8)}$ or $e_{8(-24)}$ acts on the $64^{s+}_{+\nfr{1}{2}}$ fermions as a $\nfr{2 \pi}{3}$ rotation in the planes spanned by $\{\ps^\mathbb{R}_i, \ps^{m\mathbb{I}}_i\}$ and $\{\ps^{m \mathbb{R}}_i, \ps^{\mathbb{I}}_i\}$. This suggests the triality rotation of these fermion degrees of freedom within the $128^{s+}$ may relate to the existence of three generations of Standard Model fermions.

\section{\texorpdfstring{$e_6$}{e6}}

The compact $e_6$ Lie algebra has a compound triality decomposition from combining $su(3)$ and $f_{4}$,
\beq
\ba{rcl}
e_{6(-78)} &=& u(1) + u(1) + so(8) + (1\!+\!\bar{1})_v \otimes 8_v + (1\!+\!\bar{1})_m \otimes 8_{s-} + (1\!+\!\bar{1})_p \otimes 8_{s+} \\[6pt]
&\!\!\sim\!\!& su(3,\mathbb{C} \otimes \mathbb{O}) \\[-14pt]
\ea
\label{e6decomp}
\vspace{8pt}
\eeq
and a triality automorphism that maps between the triplet of complex octonions, $\mathbb{C} \otimes \mathbb{O}$. The corresponding set of $e_6$ basis elements is
$$
\{ \, T'_1, \, T'_2, \, \ga_{ab}, \, \ga_{a' a}, \, Q^-_{a' a}, \, Q^+_{a' a} \, \}
$$
with primed index, $a'$, ranging over complex indices, $\{ 0,1 \}$, un-primed index, $a$, ranging over octonion indices, $\{ 0, ..., 7 \}$, and the bivector index, $ab$, ranging over the $28$ $so(8)$ basis generator permutations with $a < b$. The non-vanishing Lie algebra brackets between these basis elements come from combining the Lie brackets of $su(3)$ and $f_{4(-52)}$, (\ref{su3L}) and (\ref{f4L}):\footnotemark[2]
$$
\scalebox{.98}{$
\ba{rclcrcl}
\lb \ga_{a b}, \ga_{c d} \rb &=& 2 \left\{  n_{a c} \ga_{b d} - n_{a d} \ga_{b c} - n_{b c} \ga_{a d} + n_{b d} \ga_{a c} \right\} 
\!\!\!\! \!\!\!\! \!\!\!\! \!\!\!\! \!\!\!\! \!\!\!\!  & & & & \\[5pt]
\lb \ga_{ab}, \ga_{a'c} \rb &=& 2 \left\{ - n_{bc} \ga_{a'a} + n_{ac} \ga_{a'b} \right\} & \s &
\lb T'_2, Q^-_{0'a} \rb &=& +\sqrt{3} \, Q^-_{1'a} \\[5pt]
\lb \ga_{ab}, Q^-_{a'c} \rb &=&  Q^-_{a'd} ( - \bar{\Ga}_a \Ga_b )^d{}_c & &
\lb T'_2, Q^+_{0'a} \rb &=& -\sqrt{3} \, Q^+_{1'a} \\[5pt]
\lb \ga_{ab}, Q^+_{a'c} \rb &=&  Q^+_{a'd} ( - \Ga_a \bar{\Ga}_b )^d{}_c   & &
 \lb T'_2, Q^-_{1'a} \rb &=& -\sqrt{3} \, Q^-_{0'a} \\[5pt]
\lb T'_1, \ga_{0'a} \rb &=& -2 \, \ga_{1'a} & &
\lb T'_2, Q^+_{1'a} \rb &=& +\sqrt{3} \, Q^+_{0'a} \\[5pt]
\lb T'_1, Q^-_{0'a} \rb &=& +Q^-_{1'a} & &
 \lb \ga_{0'a}, \ga_{1'b} \rb &=& -2 \, T'_1 n_{ab} \\[5pt]
\lb T'_1, Q^+_{0'a} \rb &=& +Q^+_{1'a}   & &
   \lb Q^-_{0'a}, Q^-_{1'b} \rb &=& +(T'_1 + \sqrt{3} \, T'_2) n^-_{ab} \\[5pt]
\lb T'_1, \ga_{1'a} \rb &=& +2 \, \ga_{0'a}  & &
 \lb Q^+_{0'a}, Q^+_{1'b} \rb &=& +(T'_1 - \sqrt{3} \, T'_2) n^+_{ab} \\[5pt]
\lb T'_1, Q^-_{1'a} \rb &=& -Q^-_{0'a} & &
  \lb \ga_{a'a}, \ga_{a'b} \rb &=& +2 \, n'_{a'a'} \ga_{ab} \\[5pt]
\lb T'_1, Q^+_{1'a} \rb &=& -Q^+_{0'a}   & &
  \lb Q^-_{a'a}, Q^-_{a'b} \rb &=&  \ha n'^-_{a'a'} \ga_{c d}  ( \bar{\Ga}{}^c \Ga^d )_{ab} \\[5pt]
  \lb \ga_{a'a}, Q^-_{b'b} \rb &=& - M'_{\os{a}' \os{b}'}{}^{c'}  ( \Ga_a )^c{}_b Q^+_{c'c} & &
 \lb Q^+_{a'a}, Q^+_{a'b} \rb &=& \ha n'^+_{a'a'} \ga_{cd}  ( \Ga^c \bar{\Ga}{}^d )_{ab} \\[5pt]
  \lb Q^-_{a'a}, Q^+_{b'b} \rb &=& - M'_{\os{a}' \os{b}'}{}^{c'}   ( \bar{\Ga}{}^c )_{ab}  \ga_{c'c} & & 
   \lb \ga_{a'a}, Q^+_{b'b} \rb &=& M'_{\os{a}' \os{b}'}{}^{c'}  (  \bar{\Ga}_a )^c{}_b  Q^-_{c'c}
\ea$}
$$
in which $M'_{\os{a}' \os{b}'}{}^{c'}$ and $(\Ga_c)^b{}_a = M_{ca}{}^\os{b}$ are complex and octonion multiplication tables with conjugations, and $\{ n', n'^\pm, n, n^\pm \}$ are complex and octonion metrics, also used to raise or lower indices. Different real forms of $e_6$ come from combining different real forms of $su(3)$ and $f_4$, using correspondingly different multiplication tables and metrics. A canonical triality automorphism of $e_{6(-78)}$ is
$$
\scalebox{.98}{$
\ba{rclcrcl}
 t  \q : \q T'_1 &\;\mapsto\;& T''_1 = - \ha T'_1 + \fr{\sqrt{3}}{2} T'_2
 & \s\q &
 \ga_{a'a} &\;\mapsto\;& \ga'_{a'a} = Q^+_{a'a} \\[2pt]
 T'_2 &\;\mapsto\;& T''_2 = - \fr{\sqrt{3}}{2} T'_1 - \ha T'_2
 & &
 Q^-_{a'a} &\;\mapsto\;& Q'^-_{a'a} = \ga_{a'a} \\[2pt]
 \ga_{ab} &\;\mapsto\;& \ga'_{ab} = \ha \ga_{cd} t^{cd}{}_{ab} = \ga_{cd} \lp \fr{1}{4} \Ga^c \bar{\Ga}^d \rp_{ab}
 & &
 Q^+_{a'a} &\;\mapsto\;& Q'^+_{a'a} = Q^-_{a'a}
\ea
$}
$$
from combining the triality automorphisms of the constituent $su(3)$ and $f_{4(-52)}$.

The triality decomposition of $e_6$, (\ref{e6decomp}), may be concatenated, using $so(10) = u(1) + so(8) + (1\!+\!\bar{1})_v \otimes 8_v$ to get
$$
e_{6(-78)} = so(10) + u(1) + 16_{s+} + 16_{s-}
$$
which includes the algebra of the $SO(10)$ Grand Unified Theory of particle physics, including fermion states, without spin. Decomposing this into Standard Model representation spaces, we have:
$$
\ba{rcl}
e_{6(-78)} &=& su(2)_{W'} + su(2)_W + su(3)_g + u(1)_B + u(1)_H \\[4pt]
&+& \lp (2, 2; 3_{-\nfr{2}{3}} + \bar{3}_{+\nfr{2}{3}}) + (1, 1; 3_{\nfr{4}{3}} + \bar{3}_{-\nfr{4}{3}}) \rp_0 \\[4pt]
&+&  \lp (1, 2; 1_{-1} + 3_{+\nfr{1}{3}}) + (2, 1; 1_{+1} + \bar{3}_{-\nfr{1}{3}}) \rp_{-\nfr{\sqrt{3}}{2}} \\[4pt]
&+&  \lp (2, 1; 1_{-1} + 3_{+\nfr{1}{3}}) + (1, 2; 1_{+1} + \bar{3}_{-\nfr{1}{3}}) \rp_{+\nfr{\sqrt{3}}{2}} 
\ea
$$
The Cartan subalgebra generators, $\{H,U,V,x,y,z\}$, of $u(1)_H + so(10)$ or of $u(1)_H + u(1)_U + so(8)$, correspond to root coordinates, or charges, which combine to give the familiar strong, $g$, weak, $W$, hypercharge, $Y$, and electric charge, $Q$, of the Standard Model, as well as chirality (or helicity), $H$, the weaker charge, $W'$, baryon minus lepton number, $B$, and other GUT charges:
\beq
\ba{rcl}
W & = &  \ha (- U + V)    \\[4pt]  
W' & = &  \ha (U+V)        \\[4pt]
B & = & -\fr{2}{3} (x + y + z)   \\[4pt]
g^3 & = &   \ha (x - y)          \\[4pt]  
g^8 & = &  \fr{1}{2 \sqrt{3}} (x + y - 2 z)  \\[8pt]  
Y & = &  2 W' +  B =  U + V - \fr{2}{3} (x + y + z) \\[4pt]
X & = &  2 W' -  \fr{3}{2} B =  U + V + (x + y + z) \\[8pt]
Q & = &  W +  \ha Y =  V + \ha B =  V - \fr{1}{3} (x + y + z) \\[4pt]
Z & = &   W - \fr{3}{10} Y =   - \fr{4}{5} U + \fr{1}{5}( V + x + y + z )
\ea
\label{charges}
\eeq
The $e_6$ roots, labeled as elementary particle states, are shown in Table \ref{table:pE6}. The $su(3)_g$ gluons, $g$, and Pati-Salam bosons, $X^{PS}_{\pm \nfr{2}{3}}$, are in a $so(6) \!=\! su(4)$ subalgebra of $so(10)$, which has $so(4) = su(2)_W + su(2)_{W'}$ in its complement, with $su(2)_{W'}$ the $W'$ bosons. The Georgi-Glashow bosons, $X^{GG}_{\pm \nfr{4}{3}}$ and $Y^{GG}_{\pm \nfr{1}{3}}$, are in an $su(5)$ subalgebra of $so(10)$, and the GUT bosons, $X'_{\pm \nfr{2}{3}}$ and $Y'_{\pm \nfr{1}{3}}$, are in the rest of $so(10)$. The left-handed fermions and antifermions, with scaled helicity, $H = -\nfr{\sqrt{3}}{2}$, are in a $16_{s-}$ spinor of $so(10)$, while their conjugates, the right-handed fermions and antifermions, with $H = +\nfr{\sqrt{3}}{2}$, are in a $16_{s+}$.

With respect to the compound division algebra decomposition, (\ref{e6decomp}), we have $\{ W^\pm, W'{}^\pm,$ $Y^{GG}_{\pm \nfr{1}{3}}, Y'_{\pm \nfr{1}{3}}\}$ in $(1\!+\!\bar{1})_v \otimes 8_v$, the down-type fermions and antifermions in $(1\!+\!\bar{1})_m \otimes 8_{s-}$, and the up-type fermions and antifermions in $(1\!+\!\bar{1})_p \otimes 8_{s+}$. A canonical triality automorphism of $e_6$, although pretty, does not appear to correspond to anything especially interesting.

\begin{center}
    \vspace{0pt}
    \includegraphics[height=2.4in]{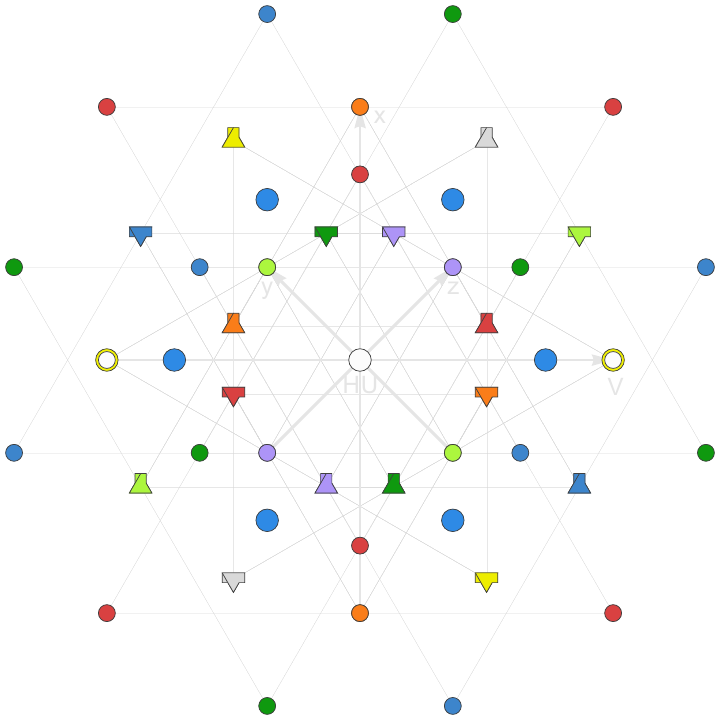}
    \vspace{-6pt}
\end{center}
\begin{table}[h!t]
\centering
\scalebox{.78}{
\renewcommand{\arraystretch}{1.05}
\begin{tabular}
{@{\vrule width1.0pt}c@{\vrule width0.2pt}c@{\vrule width1.0pt}c@{\vrule width0.0pt}c@{\vrule width0.2pt}c@{\vrule width0.0pt}c@
{\vrule width0.0pt}c@{\vrule width0.0pt}c@{\vrule width1.0pt}}
\noalign{\hrule height 1.0pt}

\multicolumn{2}{@{\vrule width1.0pt}c@{\vrule width1.0pt}}{\raisebox{1pt}{$e_6$}} & $\q\; H \;\;\;\;$ {\vrule width0.2pt} & $\q\; U \;\q$ & $\q\; V \;\;\;\;$ {\vrule width0.2pt} & $\q\; x \;\;\;\;$ {\vrule width0.2pt} & $\q\; y \;\;\;\;$ {\vrule width0.2pt} & $\q\; z \;\q$ \\
\noalign{\hrule height 1pt}

\raisebox{-1.5pt}{\gcirs{lblue}{2}} & \raisebox{1pt}{$\;\; g \;\;$} & $0 $ & $0 $ & $0 $ & $\! (+1 $ & $-1 $ & $\;\; 0 \, ) $  \\
\noalign{\hrule height 0.2pt}

$\;$ \raisebox{-1.0pt}{\gcirs{mred}{1.5} $\!\!$ \gcirs{mgree}{1.5} $\!\!$ \gcirs{mblue}{1.5}} $\;$ & $\, X_{\pm \nfr{2}{3}}^{PS} \,$ & $0$ & $0$ & $0$ & $\!(\mp1 $ & $\mp1$ & $\;\; 0 \, ) $  \\
\noalign{\hrule height 0.2pt}
\raisebox{-1.0pt}{\gcirs{mred}{1.5} $\!\!$ \gcirs{mgree}{1.5} $\!\!$ \gcirs{mblue}{1.5}} & $\, X_{\pm \nfr{4}{3}}^{GG} \,$ & $0 $ & $0 $ & $\pm1 $ & $\!(\mp1 $ & $0 $ & $\;\;0 \,) $  \\
\noalign{\hrule height 0.2pt}
\raisebox{-1.0pt}{\gcirs{mred}{1.5} $\!\!$ \gcirs{mgree}{1.5} $\!\!$ \gcirs{mblue}{1.5}} & $\, X'_{\pm \nfr{2}{3}} \,$ & $0 $ & $0$ & $\pm1 $ & $\!(\pm1 $ & $0 $ & $\;\;0 \,) $  \\
\noalign{\hrule height 0.8pt}

$\q$ \raisebox{-1.5pt}{\gcirs{lyell}{2}} $\q$ & $\q W^\pm \q$ & $ 0 $ & $ \mp 1 $ & $\pm 1$ & $0$ & $0$ & $0$  \\
\noalign{\hrule height 0.2pt}
\raisebox{-1.0pt}{\gcirs{mywhite}{1.5}} & $\, W'^\pm \,$ & $0 $ & $\pm 1 $ & $ \pm 1 $ & $0$ & $0$ & $0$  \\
\noalign{\hrule height 0.2pt}
\raisebox{-1.0pt}{\gcirs{mrora}{1.5} $\!\!$ \gcirs{mygre}{1.5} $\!\!$ \gcirs{mbvio}{1.5}} & $\, Y_{\pm \nfr{1}{3}}^{GG} \,$ & $0$ & $\pm 1 $ & $ 0 $ & $\!(\mp 1 $ & $0 $ & $\;\; 0 \, ) $  \\
\noalign{\hrule height 0.2pt}
\raisebox{-1.0pt}{\gcirs{mrora}{1.5} $\!\!$ \gcirs{mygre}{1.5} $\!\!$ \gcirs{mbvio}{1.5}} & $\, Y'_{\pm \nfr{1}{3}} \,$ & $0 $ & $\mp1 $ & $ 0 $ & $\!(\mp1 $ & $0 $ & $\;\; 0 \, ) $  \\
\noalign{\hrule height 0.8pt}

\raisebox{-1pt}{\gplus{lgray}{2}} & \raisebox{1pt}{$\, \nu_L \,$} & $-\nfr{\sqrt{3}}{2}$ & $-\nha\;$ & $+\nha\;\,$ & $+\nha \;\,$ & $+\nha\;\,$ & $+\nha\;\,$  \\
\noalign{\hrule height 0.2pt}
\raisebox{-1.0pt}{\gprus{lgray}{2}} & $\, \nu_R \,$ & $+\nfr{\sqrt{3}}{2}$ & $+\nha\;$ & $+\nha\;\,$ & $+\nha\;\,$ & $+\nha\;\,$ & $+\nha\;\,$  \\
\noalign{\hrule height 0.2pt}
\raisebox{-1.0pt}{\galus{lgray}{2}} & $\, \bar{\nu}_L \,$ & $-\nfr{\sqrt{3}}{2}$ & $-\nha\;$ & $-\nha\;\,$ & $-\nha\;\,$ & $-\nha\;\,$ & $-\nha\;\,$  \\
\noalign{\hrule height 0.2pt}
\raisebox{-1.0pt}{\garus{lgray}{2}} & $\, \bar{\nu}_R \,$ & $+\nfr{\sqrt{3}}{2}$ & $+\nha\;$ & $-\nha\;\,$ & $-\nha\;\,$ & $-\nha\;\,$ & $-\nha\;\,$  \\
\noalign{\hrule height 0.2pt}
\raisebox{-1.0pt}{\gplus{mred}{2} $\!\!$ \gplus{mgree}{2} $\!\!$ \gplus{mblue}{2}} & $u_L^{(rgb)}$ & $-\nfr{\sqrt{3}}{2}$ & $-\nha\;$ & $+\nha\;\,$ & $\!\!(+\nha\;\,$ & $-\nha\;\,$ & $\;-\nha \,)\,$  \\
\noalign{\hrule height 0.2pt}
\raisebox{-1.0pt}{\gprus{mred}{2} $\!\!$ \gprus{mgree}{2} $\!\!$ \gprus{mblue}{2}} & $u_R^{(rgb)}$ & $+\nfr{\sqrt{3}}{2}$ & $+\nha\;$ & $+\nha\;\,$ & $\!\!(+\nha\;\,$ & $-\nha\;\,$ & $\;-\nha\,)\,$  \\
\noalign{\hrule height 0.2pt}
\raisebox{-1.0pt}{\galus{mred}{2} $\!\!$ \galus{mgree}{2} $\!\!$ \galus{mblue}{2}} & $\bar{u}_L^{(rgb)}$ & $-\nfr{\sqrt{3}}{2}$ & $-\nha\;$ & $-\nha\;\,$ & $\!\!(-\nha\;\,$ & $+\nha\;\,$ & $\;+\nha\,)\,$  \\
\noalign{\hrule height 0.2pt}
\raisebox{-1.0pt}{\garus{mred}{2} $\!\!$ \garus{mgree}{2} $\!\!$ \garus{mblue}{2}} & $ \bar{u}_R^{(rgb)}$ & $+\nfr{\sqrt{3}}{2}$ & $+\nha\;$ & $-\nha\;\,$ & $\!\!(-\nha\;\,$ & $+\nha\;\,$ & $\;+\nha\,)\,$  \\
\noalign{\hrule height 0.8pt}

\raisebox{-1.0pt}{\gplus{myell}{2}} & $\, e_L \,$ & $-\nfr{\sqrt{3}}{2}$ & $+\nha\;$ & $-\nha\;\,$ & $+\nha \;\,$ & $+\nha\;\,$ & $+\nha\;\,$  \\
\noalign{\hrule height 0.2pt}
\raisebox{-1.0pt}{\gprus{myell}{2}} & $\, e_R \,$ & $+\nfr{\sqrt{3}}{2}$ & $-\nha\;$ & $-\nha\;\,$ & $+\nha\;\,$ & $+\nha\;\,$ & $+\nha\;\,$  \\
\noalign{\hrule height 0.2pt}
\raisebox{-1.0pt}{\galus{myell}{2}} & $\, \bar{e}_L \,$ & $-\nfr{\sqrt{3}}{2}$ & $+\nha\;$ & $+\nha\;\,$ & $-\nha\;\,$ & $-\nha\;\,$ & $-\nha\;\,$  \\
\noalign{\hrule height 0.2pt}
\raisebox{-1.0pt}{\garus{myell}{2}} & $\, \bar{e}_R \,$ & $+\nfr{\sqrt{3}}{2}$ & $-\nha\;$ & $+\nha\;\,$ & $-\nha\;\,$ & $-\nha\;\,$ & $-\nha\;\,$  \\
\noalign{\hrule height 0.2pt}
\raisebox{-1.0pt}{\gplus{mrora}{2} $\!\!$ \gplus{mygre}{2} $\!\!$ \gplus{mbvio}{2}} & $d_L^{(rgb)}$ & $-\nfr{\sqrt{3}}{2}$ & $+\nha\;$ & $-\nha\;\,$ & $\!\!(+\nha\;\,$ & $-\nha\;\,$ & $\;-\nha\,)\,$  \\
\noalign{\hrule height 0.2pt}
\raisebox{-1.0pt}{\gprus{mrora}{2} $\!\!$ \gprus{mygre}{2} $\!\!$ \gprus{mbvio}{2}} & $d_R^{(rgb)}$ & $+\nfr{\sqrt{3}}{2}$ & $-\nha\;$ & $-\nha\;\,$ & $\!\!(+\nha\;\,$ & $-\nha\;\,$ & $\;-\nha\,)\,$  \\
\noalign{\hrule height 0.2pt}
\raisebox{-1.0pt}{\galus{mrora}{2} $\!\!$ \galus{mygre}{2} $\!\!$ \galus{mbvio}{2}} & $\bar{d}_L^{\,(rgb)} $ & $-\nfr{\sqrt{3}}{2}$ & $+\nha\;$ & $+\nha\;\,$ & $\!\!(-\nha\;\,$ & $+\nha\;\,$ & $\;+\nha\,)\,$  \\
\noalign{\hrule height 0.2pt}
\raisebox{-1.0pt}{\garus{mrora}{2} $\!\!$ \garus{mygre}{2} $\!\!$ \garus{mbvio}{2}} & $ \bar{d}_R^{\,(rgb)}$ & $+\nfr{\sqrt{3}}{2}$ & $-\nha\;$ & $+\nha\;\,$ & $\!\!(-\nha\;\,$ & $+\nha\;\,$ & $\;+\nha\,)\,$  \\
\noalign{\hrule height 1.0pt}

\end{tabular}
}
\vspace{8pt}
\caption{The 72 roots of $e_6$ labeled as elementary particles. There are gluons, $g$, Pati-Salam X bosons, $X^{PS}$, Georgi-Glashow X and Y bosons, $X^{GG}$ and $Y^{GG}$, and remaining SO(10) GUT $X'$ and $Y'$ bosons. The $W$ and $W'$ are the weak and weaker (GUT) bosons. One generation of fermions is included: left and right-chiral neutrinos and anti-neutrinos, $\nu_{L/R}$ and $\bar{\nu}_{L/R}$, electrons and positrons, $e_{L/R}$ and $\bar{e}_{L/R}$, and up and down quarks and anti-quarks of different colors, $u_{L/R}$, $\bar{u}_{L/R}$, $d_{L/R}$, and $\bar{d}_{L/R}$. For the charges, see (\ref{charges}).}
\label{table:pE6}
\end{table}

\section{\texorpdfstring{$e_7$}{e7}}

The $e_7$ Lie algebra has a compound triality decomposition from combining $sp(3)$ and $f_4$,
\vspace{-3pt}
\beq
\scalebox{.95}{$
\ba{rcl}
e_7 &=& su(2)_M + su(2)_P + su(2)_V + so(8) \\[4pt]
 & & + \, (2,2,1)_v \otimes 8_v + (2,1,2)_m \otimes 8_{s-} + (1,2,2)_p \otimes 8_{s+} \\[6pt]
&\!\! \sim \!\!&\!\!  su(3,\mathbb{H} \otimes \mathbb{O})
\ea
$}
\label{e7decomp}
\vspace{-3pt}
\eeq
and a triality automorphism that maps between the triplet of quaterni-octonions, $\mathbb{H} \otimes \mathbb{O}$. The corresponding set of $e_7$ basis elements is
\vspace{-3pt}
$$
\scalebox{.95}{$
\{ \, T^M_{A'}, \, T^P_{A'}, \, T^V_{A'}, \, \ga_{ab}, \, \ga_{a' a}, \, Q^-_{a' a}, \, Q^+_{a' a} \, \}
$}
\vspace{-3pt}
$$
with primed index, $a'$, ranging over quaternion indices, $\{ 0, ..., 3 \}$, primed capitals, $A'$, ranging over imaginary quaternion indices, $\{1, ..., 3\}$, un-primed index, $a$, ranging over octonion indices, $\{ 0, ..., 7 \}$, and the bivector index, $ab$, ranging over the $28$ $so(8)$ basis generator permutations with $a < b$. The non-vanishing Lie algebra brackets between these basis elements come from combining the Lie brackets of $sp(3)$ and $f_{4(-52)}$, (\ref{sp3L}) and (\ref{f4L}):\footnotemark[2]
\vspace{-3pt}
$$
\scalebox{0.95}{$
\ba{rclcrcl}
\lb \ga_{a b}, \ga_{c d} \rb &=& 2 \left\{ n_{a c} \ga_{b d} - n_{a d} \ga_{b c} - n_{b c} \ga_{a d} + n_{b d} \ga_{a c} \right\} 
\!\!\!\! \!\!\!\! \!\!\!\! \!\!\!\! \!\!\!\! \!\!\!\! \!\!\!\! \!\!\!\! \!\!\!\! \!\!\!\! \!\!\!\! \!\! & & & & \\[5pt]
\lb \ga_{ab}, \ga_{a'c} \rb &=& 2 \left\{ - n_{bc} \ga_{a'a} + n_{ac} \ga_{a'b} \right\} \!\!\!\! & & & & \\[5pt]
\lb \ga_{ab}, Q^-_{a'c} \rb &=&  Q^-_{a'd} ( - \bar{\Ga}_a \Ga_b )^d{}_c & \s \s &
\lb \ga_{ab}, Q^+_{a'c} \rb &=&  Q^+_{a'd} ( - \Ga_a \bar{\Ga}_b )^d{}_c \\[5pt]

\lb T^M_{A'}, T^M_{B'} \rb &=& T^M_{C'} (2 M'_{[A'B']}{}^{C'}) & &
\lb T^M_{A'}, \ga_{b'a} \rb &=& \ga_{c'a} \, (- M'_{b'A'}{}^{c'}) \\[5pt]
\lb T^P_{A'}, T^P_{B'} \rb &=& T^P_{C'} (2 M'_{[A'B']}{}^{C'}) &  &
\lb T^P_{A'}, \ga_{b'a} \rb &=& \ga_{c'a} \, (M'_{A'b'}{}^{c'}) \\[5pt]
\lb T^V_{A'}, T^V_{B'} \rb &=& T^V_{C'} (2 M'_{[A'B']}{}^{C'})  & &
\lb T^V_{A'}, Q^-_{b'a} \rb &=& Q^-_{c'a} \,(- M'_{b'A'}{}^{c'}) \\[5pt]

\lb T^M_{A'}, Q^-_{b'a} \rb &=& Q^-_{c'a} \, (M'_{A'b'}{}^{c'})  &  &
\lb \ga_{a'a}, Q^-_{b'b} \rb &=& - \fr{1}{\sqrt{2}} M'_{\os{b}' \os{a}'}{}^{c'} \! ( \Ga_a )^c{}_b Q^+_{c'c} \\[5pt]
\lb T^P_{A'}, Q^+_{b'a} \rb &=& Q^+_{c'a} \,(- M'_{b'A'}{}^{c'}) & &
\lb \ga_{a'a}, Q^+_{b'b} \rb &=& \fr{1}{\sqrt{2}} M'_{\os{a}' \os{b}'}{}^{c'}  \! (  \bar{\Ga}_a )^c{}_b  Q^-_{c'c} \\[5pt]
\lb T^V_{A'}, Q^+_{b'a} \rb &=& Q^+_{c'a} \,( M'_{A'b'}{}^{c'})   & &
\lb Q^-_{a'a}, Q^+_{b'b} \rb &=& - \fr{1}{\sqrt{2}} M'_{\os{b}' \os{a}'}{}^{c'}  \! ( \bar{\Ga}{}^c )_{ab}  \ga_{c'c} \\[5pt]
 
\!\!\!\! \lb \ga_{a'a}, \ga_{b'b} \rb &=& \lp T^M_{C'} \ha (M'_{\os{b}'a'}{}^{C'} \!-\! M'_{\os{a}'b'}{}^{C'}) + T^P_{C'} \ha (M'_{b'\os{a}'}{}^{C'} \!-\! M'_{a'\os{b}'}{}^{C'}) \rp n_{ab} 
+ \, n'_{a'b'} \ga_{ab}
\!\!\!\! \!\!\!\! \!\!\!\! \!\!\!\! \!\!\!\! \!\!\!\! \!\!\!\! \!\!\!\! \!\!\!\! \!\!\!\! \!\!\!\! \!\!\!\! \!\!\!\! \!\!\!\! \!\!\!\! \!\!\!\! \!\!\!\!
\!\!\!\! \!\!\!\! \!\!\!\! \!\!\!\! \!\!\!\! \!\!\!\! \!\!\!\! \!\!\!\! \!\!\!\! \!\!\!\! \!\!\!\! \!\!\!\! \!\!\!\! \!\!\!\! \!\!\!\! \!\!\!\! \!\!\!\! \!\!\!\! \!\!\!\! \!\!\!\! \!\!\!\! \!\!\!\! \!\!\!\! \!\!\!\! \!\!\!\!
 & &  & & \\[5pt]
\!\!\!\! \lb Q^-_{a'a}, Q^-_{b'b} \rb &=& \lp T^V_{C'} \ha (M'_{\os{b}'a'}{}^{C'} \!-\! M'_{\os{a}'b'}{}^{C'}) + T^M_{C'} \ha (M'_{b'\os{a}'}{}^{C'} \!-\! M'_{a'\os{b}'}{}^{C'}) \rp n^-_{ab} 
+ \, \fr{1}{4} n'^-_{a'b'} \ga_{c d}  ( \bar{\Ga}{}^c \Ga^d )_{ab}
\!\!\!\! \!\!\!\! \!\!\!\! \!\!\!\! \!\!\!\! \!\!\!\! \!\!\!\! \!\!\!\! \!\!\!\! \!\!\!\! \!\!\!\! \!\!\!\! \!\!\!\! \!\!\!\! \!\!\!\! \!\!\!\! \!\!\!\!
\!\!\!\! \!\!\!\! \!\!\!\! \!\!\!\! \!\!\!\! \!\!\!\! \!\!\!\! \!\!\!\! \!\!\!\! \!\!\!\! \!\!\!\! \!\!\!\! \!\!\!\! \!\!\!\! \!\!\!\! \!\!\!\! \!\!\!\! \!\!\!\! \!\!\!\! \!\!\!\! \!\!\!\! \!\!\!\! \!\!\!\! \!\!\!\! \!\!\!\!
 & & & &   \\[5pt]
\!\!\!\! \lb Q^+_{a'a}, Q^+_{b'b} \rb &=& \lp T^P_{C'} \ha (M'_{\os{b}'a'}{}^{C'} \!-\! M'_{\os{a}'b'}{}^{C'}) + T^V_{C'} \ha (M'_{b'\os{a}'}{}^{C'} \!-\! M'_{a'\os{b}'}{}^{C'}) \rp n^+_{ab} 
+ \, \fr{1}{4} n'^+_{a'b'} \ga_{c d}  ( \Ga^c \bar{\Ga}{}^d )_{ab}
\!\!\!\! \!\!\!\! \!\!\!\! \!\!\!\! \!\!\!\! \!\!\!\! \!\!\!\! \!\!\!\! \!\!\!\! \!\!\!\! \!\!\!\! \!\!\!\! \!\!\!\! \!\!\!\! \!\!\!\! \!\!\!\! \!\!\!\!
\!\!\!\! \!\!\!\! \!\!\!\! \!\!\!\! \!\!\!\! \!\!\!\! \!\!\!\! \!\!\!\! \!\!\!\! \!\!\!\! \!\!\!\! \!\!\!\! \!\!\!\! \!\!\!\! \!\!\!\! \!\!\!\! \!\!\!\! \!\!\!\! \!\!\!\! \!\!\!\! \!\!\!\! \!\!\!\! \!\!\!\! \!\!\!\! \!\!\!\!
 & & & & 
\ea$}
\vspace{-3pt}
$$
in which $M'_{\os{a}' \os{b}'}{}^{c'}$ and $(\Ga_c)^b{}_a = M_{ca}{}^\os{b}$ are quaternion and octonion multiplication tables with conjugations, and $\{ n', n'^\pm, n, n^\pm \}$ are quaternion and octonion metrics, also used to raise or lower indices. Different real forms of $e_7$ come from combining different real forms of $sp(3)$ and $f_4$, using correspondingly different multiplication tables and metrics.

A canonical triality automorphism of $e_{7(-133)}$ is
\vspace{-3pt}
$$
\scalebox{.95}{$
\ba{rclcrclcrcl}
 t \q : \q   
  \ga_{ab} &\;\mapsto\;& \ga'_{ab} = \ha \ga_{cd} t^{cd}{}_{ab} = \ga_{cd} \lp \fr{1}{4} \Ga^c \bar{\Ga}^d \rp_{ab}
 \!\!\!\! \!\!\!\! \!\!\!\! \!\!\!\! \!\!\!\! \!\!\!\! \!\!\!\! \!\!\!\! \!\!\!\! \!\!\!\! \!\!\!\! \!\!\!\! \!\!\!\! \!\!\!\! \!\!\!\! \!\!\!\! \!\!\!\! \!\!\!\! \!\!\!\! \!\!\!\! \!\!\!\! \!\!\!\!
   & & & &  \\[5pt]
    T^M_{A'} &\;\mapsto\;& T'^M_{A'} = T^P_{A'} & \s\q & T^V_{A'} &\;\mapsto\;& T'^V_{A'} = T^M_{A'} & \s\q & T^P_{A'} &\;\mapsto\;& T'^P_{A'} = T^V_{A'} \\[5pt]
    \ga_{a'a} &\;\mapsto\;& \ga'_{a'a} = Q^+_{a'a} & & Q^-_{a'a} &\;\mapsto\;& Q'^-_{a'a} = \ga_{a'a} & &  Q^+_{a'a} &\;\mapsto\;& Q'^+_{a'a} = Q^-_{a'a} \\
\ea
$}
\vspace{-3pt}
$$
from combining the triality automorphisms of $sp(3)$ and $f_{4(-52)}$. 

The triality decomposition of $e_7$, (\ref{e7decomp}), may be concatenated using
$$
so(12) = su(2)_M + su(2)_P  + so(8) + (2,2)_v \otimes 8_v
$$
to get
$$
e_{7(-133)} = so(12) + su(2) + 2 \otimes 32_{s+}
$$
which includes the algebra of the $SO(10)$ Grand Unified Theory of particle physics, including fermion states with spin. For the following eigenspace description, we use a conjugate $so(12)+su(2)$ embedding containing the triality invariant subalgebra. The Cartan subalgebra generators, $\{\om, U, V, h, x, y, z\}$, of $su(2)_\om + so(12)$ or of $su(2)_\om + su(2)_{W} + su(2)_{W'} + so(8)$, correspond to charges which combine to give the strong, $g$, weak, $W$, hypercharge, $Y$, and electric charge, $Q$, of the Standard Model, as well as the weaker charge, $W'$, and reversed chirality (or helicity), $h$. The triality invariant subalgebra of $e_{7(-133)}$ is
$$
\mathfrak{g}_0 = so(10) + su(2)_\om + u(1)_h
$$
which is a subalgebra of $so(12) + su(2)$. With respect to this subalgebra, the $+1$ triality eigenspace is
$$
\mathfrak{g}_1 = 10^v_{+1} + 2 \otimes 16^{s-}_{-\nfr{1}{2}}
$$
which includes vector and spinor representation spaces with $h \in \{+1,-\nfr{1}{2}\}$. The $2 \otimes 16^{s-}_{-\nfr{1}{2}}$ representation space matches a set of left-chiral Standard Model fermions and antifermions, while the $10^v_{+1}$ matches a set of GUT Higgs fields, $\Phi$. For $e_{7(-133)}$ and triality, the semidirect-product algebra associated with the Vinberg $\Th$-group matches the $SO(10)$ GUT with a Higgs and one left-chiral generation of fermions with spin acted on by a $su(2)_\om$. Recall that the triality automorphism rotates this representation space within the $2 \otimes 32_{s+}$ representation space of $so(12) + su(2)$, permuting $3 \otimes 32$ non-linearly-independent sets of vectors. It is compelling to consider these spaces as three different left-chiral generations of fermions. The $2 \otimes 32_{s+}$ representation space is spanned by the $[2 \otimes 16^{s-}_{-\nfr{1}{2}} ]_{+1}$ eigenvectors and the $[ 2 \otimes 16^{s+}_{+\nfr{1}{2}} ]_{-1}$ conjugate eigenvectors. These conjugate fermions, such as $(e_L^\wedge)^* = \bar{e}_R^\vee$, can be interpreted in $e_7$ as either the right-chiral fermions and antifermions, or as the conjugate partner directions to the Standard Model fermions, allowing three triality-related generations to inhabit the $64$-dimensional space. The geometry of the Higgs states, with $10^v_{+1}$ rotated by triality in a $20$-dimensional space spanned by $10^v_{-1}$ and $10^v_{+1}$, can be treated similarly.

The $e_7$ roots, labeled suggestively as elementary particles, with conjugate states related by triality, are shown in Table \ref{table:pE7}. The tabulated $W,W'$ are $\sqrt{2}$ times the charges in (\ref{charges}). Note that, to incorporate three generations, the right-chiral fermions and antifermions here need to be interpreted as the conjugate root vectors of the left-chiral fermions and antifermions, with the triality automorphism intra-rotating their real and imaginary parts. Alternatively, $e_7$ could have just a single generation of fermions, including the right-chiral fermions and antifermions.

\begin{center}
    \vspace{0pt}
    \includegraphics[height=2.3in]{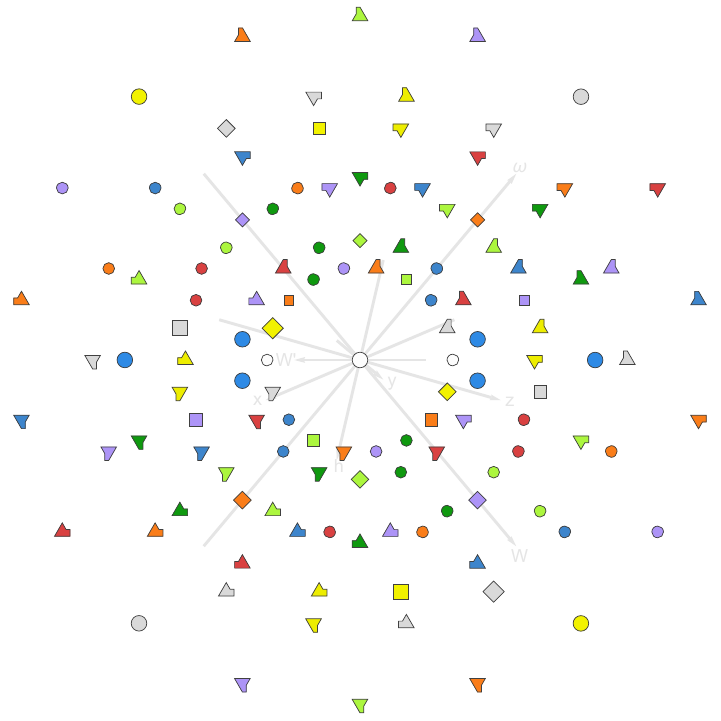}
    \vspace{-10pt}
\end{center}
\begin{table}[h!t]
\centering
\scalebox{0.64}{
\renewcommand{\arraystretch}{1.1}
\begin{tabular}
{@{\vrule width1.0pt}c@{\vrule width0.2pt}c@{\vrule width1.0pt}c@{\vrule width0.0pt}c@{\vrule width0.0pt}c@{\vrule width0.2pt}c@{\vrule width0.0pt}c@
{\vrule width0.0pt}c@{\vrule width0.0pt}c@{\vrule width1.0pt}}
\noalign{\hrule height 1.0pt}

\multicolumn{2}{@{\vrule width1.0pt}c@{\vrule width1.0pt}}{\raisebox{1pt}{$e_7$}} &$\q\;\; \om \;\;\q$ {\vrule width0.2pt} & $\q\; W \;\q$ {\vrule width0.2pt} & $\q\; W' \q$ & $\q\;\; h \;\;\;\;\;$ {\vrule width0.2pt} & $\q\;\; x \;\;\;\;\;$ {\vrule width0.2pt} & $\q\;\; y \;\;\;\;\;$ {\vrule width0.2pt} & $\q\;\; z \;\;\q$ \\
\noalign{\hrule height 1pt}

 \raisebox{-2pt}{\gcirs{lgray}{2}}  & $\;\; \om^{\wedge/\vee} \;\;$ & $\; \pm \sqrt{2} \;$ & $\; 0 \;$ & $\; 0 \;$ & $0 $ & $0 $ & $0 $ & $0 $ \\
\noalign{\hrule height 0.2pt}
 \raisebox{-2pt}{\gcirs{lyell}{2}}  & $\;\; W^{\pm} \;\;$ & $\; 0 \;$ & $\; \pm \sqrt{2} \;$  & $\; 0 \;$ & $0 $ & $0 $ & $0 $ & $0 $ \\
\noalign{\hrule height 0.2pt}
\raisebox{-1pt}{\gcirs{mywhite}{1.5}}  & $\;\; W'{}^{\pm}_{\phantom{R}} \;\;$ & $\; 0 \;$ & $\; 0 \;$  & $\; \pm \sqrt{2} \;$ & $0 $ & $0 $ & $0 $ & $0 $ \\
\noalign{\hrule height .2pt}

\raisebox{-2pt}{\gcirs{lblue}{2}} & \raisebox{1pt}{$\;\; g \;\;$} & $0 $ & $0 $ & $0 $ & $0 $ & $\! ( \, +1 \, \;$ & $-1 \;$ & $\;\;\; 0 \; ) $  \\
\noalign{\hrule height 0.2pt}
$\;\!$ \raisebox{-1pt}{\gcirs{mred}{1.5} $\!$ \gcirs{mgree}{1.5} $\!$ \gcirs{mblue}{1.5}} $\;\!$ & $\;\, X_{\pm \nfr{2}{3}}^{PS}  \,\;$ & $0 $ & $0 $ & $0 $ & $ 0 $ & $\! ( \, \mp 1 \, \;$ & $ \mp 1 \;$ & $\;\;\; 0 \; ) $ \\
\noalign{\hrule height 0.2pt}
$\;\!$ \raisebox{-1pt}{\gcirs{mred}{1.5} $\!$ \gcirs{mgree}{1.5} $\!$ \gcirs{mblue}{1.5}} $\;\!$ & $\;\, X_{\pm \nfr{4}{3}}^{GG}  \,\;$ & $0 $ & $\pm \nicefrac{1}{\sqrt{2}} \;$ & $\pm \nicefrac{1}{\sqrt{2}} \;$ & $ 0 $ & $\! ( \, \mp 1 \, \;$ & $ 0 $ & $\;\;\; 0 \; ) $ \\
\noalign{\hrule height 0.2pt}
$\;\!$ \raisebox{-1pt}{\gcirs{mred}{1.5} $\!$ \gcirs{mgree}{1.5} $\!$ \gcirs{mblue}{1.5}} $\;\!$ & $\;\, X'_{\pm \nfr{2}{3}}  \,\;$ & $0 $ & $\pm \nicefrac{1}{\sqrt{2}} \;$ & $\pm \nicefrac{1}{\sqrt{2}} \;$ & $ 0 $ & $\! ( \, \pm 1 \, \;$ & $ 0 $ & $\;\;\; 0 \; ) $ \\
\noalign{\hrule height 0.2pt}
$\;\!$ \raisebox{-1pt}{\gcirs{mrora}{1.5} $\!$ \gcirs{mygre}{1.5} $\!$ \gcirs{mbvio}{1.5}} $\;\!$ & $\;\, Y_{\pm \nfr{1}{3}}^{GG}  \,\;$ & $0 $ & $\mp \nicefrac{1}{\sqrt{2}} \;$ & $\pm \nicefrac{1}{\sqrt{2}} \;$ & $ 0 $ & $\! ( \, \mp 1 \, \;$ & $ 0 $ & $\;\;\; 0 \; ) $ \\
\noalign{\hrule height 0.2pt}
$\;\!$ \raisebox{-1pt}{\gcirs{mrora}{1.5} $\!$ \gcirs{mygre}{1.5} $\!$ \gcirs{mbvio}{1.5}} $\;\!$ & $\;\, Y'_{\pm \nfr{1}{3}}  \,\;$ & $0 $ & $\pm \nicefrac{1}{\sqrt{2}} \;$ & $\mp \nicefrac{1}{\sqrt{2}} \;$ & $ 0 $ & $\! ( \, \mp 1 \, \;$ & $ 0 $ & $\;\;\; 0 \; ) $ \\
\noalign{\hrule height 0.8pt}

\raisebox{-1.5pt}{\gsqus{lgray}{2} \! \gsqus{lgray}{1.7}} & $\! \Ph_{0} \;\;\;\; \Ph_{0}^{m}$ & $\; 0 \;$ & $\mp \nicefrac{1}{\sqrt{2}} \;$ & $\pm \nicefrac{1}{\sqrt{2}} \;$ & $\pm 1  $ & $\; 0  \;$ & $\; 0 \;$ & $\; 0 \;$ \\
\noalign{\hrule height 0.2pt}
\raisebox{-2pt}{\gdias{lgray}{2} \!\! \gdias{lgray}{1.7}} & $\, \Ph_{0}^{*} \;\;\;\; \Ph_{0}^{* m}$ & $\; 0 \;$ & $\pm \nicefrac{1}{\sqrt{2}} \;$ & $\mp \nicefrac{1}{\sqrt{2}} \;$ & $\pm 1  $ & $\; 0  \;$ & $\; 0 \;$ & $\; 0 \;$ \\
\noalign{\hrule height 0.2pt}
\raisebox{-1.5pt}{\gsqus{lyell}{2} \! \gsqus{lyell}{1.7}} & $\Ph_{+1} \; \Ph_{-1}^{m}$ & $\; 0 \;$ & $\pm \nicefrac{1}{\sqrt{2}} \;$ & $\pm \nicefrac{1}{\sqrt{2}} \;$ & $\pm 1  $ & $\; 0  \;$ & $\; 0 \;$ & $\; 0 \;$ \\
\noalign{\hrule height 0.2pt}
\raisebox{-2pt}{\gdias{lyell}{2} \!\! \gdias{lyell}{1.7}} & $\Ph_{-1} \; \Ph_{+1}^{m}$ & $\; 0 \;$ & $\mp \nicefrac{1}{\sqrt{2}} \;$ & $\mp \nicefrac{1}{\sqrt{2}} \;$ & $\pm 1  $ & $\; 0  \;$ & $\; 0 \;$ & $\; 0 \;$ \\
\noalign{\hrule height 0.2pt}
\raisebox{-2pt}{\gsqus{mrora}{1.7} \!\!\!\!\!\! \gsqus{mygre}{1.7} \!\!\!\!\!\! \gsqus{mbvio}{1.7} \! \gsqus{mrora}{1.5} \!\!\!\!\!\! \gsqus{mygre}{1.5} \!\!\!\!\!\! \gsqus{mbvio}{1.5}} & $\; \Ph_{+ \nfr{1}{3}} \; \Ph_{- \nfr{1}{3}}^{m} \;$ & $0 $ & $0 $ & $0 $ & $ \pm 1 $ & $\! ( \, \mp 1 \, \;$ & $ 0 $ & $\;\;\; 0 \; )$ \\
\noalign{\hrule height 0.2pt}
\raisebox{-2pt}{\gdias{mrora}{1.7} \!\!\!\!\!\!\! \gdias{mygre}{1.7} \!\!\!\!\!\!\! \gdias{mbvio}{1.7} \gdias{mrora}{1.5} \!\!\!\!\!\!\! \gdias{mygre}{1.5} \!\!\!\!\!\!\! \gdias{mbvio}{1.5}} & $\Ph_{- \nfr{1}{3}} \; \Ph_{+ \nfr{1}{3}}^{m}$ & $0 $ & $0 $ & $0 $ & $ \pm 1 $ & $\! ( \, \pm 1 \, \;$ & $ 0 $ & $\;\;\; 0 \; )$ \\
\noalign{\hrule height 0.8pt}

\raisebox{-.5pt}{\gplus{lgray}{2} \gplds{lgray}{2}} & $\nu_{eL}^{\wedge/\vee}$ & $\pm \nicefrac{1}{\sqrt{2}} \;$ & $+ \nicefrac{1}{\sqrt{2}} \;$ & $\; 0 \;$ & $ -\nha \;$ & $\!\! \, + \nha \;$ & $\!\! +\nha \,\;$ & $\!\!\! +\nha \,\,$ \\
\noalign{\hrule height 0.2pt}
\raisebox{-.5pt}{\galus{lgray}{2} \galds{lgray}{2}} & $\bar{\nu}_{eL}^{\wedge/\vee}$ & $\pm \nicefrac{1}{\sqrt{2}} \;$ & $\; 0 \;$ & $- \nicefrac{1}{\sqrt{2}} \;$ & $ -\nha \;$ & $\!\! \, -\nha \;$ & $\!\! -\nha \,\;$ & $\!\!\! -\nha \,\,$ \\
\noalign{\hrule height 0.2pt}
\raisebox{-.5pt}{\gplus{myell}{2} \gplds{myell}{2}}  & $e_L^{\wedge/\vee}$ & $\pm \nicefrac{1}{\sqrt{2}} \;$ & $- \nicefrac{1}{\sqrt{2}} \;$ & $\; 0 \;$ & $ -\nha \;$ & $\!\! \, + \nha \;$ & $\!\! +\nha \,\;$ & $\!\!\! +\nha \,\,$ \\
\noalign{\hrule height 0.2pt}
\raisebox{-.5pt}{\galus{myell}{2} \galds{myell}{2}}  & $\bar{e}_L^{\wedge/\vee}$ & $\pm \nicefrac{1}{\sqrt{2}} \;$ & $\; 0 \;$ & $+ \nicefrac{1}{\sqrt{2}} \;$ & $ -\nha \;$ & $\!\! \, -\nha \;$ & $\!\! -\nha \,\;$ & $\!\!\! -\nha \,\,$ \\
\noalign{\hrule height 0.2pt}
$\;$ \raisebox{-.5pt}{\gplus{mred}{2} \!\!\!\!\!\!\! \gplus{mgree}{2} \!\!\!\!\!\!\! \gplus{mblue}{2} \gplds{mred}{2} \!\!\!\!\!\!\! \gplds{mgree}{2} \!\!\!\!\!\!\! \gplds{mblue}{2}} $\;$& $u_{L}^{\wedge/\vee}$ & $\pm \nicefrac{1}{\sqrt{2}} \;$ & $+ \nicefrac{1}{\sqrt{2}} \;$ & $\; 0 \;$ & $ -\nha \;$ & $\!\!\! ( + \nha \;$ & $\!\! -\nha \,\;$ & $\!\! -\nha \, )$ \\
\noalign{\hrule height 0.2pt}
\raisebox{-1pt}{\galus{mred}{2} \!\!\!\!\!\!\! \galus{mgree}{2} \!\!\!\!\!\!\! \galus{mblue}{2} \galds{mred}{2} \!\!\!\!\!\!\! \galds{mgree}{2} \!\!\!\!\!\!\! \galds{mblue}{2}} & $\bar{u}_{L}^{\wedge/\vee}$ & $\pm \nicefrac{1}{\sqrt{2}} \;$ & $\; 0 \;$ & $- \nicefrac{1}{\sqrt{2}} \;$ & $ -\nha \;$ & $\!\!\! ( - \nha \;$ & $\!\! +\nha \,\;$ & $\!\! +\nha \, )$ \\
\noalign{\hrule height 0.2pt}
\raisebox{-.5pt}{\gplus{mrora}{2} \!\!\!\!\!\!\! \gplus{mygre}{2} \!\!\!\!\!\!\! \gplus{mbvio}{2} \gplds{mrora}{2} \!\!\!\!\!\!\! \gplds{mygre}{2} \!\!\!\!\!\!\! \gplds{mbvio}{2}} & $d_{L}^{\wedge/\vee}$ & $\pm \nicefrac{1}{\sqrt{2}} \;$ & $- \nicefrac{1}{\sqrt{2}} \;$ & $\; 0 \;$ & $ -\nha \;$ & $\!\!\! ( + \nha \;$ & $\!\! -\nha \,\;$ & $\!\! -\nha \, )$ \\
\noalign{\hrule height 0.2pt}
\raisebox{-1pt}{\galus{mrora}{2} \!\!\!\!\!\!\! \galus{mygre}{2} \!\!\!\!\!\!\! \galus{mbvio}{2} \galds{mrora}{2} \!\!\!\!\!\!\! \galds{mygre}{2} \!\!\!\!\!\!\! \galds{mbvio}{2}} & $\bar{d}_{L}^{\wedge/\vee}$ & $\pm \nicefrac{1}{\sqrt{2}} \;$ & $\; 0 \;$ & $+ \nicefrac{1}{\sqrt{2}} \;$ & $ -\nha \;$ & $\!\!\! ( - \nha \;$ & $\!\! +\nha \,\;$ & $\!\! +\nha \, )$ \\
\noalign{\hrule height 0.8pt}

\raisebox{-.5pt}{\gprus{lgray}{1.7} \gprds{lgray}{1.7}} & $\nu_{eR}^{\wedge/\vee}$ & $\pm \nicefrac{1}{\sqrt{2}} \;$ & $\; 0 \;$ & $+\nicefrac{1}{\sqrt{2}} \;$ & $ +\nha \;$ & $\!\! \, + \nha \;$ & $\!\! +\nha \,\;$ & $\!\!\! +\nha \,\,$ \\
\noalign{\hrule height 0.2pt}
\raisebox{-.5pt}{\garus{lgray}{1.7} \gards{lgray}{1.7}} & $\bar{\nu}_{eR}^{\wedge/\vee}$ & $\pm \nicefrac{1}{\sqrt{2}} \;$ & $-\nicefrac{1}{\sqrt{2}} \;$ & $\; 0 \;$ & $ +\nha \;$ & $\!\! \, -\nha \;$ & $\!\! -\nha \,\;$ & $\!\!\! -\nha \,\,$ \\
\noalign{\hrule height 0.2pt}
\raisebox{-.5pt}{\gprus{myell}{1.7} \gprds{myell}{1.7}}  & $e_R^{\wedge/\vee}$ & $\pm \nicefrac{1}{\sqrt{2}} \;$ & $\; 0 \;$ & $-\nicefrac{1}{\sqrt{2}} \;$ & $ +\nha \;$ & $\!\! \, + \nha \;$ & $\!\! +\nha \,\;$ & $\!\!\! +\nha \,\,$ \\
\noalign{\hrule height 0.2pt}
\raisebox{-.5pt}{\garus{myell}{1.7} \gards{myell}{1.7}}  & $\bar{e}_R^{\wedge/\vee}$ & $\pm \nicefrac{1}{\sqrt{2}} \;$ & $+\nicefrac{1}{\sqrt{2}} \;$ & $\; 0 \;$ & $ +\nha \;$ & $\!\! \, -\nha \;$ & $\!\! -\nha \,\;$ & $\!\!\! -\nha \,\,$ \\
\noalign{\hrule height 0.2pt}
$\;$ \raisebox{-.5pt}{\gprus{mred}{1.7} \!\!\!\!\!\!\! \gprus{mgree}{1.7} \!\!\!\!\!\!\! \gprus{mblue}{1.7} \gprds{mred}{1.7} \!\!\!\!\!\!\! \gprds{mgree}{1.7} \!\!\!\!\!\!\! \gprds{mblue}{1.7}} $\;$& $u_{R}^{\wedge/\vee}$ & $\pm \nicefrac{1}{\sqrt{2}} \;$ & $\; 0 \;$ & $+\nicefrac{1}{\sqrt{2}} \;$ & $ +\nha \;$ & $\!\!\! ( +\nha \;$ & $\!\! -\nha \,\;$ & $\!\! -\nha \, )$ \\
\noalign{\hrule height 0.2pt}
\raisebox{-1pt}{\garus{mred}{1.7} \!\!\!\!\!\!\! \garus{mgree}{1.7} \!\!\!\!\!\!\! \garus{mblue}{1.7} \gards{mred}{1.7} \!\!\!\!\!\!\! \gards{mgree}{1.7} \!\!\!\!\!\!\! \gards{mblue}{1.7}} & $\bar{u}_{R}^{\wedge/\vee}$ & $\pm \nicefrac{1}{\sqrt{2}} \;$ & $-\nicefrac{1}{\sqrt{2}} \;$ & $\; 0 \;$ & $ +\nha \;$ & $\!\!\! ( -\nha \;$ & $\!\! +\nha \,\;$ & $\!\! +\nha \, )$ \\
\noalign{\hrule height 0.2pt}
\raisebox{-.5pt}{\gprus{mrora}{1.7} \!\!\!\!\!\!\! \gprus{mygre}{1.7} \!\!\!\!\!\!\! \gprus{mbvio}{1.7} \gprds{mrora}{1.7} \!\!\!\!\!\!\! \gprds{mygre}{1.7} \!\!\!\!\!\!\! \gprds{mbvio}{1.7}} & $d_{R}^{\wedge/\vee}$ & $\pm \nicefrac{1}{\sqrt{2}} \;$ & $\; 0 \;$ & $-\nicefrac{1}{\sqrt{2}} \;$ & $ +\nha \;$ & $\!\!\! ( +\nha \;$ & $\!\! -\nha \,\;$ & $\!\! -\nha \, )$ \\
\noalign{\hrule height 0.2pt}
\raisebox{-1pt}{\garus{mrora}{1.7} \!\!\!\!\!\!\! \garus{mygre}{1.7} \!\!\!\!\!\!\! \garus{mbvio}{1.7} \gards{mrora}{1.7} \!\!\!\!\!\!\! \gards{mygre}{1.7} \!\!\!\!\!\!\! \gards{mbvio}{1.7}} & $\bar{d}_{R}^{\wedge/\vee}$ & $\pm \nicefrac{1}{\sqrt{2}} \;$ & $+\nicefrac{1}{\sqrt{2}} \;$ & $\; 0 \;$ & $ +\nha \;$ & $\!\!\! ( -\nha \;$ & $\!\! +\nha \,\;$ & $\!\! +\nha \, )$ \\
\noalign{\hrule height 1.0pt}

\end{tabular}
}
\vspace{8pt}
\caption{The $126$ roots of $e_7$, suggestively labeled as elementary particles, with triality eigenspaces corresponding to $h=0$, $h \in \{-1,+\nha \}$, and $h \in \{+1,-\nha \}$. The bosons are the same as for $e_6$ plus a spatial spin connection, \scalebox{.9}{$\om^{\wedge/\vee}$}, and some more Higgs fields, $\Ph$. There is a full set of left-chiral fermions and antifermions with spin, \scalebox{.9}{$f^{\wedge/\vee}_L$} and \scalebox{.9}{$\bar{f}^{\wedge/\vee}_L$}. The fermions labeled \scalebox{.9}{$f^{\wedge/\vee}_R$} and \scalebox{.9}{$\bar{f}^{\wedge/\vee}_R$} can be considered either the right-chiral partners, or as conjugates related to the left-chiral fermions by triality.}
\label{table:pE7}
\end{table}

\section{\texorpdfstring{$e_8$}{e8}}

The $e_8$ Lie algebra has a compound triality decomposition from combining two $f_4$'s,
$$
\ba{rcl}
e_8 &=& so(8)' \,+\, so(8) \,+\, 8'_v \otimes 8_v \,+\, 8'_{s-} \otimes 8_{s-} \,+\, 8'_{s+} \otimes 8_{s+} \\[4pt]
& \sim & su(3,\mathbb{O} \otimes \mathbb{O})
\ea
$$
and a triality automorphism that maps between the triplet of octo-octonions, $\mathbb{O} \otimes \mathbb{O}$. The corresponding set of $e_8$ basis elements is
\beq
\{ \, \ga_{a'b'}, \, \ga_{ab}, \, \ga_{a' a}, \, Q^-_{a' a}, \, Q^+_{a' a} \, \}
\label{e8tri}
\eeq
with primed and unprimed indices, $a'$ and $a$, ranging over octonion indices, $\{ 0, ...,7 \}$, and the bivector indices, $a'b'$ and $ab$, each ranging over the $28$ $so(8)$ basis generator permutations with $a < b$. The non-vanishing Lie algebra brackets between these basis elements come from combining the Lie brackets of two $f_{4(-52)}$'s, (\ref{f4L}):\footnotemark[2]
$$
\ba{rclcrcl}
\!\! \lb \ga_{a' b'}, \ga_{c' d'} \rb &=& 2 \left\{  n'_{a' c'} \ga_{b' d'} - n'_{a' d'} \ga_{b' c'} - n'_{b' c'} \ga_{a' d'} + n'_{b' d'} \ga_{a' c'} \right\} 
\!\!\!\! \!\!\!\! \!\!\!\! \!\!\!\! \!\!\!\! \!\!\!\! \!\!\!\! \!\!\!\! \!\!\!\! \!\!\!\! \!\!\!\! \!\!\!\! \!\!\!\! \!\!\!\! \!\!\!\! \!\!\!\! & & & & \\[4pt]
\!\! \lb \ga_{a b}, \ga_{c d} \rb &=& 2 \left\{ n_{a c} \ga_{b d} - n_{a d} \ga_{b c} - n_{b c} \ga_{a d} + n_{b d} \ga_{a c} \right\}  \!\!\!\! \!\!\!\! \!\!\!\! \!\!\!\! \!\!\!\! \!\!\!\! \!\!\!\! \!\!\!\! \!\!\!\! \!\!\!\! & & & \\[4pt]
\!\! \lb \ga_{a'b'}, \ga_{c'a} \rb &=& 2 \left\{ - n'_{b'c'} \ga_{a'a} + n'_{a'c'} \ga_{b'a} \right\} & & & & \\[4pt]
\!\! \lb \ga_{ab}, \ga_{a'c} \rb &=& 2 \left\{ - n_{bc} \ga_{a'a} + n_{ac} \ga_{a'b} \right\} & \s\s &
\lb \ga_{a'b'}, Q^-_{c'a} \rb &=&  Q^-_{d'a} ( - \bar{\Ga}_{a'} \Ga_{b'} )^{d'}{}_{c'} \\[4pt]
\!\! \lb \ga_{a'a}, \ga_{b'b} \rb &=& + 2 \, n_{ab} \ga_{a'b'} + 2 \, n'_{a'b'} \ga_{ab} & &
\lb \ga_{a'b'}, Q^+_{c'a} \rb &=&  Q^+_{d'a} ( - \Ga_{a'} \bar{\Ga}_{b'} )^{d'}{}_{c'} \\[4pt]
\!\! \lb \ga_{a'a}, Q^-_{b'b} \rb &=&  ( \Ga_{a'} )^{c'}{}_{b'} ( \Ga_a )^c{}_b Q^+_{c'c}   & & 
 \lb \ga_{ab}, Q^-_{a'c} \rb &=&  Q^-_{a'd} ( - \bar{\Ga}_a \Ga_b )^d{}_c \\[4pt]
\!\! \lb \ga_{a'a}, Q^+_{b'b} \rb &=& - ( \bar{\Ga}_{a'} )^{c'}{}_{b'} ( \bar{\Ga}_{a} )^{c}{}_{b}  Q^-_{c'c}  & &
\lb \ga_{ab}, Q^+_{a'c} \rb &=&  Q^+_{a'd} ( - \Ga_a \bar{\Ga}_b )^d{}_c \\[4pt]
\!\! \lb Q^-_{a'a}, Q^-_{b'b} \rb &=& \ha n^-_{ab} \ga_{c'd'}  ( \bar{\Ga}{}^{c'} \Ga^{d'} )_{a'b'} + \ha n'^-_{a'b'} \ga_{c d}  ( \bar{\Ga}{}^c \Ga^d )_{ab} \!\!\!\! \!\!\!\! \!\!\!\! \!\!\!\! \!\!\!\! \!\!\!\! \!\!\!\! \!\!\!\!& & & & \\[4pt]
\!\! \lb Q^+_{a'a}, Q^+_{b'b} \rb &=& \ha n^+_{ab} \ga_{c'd'}  ( \Ga^{c'} \bar{\Ga}{}^{d'} )_{a'b'} + \ha n'^+_{a'b'} \ga_{c d}  ( \Ga^c \bar{\Ga}{}^d )_{ab}  \!\!\!\! \!\!\!\! \!\!\!\! \!\!\!\! \!\!\!\! \!\!\!\! \!\!\!\! \!\!\!\!& & & & \\[4pt]
\!\! \lb Q^-_{a'a}, Q^+_{b'b} \rb &=& ( \bar{\Ga}{}^{c'} )_{a'b'} ( \bar{\Ga}{}^c )_{ab}  \ga_{c'c} & & & &\\[0pt]
\ea 
$$
in which $(\Ga_{c'})^{b'}{}_{a'} = M'_{c'a'}{}^{\os{b}'}$ and $(\Ga_c)^b{}_a = M_{ca}{}^\os{b}$ are octonion multiplication tables with conjugations, and $\{ n', n'^\pm, n, n^\pm \}$ are octonion metrics, also used to raise or lower indices. Different real forms of $e_8$ come from using one or two copies of $f_{4(4)}$, incorporating split-octonion multiplication tables and metrics, or using two $f_{4(-52)}$'s and flipping some signs to get split real $e_{8(8)}$.

\begin{figure}[p]
    \centering
    \vspace{40pt}
    \includegraphics[height=5.0in]{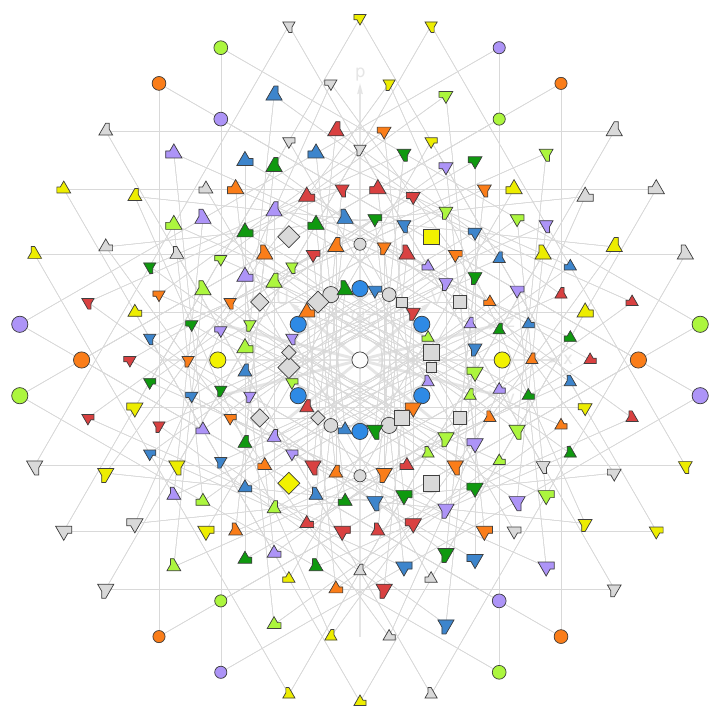}
    \vspace{60pt}
    \caption{The $240$ roots of $e_8$, suggestively labeled as elementary particles, with ``generations'' related by triality.}
    \label{fig:fE8p}
\end{figure}

A canonical triality automorphism of compact real $e_{8(-248)}$ is
$$
\ba{rclcrclcrcl}
t \q : \q   
\ga_{a'b'} &\;\mapsto\;& \ga'_{a'b'} = \ha \ga_{c'd'} t^{c'd'}{}_{a'b'} = \ga_{c'd'} \lp \fr{1}{4} \Ga^{c'} \bar{\Ga}^{d'} \rp_{a'b'}
 \!\!\!\! \!\!\!\! \!\!\!\! \!\!\!\! \!\!\!\! \!\!\!\! \!\!\!\! \!\!\!\! \!\!\!\! \!\!\!\! \!\!\!\! \!\!\!\! \!\!\!\! \!\!\!\! \!\!\!\! \!\!\!\! \!\!\!\! \!\!\!\! \!\!\!\!  \!\!\!\! \!\!\!\!
   & & & & \\[5pt]
\ga_{ab} &\;\mapsto\;& \ga'_{ab} = \ha \ga_{cd} t^{cd}{}_{ab} = \ga_{cd} \lp \fr{1}{4} \Ga^c \bar{\Ga}^d \rp_{ab}
 \!\!\!\! \!\!\!\! \!\!\!\! \!\!\!\! \!\!\!\! \!\!\!\! \!\!\!\! \!\!\!\! \!\!\!\! \!\!\!\! \!\!\!\! \!\!\!\! \!\!\!\! \!\!\!\! \!\!\!\! \!\!\!\! \!\!\!\! \!\!\!\! \!\!\!\! \!\!\!\! \!\!\!\!
   & & & & \\[4pt]
   \ga_{a'a} &\;\mapsto\;& \ga'_{a'a} = Q^+_{a'a} & \s & Q^-_{a'a} &\;\mapsto\;& Q'^-_{a'a} = \ga_{a'a} & \s & Q^+_{a'a} &\;\mapsto\;& Q'^+_{a'a} = Q^-_{a'a} \\
\ea
$$
from combining the triality automorphisms of $f_{4(-52)}$'s. For quaternionic $e_{8(-24)}$ or split real $e_{8(8)}$, some real triality automorphisms --- obtained by mixing triality automorphisms of $f_4$'s --- may contain explicit $i$'s. The $e_8$ roots and canonical triality automorphism, with a triality matrix (\ref{tri}) from combining triality matrices of two $f_4$'s, are shown in Table \ref{table:pE8p} and Figure \ref{fig:fE8p}.

\begin{table}[h!t]
\centering
\scalebox{.54}{
\renewcommand{\arraystretch}{0.95}
\begin{tabular}
{@{\vrule width1.0pt}c@{\vrule width0.2pt}c@{\vrule width1.0pt}c@{\vrule width0.0pt}c@{\vrule width0.0pt}c@{\vrule width0.0pt}c@{\vrule width0.2pt}c@{\vrule width0.0pt}c@
{\vrule width0.0pt}c@{\vrule width0.0pt}c@{\vrule width1.0pt}}
\noalign{\hrule height 1.0pt}
\multicolumn{2}{@{\vrule width1.0pt}c@{\vrule width1.0pt}}{\raisebox{1pt}{$e_8$}} & $\q\; \om_T \;\;\;$ {\vrule width0.2pt} & $\q \om_s \;\;\;$ {\vrule width0.2pt} & $\q U \;\;\;$ {\vrule width0.2pt} & $\q V \q$ & $\q\; p \;\;\;\;$ {\vrule width0.2pt} & $\q\; x \;\;\;\;$ {\vrule width0.2pt} & $\q\; y \;\;\;\;$ {\vrule width0.2pt} & $\q\; z \;\q$ \\
\noalign{\hrule height 1pt}

 \raisebox{-1pt}{\gcirs{lgray}{2}}  & $\;\; \om_I^{\wedge/\vee} \;\;$ & $\; \mp 1 \;$ & $\; \pm 1 \;$ & $\; 0 \;$ & $\; 0 \;$ & $0 $ & $0 $ & $0 $ & $0 $ \\
\noalign{\hrule height 0.2pt}
 \raisebox{-.5pt}{\gcirs{lgray}{1.7}}  & $\;\; \om_{II}^{\wedge/\vee} \;\;$ & $\; \mp 1 \;$ & $\; \mp 1 \;$ & $\; 0 \;$ & $\; 0 \;$  & $0 $ & $0 $ & $0 $ & $0 $ \\
\noalign{\hrule height 0.2pt}
 \raisebox{0pt}{\gcirs{lgray}{1.5}}  & $\;\; \om_{III}^{\wedge/\vee} \;\;$ & $\; 0 \;$ & $\; 0 \;$ & $\; \mp 1 \;$ & $\; \mp 1 \;$  & $0 $ & $0 $ & $0 $ & $0 $ \\
\noalign{\hrule height 0.2pt}
\raisebox{-2pt}{\gcirs{lyell}{2}}  & $\;\; W^{\pm\phantom{R}}_{\phantom{R}}\!\!\! \;\;$ & $\; 0 \;$ & $\; 0 \;$ & $\; \mp 1 \;$ & $\; \pm 1 \;$  & $0 $ & $0 $ & $0 $ & $0 $ \\
\noalign{\hrule height 0.2pt}

 \raisebox{-1pt}{\gsqus{lgray}{2} \gdias{lgray}{2}} & $\;\; e_T^{\wedge/\vee} \ph_{I}^{/*}  \;\;$ & $\; \pm 1 \;$ & $\; 0 \;$ & $\; \mp 1 \;$ & $\; 0 \;$  & $0 $ & $0 $ & $0 $ & $0 $ \\
\noalign{\hrule height 0.2pt}
\gsqus{lgray}{1.7} \gdias{lgray}{1.7} & $\;\; e_T^{\wedge/\vee} \ph_{II}^{/*}  \;\;$ & $\; 0 \;$ & $\; \pm 1 \;$ & $\; 0 \;$ & $\; \pm 1 \;$  & $0 $ & $0 $ & $0 $ & $0 $ \\
\noalign{\hrule height 0.2pt}
\gsqus{lgray}{1.5} \gdias{lgray}{1.5} & $\;\; e_T^{\wedge/\vee} \ph_{III}^{/*}  \;\;$ & $\; 0 \;$ & $\; \mp 1 \;$ & $\; 0 \;$ & $\; \pm 1 \;$  & $0 $ & $0 $ & $0 $ & $0 $ \\
\noalign{\hrule height 0.2pt}

 \raisebox{-1pt}{\gdias{lgray}{2} \gsqus{lgray}{2}} & $\;\; e_s^{\wedge/\vee} \ph_{I}^{*/}  \;\;$ & $\; 0 \;$ & $\; \pm 1 \;$ & $\; \pm 1 \;$ & $\; 0 \;$  & $0 $ & $0 $ & $0 $ & $0 $ \\
\noalign{\hrule height 0.2pt}
\gdias{lgray}{1.7} \gsqus{lgray}{1.7} & $\;\; e_s^{\wedge/\vee} \ph_{II}^{*/}  \;\;$ & $\; 0 \;$ & $\; \mp 1 \;$ & $\; \pm 1 \;$ & $\; 0 \;$  & $0 $ & $0 $ & $0 $ & $0 $ \\
\noalign{\hrule height 0.2pt}
\gdias{lgray}{1.5} \gsqus{lgray}{1.5} & $\;\; e_s^{\wedge/\vee} \ph_{III}^{*/}  \;\;$ & $\; \pm 1 \;$ & $\; 0 \;$ & $\; 0 \;$ & $\; \mp 1 \;$  & $0 $ & $0 $ & $0 $ & $0 $ \\
\noalign{\hrule height 0.2pt}

\raisebox{-1pt}{\gdias{lgray}{2} \gsqus{lgray}{2}} & $\;\; e_T^{\wedge/\vee} \ph^{*/}  \;\;$ & $\; \pm 1 \;$ & $\; 0 \;$ & $\; \pm 1 \;$ & $\; 0 \;$  & $0 $ & $0 $ & $0 $ & $0 $ \\
\noalign{\hrule height 0.2pt}
\raisebox{-1pt}{\gsqus{lyell}{2} \gdias{lyell}{2}} & $\;\; e_T^{\wedge/\vee} \ph_{\pm}  \;\;$ & $\; \pm 1 \;$ & $\; 0 \;$ & $\; 0 \;$ & $\; \pm 1 \;$  & $0 $ & $0 $ & $0 $ & $0 $ \\
\noalign{\hrule height .8pt}

\raisebox{-1.5pt}{\gcirs{lblue}{2}} & \raisebox{1pt}{$\;\; g^{\phantom{R}}_{\phantom{R}}\!\!\! \;\;$} & $0 $ & $0 $ & $0 $ & $0 $ & $0 $ & $\! (+1 \;$ & $-1 \;$ & $\;\;\; 0 \; ) $  \\
\noalign{\hrule height 0.2pt}
$\;\!$ \raisebox{-2pt}{\gcirs{mrora}{2} $\!$ \gcirs{mygre}{2} $\!$ \gcirs{mbvio}{2}} $\;\!$ & $\;\, X_{I}^{rgb} \; \bar{X}_{I}^{rgb} \,\;$ & $0 $ & $0 $ & $0 $ & $0 $ & $ 0 $ & $\! ( \pm 1 \;$ & $ \pm 1 \;$ & $\;\;\; 0 \; ) $ \\
\noalign{\hrule height 0.2pt}
 \raisebox{-1pt}{\gcirs{mrora}{1.7} $\!$ \gcirs{mygre}{1.7} $\!$ \gcirs{mbvio}{1.7}}  & $ X_{II}^{rgb} \; \bar{X}_{II}^{rgb} $ & $0 $ & $0 $ & $0 $ & $0 $ & $ \pm 1 $ & $\! ( \mp 1 \;$ & $ 0 $ & $\;\;\; 0 \; )$ \\
\noalign{\hrule height 0.2pt}
 \gcirs{mrora}{1.5}  \hspace{-.2em}  \gcirs{mygre}{1.5}  \hspace{-.2em}  \gcirs{mbvio}{1.5}  $\!\!$ & $\; X_{III}^{rgb} \; \bar{X}_{III}^{rgb} \;$ & $0 $ & $0 $ & $0 $ & $0 $ & $ \mp 1 $ & $\! ( \mp 1 \;$ & $ 0 $ & $\;\;\; 0 \; )$ \\
\noalign{\hrule height .8pt}

\raisebox{-.5pt}{\gplus{lgray}{2} \gplds{lgray}{2}} & $\nu_{eL}^{\wedge/\vee}$ & $\mp \nha \;$ & $\pm \nha \;$ & $- \nha \;$ & $+ \nha \;$ & $ + \nha \;$ & $\!\! \, + \nha \;$ & $\!\! +\nha \,\;$ & $\!\!\! +\nha \,\,$ \\
\noalign{\hrule height 0.2pt}
\raisebox{-.5pt}{\gprus{lgray}{2} \gprds{lgray}{2}}  & $\nu_{eR}^{\wedge/\vee}$ & $\pm \nha \;$ & $\pm \nha \;$ & $+ \nha \;$ & $+ \nha \;$ & $ + \nha \;$ & $\!\! \, + \nha \;$ & $\!\! +\nha \,\;$ & $\!\!\! +\nha \,\,$ \\
\noalign{\hrule height 0.2pt}
\raisebox{-1pt}{\galus{lgray}{2} \galds{lgray}{2}} & $\bar{\nu}_{eL}^{\wedge/\vee}$ & $\pm \nha \;$ & $\pm \nha \;$ & $- \nha \;$ & $- \nha \;$ & $ -\nha \;$ & $\!\! \, - \nha \;$ & $\!\! -\nha \,\;$ & $\!\!\! -\nha \,\,$ \\
\noalign{\hrule height 0.2pt}
\raisebox{-1pt}{\garus{lgray}{2} \gards{lgray}{2}} & $\bar{\nu}_{eR}^{\wedge/\vee}$ & $\mp \nha \;$ & $\pm \nha \;$ & $+ \nha \;$ & $- \nha \;$ & $ -\nha \;$ & $\!\! \, - \nha \;$ & $\!\! -\nha \,\;$ & $\!\!\! -\nha \,\,$ \\
\noalign{\hrule height 0.2pt}
\raisebox{-.5pt}{\gplus{myell}{2} \gplds{myell}{2}}  & $e_L^{\wedge/\vee}$ & $\mp \nha \;$ & $\pm \nha \;$ & $+ \nha \;$ & $- \nha \;$ & $ + \nha \;$ & $\!\! \, + \nha \;$ & $\!\! +\nha \,\;$ & $\!\!\! +\nha \,\,$ \\
\noalign{\hrule height 0.2pt}
\raisebox{-.5pt}{\gprus{myell}{2} \gprds{myell}{2}}  & $e_R^{\wedge/\vee}$ & $\pm \nha \;$ & $\pm \nha \;$ & $- \nha \;$ & $- \nha \;$ & $ + \nha \;$ & $\!\! \, + \nha \;$ & $\!\! +\nha \,\;$ & $\!\!\! +\nha \,\,$ \\
\noalign{\hrule height 0.2pt}
\raisebox{-1pt}{\galus{myell}{2} \galds{myell}{2}} & $\bar{e}_{L}^{\wedge/\vee}$ & $\pm \nha \;$ & $\pm \nha \;$ & $+ \nha \;$ & $+ \nha \;$ & $ -\nha \;$ & $\!\! \, - \nha \;$ & $\!\! -\nha \,\;$ & $\!\!\! -\nha \,\,$ \\
\noalign{\hrule height 0.2pt}
\raisebox{-1pt}{\garus{myell}{2} \gards{myell}{2}} & $\bar{e}_{R}^{\wedge/\vee}$ & $\mp \nha \;$ & $\pm \nha \;$ & $- \nha \;$ & $+ \nha \;$ & $ -\nha \;$ & $\!\! \, - \nha \;$ & $\!\! -\nha \,\;$ & $\!\!\! -\nha \,\,$ \\
\noalign{\hrule height 0.2pt}
$\;$ \raisebox{-.5pt}{\gplus{mred}{2} \!\!\!\!\!\!\! \gplus{mgree}{2} \!\!\!\!\!\!\! \gplus{mblue}{2} \gplds{mred}{2} \!\!\!\!\!\!\! \gplds{mgree}{2} \!\!\!\!\!\!\! \gplds{mblue}{2}} $\;$& $u_{L}^{\wedge/\vee}$ & $\mp \nha \;$ & $\pm \nha \;$ & $- \nha \;$ & $+ \nha \;$ & $ + \nha \;$ & $\!\!\! ( + \nha \;$ & $\!\! -\nha \,\;$ & $\!\! -\nha \, )$ \\
\noalign{\hrule height 0.2pt}
\raisebox{-.5pt}{\gprus{mred}{2} \!\!\!\!\!\!\! \gprus{mgree}{2} \!\!\!\!\!\!\! \gprus{mblue}{2} \gprds{mred}{2} \!\!\!\!\!\!\! \gprds{mgree}{2} \!\!\!\!\!\!\! \gprds{mblue}{2}} & $u_{R}^{\wedge/\vee}$ & $\pm \nha \;$ & $\pm \nha \;$ & $+ \nha \;$ & $+ \nha \;$ & $ + \nha \;$ & $\!\!\! ( + \nha \;$ & $\!\! -\nha \,\;$ & $\!\! -\nha \, )$ \\
\noalign{\hrule height 0.2pt}
\raisebox{-1pt}{\galus{mred}{2} \!\!\!\!\!\!\! \galus{mgree}{2} \!\!\!\!\!\!\! \galus{mblue}{2} \galds{mred}{2} \!\!\!\!\!\!\! \galds{mgree}{2} \!\!\!\!\!\!\! \galds{mblue}{2}} & $\bar{u}_{L}^{\wedge/\vee}$ & $\pm \nha \;$ & $\pm \nha \;$ & $- \nha \;$ & $- \nha \;$ & $ - \nha \;$ & $\!\!\! ( - \nha \;$ & $\!\! +\nha \,\;$ & $\!\! +\nha \, )$ \\
\noalign{\hrule height 0.2pt}
\raisebox{-1pt}{\garus{mred}{2} \!\!\!\!\!\!\! \garus{mgree}{2} \!\!\!\!\!\!\! \garus{mblue}{2} \gards{mred}{2} \!\!\!\!\!\!\! \gards{mgree}{2} \!\!\!\!\!\!\! \gards{mblue}{2}} & $\bar{u}_{R}^{\wedge/\vee}$ & $\mp \nha \;$ & $\pm \nha \;$ & $+ \nha \;$ & $- \nha \;$ & $ - \nha \;$ & $\!\!\! ( - \nha \;$ & $\!\! +\nha \,\;$ & $\!\! +\nha \, )$ \\
\noalign{\hrule height 0.2pt}
\raisebox{-.5pt}{\gplus{mrora}{2} \!\!\!\!\!\!\! \gplus{mygre}{2} \!\!\!\!\!\!\! \gplus{mbvio}{2} \gplds{mrora}{2} \!\!\!\!\!\!\! \gplds{mygre}{2} \!\!\!\!\!\!\! \gplds{mbvio}{2}} & $d_{L}^{\wedge/\vee}$ & $\mp \nha \;$ & $\pm \nha \;$ & $+ \nha \;$ & $- \nha \;$ & $ + \nha \;$ & $\!\!\! ( + \nha \;$ & $\!\! -\nha \,\;$ & $\!\! -\nha \, )$ \\
\noalign{\hrule height 0.2pt}
\raisebox{-.5pt}{\gprus{mrora}{2} \!\!\!\!\!\!\! \gprus{mygre}{2} \!\!\!\!\!\!\! \gprus{mbvio}{2} \gprds{mrora}{2} \!\!\!\!\!\!\! \gprds{mygre}{2} \!\!\!\!\!\!\! \gprds{mbvio}{2}} & $d_{R}^{\wedge/\vee}$ & $\pm \nha \;$ & $\pm \nha \;$ & $- \nha \;$ & $- \nha \;$ & $ + \nha \;$ & $\!\!\! ( + \nha \;$ & $\!\! -\nha \,\;$ & $\!\! -\nha \, )$ \\
\noalign{\hrule height 0.2pt}
\raisebox{-1pt}{\galus{mrora}{2} \!\!\!\!\!\!\! \galus{mygre}{2} \!\!\!\!\!\!\! \galus{mbvio}{2} \galds{mrora}{2} \!\!\!\!\!\!\! \galds{mygre}{2} \!\!\!\!\!\!\! \galds{mbvio}{2}} & $\bar{d}_{L}^{\wedge/\vee}$ & $\pm \nha \;$ & $\pm \nha \;$ & $+ \nha \;$ & $+ \nha \;$ & $ - \nha \;$ & $\!\!\! ( - \nha \;$ & $\!\! +\nha \,\;$ & $\!\! +\nha \, )$ \\
\noalign{\hrule height 0.2pt}
\raisebox{-1pt}{\garus{mrora}{2} \!\!\!\!\!\!\! \garus{mygre}{2} \!\!\!\!\!\!\! \garus{mbvio}{2} \gards{mrora}{2} \!\!\!\!\!\!\! \gards{mygre}{2} \!\!\!\!\!\!\! \gards{mbvio}{2}} & $\bar{d}_{R}^{\wedge/\vee}$ & $\mp \nha \;$ & $\pm \nha \;$ & $- \nha \;$ & $+ \nha \;$ & $ - \nha \;$ & $\!\!\! ( - \nha \;$ & $\!\! +\nha \,\;$ & $\!\! +\nha \, )$ \\
\noalign{\hrule height .8pt}

\raisebox{0pt}{\gplus{lgray}{1.7} \gplds{lgray}{1.7}} & $\nu_{\mu L}^{\wedge/\vee}$ & $\mp \nha \;$ & $\mp \nha \;$ & $- \nha \;$ & $+ \nha \;$ & $ +\nha \;$ & $\!\! \, -\nha \;$ & $\!\! -\nha \,\;$ & $\!\!\! -\nha \,\,$ \\
\noalign{\hrule height 0.2pt}
\raisebox{0pt}{\gprus{lgray}{1.7} \gprds{lgray}{1.7}}  & $\nu_{\mu R}^{\wedge/\vee}$ & $+ \nha \;$ & $- \nha \;$ & $\pm \nha \;$ & $\pm \nha \;$ & $ +\nha \;$ & $\!\! \, - \nha \;$ & $\!\! -\nha \,\;$ & $\!\!\! -\nha \,\,$ \\
\noalign{\hrule height 0.2pt}
\raisebox{0pt}{\galus{lgray}{1.7} \galds{lgray}{1.7}} & $\bar{\nu}_{\mu L}^{\wedge/\vee}$ & $- \nha \;$ & $+ \nha \;$ & $\pm \nha \;$ & $\pm \nha \;$ & $ -\nha \;$ & $\!\! \, + \nha \;$ & $\!\! +\nha \,\;$ & $\!\!\! +\nha \,\,$ \\
\noalign{\hrule height 0.2pt}
\raisebox{0pt}{\garus{lgray}{1.7} \gards{lgray}{1.7}} & $\bar{\nu}_{\mu R}^{\wedge/\vee}$ & $\mp \nha \;$ & $\mp \nha \;$ & $+ \nha \;$ & $- \nha \;$ & $ -\nha \;$ & $\!\! \, + \nha \;$ & $\!\! +\nha \,\;$ & $\!\!\! +\nha \,\,$ \\
\noalign{\hrule height 0.2pt}
\raisebox{0pt}{\gplus{myell}{1.7} \gplds{myell}{1.7}}  & $\mu_L^{\wedge/\vee}$ & $\mp \nha \;$ & $\mp \nha \;$ & $+ \nha \;$ & $- \nha \;$ & $ + \nha \;$ & $\!\! \, - \nha \;$ & $\!\! - \nha \,\;$ & $\!\!\! -\nha \,\,$ \\
\noalign{\hrule height 0.2pt}
\raisebox{0pt}{\gprus{myell}{1.7} \gprds{myell}{1.7}}  & $\mu_R^{\wedge/\vee}$ & $- \nha \;$ & $+ \nha \;$ & $\pm \nha \;$ & $\pm \nha \;$ & $ + \nha \;$ & $\!\! \, - \nha \;$ & $\!\! -\nha \,\;$ & $\!\!\! -\nha \,\,$ \\
\noalign{\hrule height 0.2pt}
\raisebox{0pt}{\galus{myell}{1.7} \galds{myell}{1.7}} & $\bar{\mu}_{L}^{\wedge/\vee}$ & $+ \nha \;$ & $- \nha \;$ & $\pm \nha \;$ & $\pm \nha \;$ & $ - \nha \;$ & $\!\! \, + \nha \;$ & $\!\! +\nha \,\;$ & $\!\!\! +\nha \,\,$ \\
\noalign{\hrule height 0.2pt}
\raisebox{0pt}{\garus{myell}{1.7} \gards{myell}{1.7}} & $\bar{\mu}_{R}^{\wedge/\vee}$ & $\mp \nha \;$ & $\mp \nha \;$ & $- \nha \;$ & $+ \nha \;$ & $ - \nha \;$ & $\!\! \, + \nha \;$ & $\!\! +\nha \,\;$ & $\!\!\! +\nha \,\,$ \\

\noalign{\hrule height 0.2pt}
\raisebox{0pt}{\gplus{mred}{1.7} \!\!\!\!\!\!\! \gplus{mgree}{1.7} \!\!\!\!\!\!\! \gplus{mblue}{1.7} \gplds{mred}{1.7} \!\!\!\!\!\!\! \gplds{mgree}{1.7} \!\!\!\!\!\!\! \gplds{mblue}{1.7}} & $c_{L}^{\wedge/\vee}$ & $\mp \nha \;$ & $\mp \nha \;$ & $- \nha \;$ & $+ \nha \;$ & $ - \nha \;$ & $\!\!\! ( + \nha \;$ & $\!\! -\nha \,\;$ & $\!\! -\nha \, )$ \\
\noalign{\hrule height 0.2pt}
\raisebox{0pt}{\gprus{mred}{1.7} \!\!\!\!\!\!\! \gprus{mgree}{1.7} \!\!\!\!\!\!\! \gprus{mblue}{1.7} \gprds{mred}{1.7} \!\!\!\!\!\!\! \gprds{mgree}{1.7} \!\!\!\!\!\!\! \gprds{mblue}{1.7}} & $c_{R}^{\wedge/\vee}$ & $+ \nha \;$ & $- \nha \;$ & $\pm \nha \;$ & $\pm \nha \;$ & $ - \nha \;$ & $\!\!\! ( + \nha \;$ & $\!\! -\nha \,\;$ & $\!\! -\nha \, )$ \\
\noalign{\hrule height 0.2pt}
\raisebox{0pt}{\galus{mred}{1.7} \!\!\!\!\!\!\! \galus{mgree}{1.7} \!\!\!\!\!\!\! \galus{mblue}{1.7} \galds{mred}{1.7} \!\!\!\!\!\!\! \galds{mgree}{1.7} \!\!\!\!\!\!\! \galds{mblue}{1.7}} & $\bar{c}_{L}^{\wedge/\vee}$ & $- \nha \;$ & $+ \nha \;$ & $\pm \nha \;$ & $\pm \nha \;$ & $ + \nha \;$ & $\!\!\! ( - \nha \;$ & $\!\! +\nha \,\;$ & $\!\! +\nha \, )$ \\
\noalign{\hrule height 0.2pt}
\raisebox{0pt}{\garus{mred}{1.7} \!\!\!\!\!\!\! \garus{mgree}{1.7} \!\!\!\!\!\!\! \garus{mblue}{1.7} \gards{mred}{1.7} \!\!\!\!\!\!\! \gards{mgree}{1.7} \!\!\!\!\!\!\! \gards{mblue}{1.7}} & $\bar{c}_{R}^{\wedge/\vee}$ & $\mp \nha \;$ & $\mp \nha \;$ & $+ \nha \;$ & $- \nha \;$ & $ + \nha \;$ & $\!\!\! ( - \nha \;$ & $\!\! +\nha \,\;$ & $\!\! +\nha \, )$ \\
\noalign{\hrule height 0.2pt}
\raisebox{0pt}{\gplus{mrora}{1.7} \!\!\!\!\!\!\! \gplus{mygre}{1.7} \!\!\!\!\!\!\! \gplus{mbvio}{1.7} \gplds{mrora}{1.7} \!\!\!\!\!\!\! \gplds{mygre}{1.7} \!\!\!\!\!\!\! \gplds{mbvio}{1.7}} & $s_{L}^{\wedge/\vee}$ & $\mp \nha \;$ & $\mp \nha \;$ & $+ \nha \;$ & $- \nha \;$ & $ - \nha \;$ & $\!\!\! ( + \nha \;$ & $\!\! -\nha \,\;$ & $\!\! -\nha \, )$ \\
\noalign{\hrule height 0.2pt}
\raisebox{0pt}{\gprus{mrora}{1.7} \!\!\!\!\!\!\! \gprus{mygre}{1.7} \!\!\!\!\!\!\! \gprus{mbvio}{1.7} \gprds{mrora}{1.7} \!\!\!\!\!\!\! \gprds{mygre}{1.7} \!\!\!\!\!\!\! \gprds{mbvio}{1.7}} & $s_{R}^{\wedge/\vee}$ & $- \nha \;$ & $+ \nha \;$ & $\pm \nha \;$ & $\pm \nha \;$ & $ - \nha \;$ & $\!\!\! ( + \nha \;$ & $\!\! -\nha \,\;$ & $\!\! -\nha \, )$ \\
\noalign{\hrule height 0.2pt}
\raisebox{0pt}{\galus{mrora}{1.7} \!\!\!\!\!\!\! \galus{mygre}{1.7} \!\!\!\!\!\!\! \galus{mbvio}{1.7} \galds{mrora}{1.7} \!\!\!\!\!\!\! \galds{mygre}{1.7} \!\!\!\!\!\!\! \galds{mbvio}{1.7}} & $\bar{s}_{L}^{\wedge/\vee}$ & $+ \nha \;$ & $- \nha \;$ & $\pm \nha \;$ & $\pm \nha \;$ & $ + \nha \;$ & $\!\!\! ( - \nha \;$ & $\!\! +\nha \,\;$ & $\!\! +\nha \, )$ \\
\noalign{\hrule height 0.2pt}
\raisebox{0pt}{\garus{mrora}{1.7} \!\!\!\!\!\!\! \garus{mygre}{1.7} \!\!\!\!\!\!\! \garus{mbvio}{1.7} \gards{mrora}{1.7} \!\!\!\!\!\!\! \gards{mygre}{1.7} \!\!\!\!\!\!\! \gards{mbvio}{1.7}} & $\bar{s}_{R}^{\wedge/\vee}$ & $\mp \nha \;$ & $\mp \nha \;$ & $- \nha \;$ & $+ \nha \;$ & $ + \nha \;$ & $\!\!\! ( - \nha \;$ & $\!\! +\nha \,\;$ & $\!\! +\nha \, )$ \\
\noalign{\hrule height .8pt}

\raisebox{-.5pt}{\gplus{lgray}{1.5} \gards{lgray}{1.5}} & $\nu_{\ta L}^{\wedge} \; \bar{\nu}_{\ta R}^{\vee}$ & $0$ & $0$ & $\mp 1$ & $0$ & $\mp 1$ & $0$ & $0$ & $0$ \\
\noalign{\hrule height 0.2pt}
\raisebox{-.5pt}{\gplds{lgray}{1.5} \garus{lgray}{1.5}} & $\nu_{\ta L}^{\vee} \; \bar{\nu}_{\ta R}^{\wedge}$ & $0$ & $0$ & $0$ & $\pm 1$ & $\mp 1$ & $0$ & $0$ & $0$ \\
\noalign{\hrule height 0.2pt}
\raisebox{-.5pt}{\gprus{lgray}{1.5} \galds{lgray}{1.5}} & $\nu_{\ta R}^{\wedge} \; \bar{\nu}_{\ta L}^{\vee}$ & $\pm 1$ & $0$ & $0$ & $0$ & $\mp 1$ & $0$ & $0$ & $0$ \\
\noalign{\hrule height 0.2pt}
\raisebox{-.5pt}{\gprds{lgray}{1.5} \galus{lgray}{1.5}} & $\nu_{\ta R}^{\vee} \; \bar{\nu}_{\ta L}^{\wedge}$ & $0$ & $\pm 1$ & $0$ & $0$ & $\mp 1$ & $0$ & $0$ & $0$ \\
\noalign{\hrule height 0.2pt}
\raisebox{-.5pt}{\gplus{lyell}{1.5} \gards{lyell}{1.5}} & $\ta_{L}^{\wedge} \; \bar{\ta}_{R}^{\vee}$ & $0$ & $0$ & $0$ & $\mp 1$ & $\mp 1$ & $0$ & $0$ & $0$ \\
\noalign{\hrule height 0.2pt}
\raisebox{-.5pt}{\gplds{lyell}{1.5} \garus{lyell}{1.5}} & $\ta_{L}^{\vee} \; \bar{\ta}_{R}^{\wedge}$ & $0$ & $0$ & $\pm 1$ & $0$ & $\mp 1$ & $0$ & $0$ & $0$ \\
\noalign{\hrule height 0.2pt}
\raisebox{-.5pt}{\gprus{lyell}{1.5} \galds{lyell}{1.5}} & $\ta_{R}^{\wedge} \; \bar{\ta}_{L}^{\vee}$ & $0$ & $\mp 1$ & $0$ & $0$ & $\mp 1$ & $0$ & $0$ & $0$ \\
\noalign{\hrule height 0.2pt}
\raisebox{-.5pt}{\gprds{lyell}{1.5} \galus{lyell}{1.5}} & $\ta_{R}^{\vee} \; \bar{\ta}_{L}^{\wedge}$ & $\mp 1$ & $0$ & $0$ & $0$ & $\mp 1$ & $0$ & $0$ & $0$ \\
\noalign{\hrule height 0.2pt}
\raisebox{-.5pt}{\gplus{mred}{1.5} \!\!\!\!\!\!\! \gplus{mgree}{1.5} \!\!\!\!\!\!\! \gplus{mblue}{1.5} \gards{mred}{1.5} \!\!\!\!\!\!\! \gards{mgree}{1.5} \!\!\!\!\!\!\! \gards{mblue}{1.5}} & $\;\; t_{L}^{\wedge} \; \bar{t}_{R}^{\vee} \;\,$ & $0$ & $0$ & $\mp 1$ & $0$ & $0$ & $\! (  \pm 1 \;$ & $0$ & $\;\;\; 0 \; )$ \\
\noalign{\hrule height 0.2pt}
\raisebox{-.5pt}{\gplds{mred}{1.5} \!\!\!\!\!\!\! \gplds{mgree}{1.5} \!\!\!\!\!\!\! \gplds{mblue}{1.5} \garus{mred}{1.5} \!\!\!\!\!\!\! \garus{mgree}{1.5} \!\!\!\!\!\!\! \garus{mblue}{1.5}} & $\;\; t_{L}^{\vee} \; \bar{t}_{R}^{\wedge} \;\,$ & $0$ & $0$ & $0$ & $\pm 1$ & $0$ & $\! (  \pm 1 \;$ & $0$ & $\;\;\; 0 \; )$ \\
\noalign{\hrule height 0.2pt}
\raisebox{-.5pt}{\gprus{mred}{1.5} \!\!\!\!\!\!\! \gprus{mgree}{1.5} \!\!\!\!\!\!\! \gprus{mblue}{1.5} \galds{mred}{1.5} \!\!\!\!\!\!\! \galds{mgree}{1.5} \!\!\!\!\!\!\! \galds{mblue}{1.5}} & $\;\; t_{R}^{\wedge} \; \bar{t}_{L}^{\vee} \;\,$ & $\pm 1$ & $0$ & $0$ & $0$ & $0$ & $\! (  \pm 1 \;$ & $0$ & $\;\;\; 0 \; )$ \\
\noalign{\hrule height 0.2pt}
\raisebox{-.5pt}{\gprds{mred}{1.5} \!\!\!\!\!\!\! \gprds{mgree}{1.5} \!\!\!\!\!\!\! \gprds{mblue}{1.5} \galus{mred}{1.5} \!\!\!\!\!\!\! \galus{mgree}{1.5} \!\!\!\!\!\!\! \galus{mblue}{1.5}} & $\;\; t_{R}^{\vee} \; \bar{t}_{L}^{\wedge} \;\,$ & $0$ & $\pm 1$ & $0$ & $0$ & $0$ & $\! (  \pm 1 \;$ & $0$ & $\;\;\; 0 \; )$ \\
\noalign{\hrule height 0.2pt}
\raisebox{-.5pt}{\gplus{mrora}{1.5} \!\!\!\!\!\!\! \gplus{mygre}{1.5} \!\!\!\!\!\!\! \gplus{mbvio}{1.5} \gards{mrora}{1.5} \!\!\!\!\!\!\! \gards{mygre}{1.5} \!\!\!\!\!\!\! \gards{mbvio}{1.5}} & $\;\; b_{L}^{\wedge} \; \bar{b}_{R}^{\vee} \;\,$ & $0$ & $0$ & $0$ & $\mp 1$ & $0$ & $\! (  \pm 1 \;$ & $0$ & $\;\;\; 0 \; )$ \\
\noalign{\hrule height 0.2pt}
\raisebox{-.5pt}{\gplds{mrora}{1.5} \!\!\!\!\!\!\! \gplds{mygre}{1.5} \!\!\!\!\!\!\! \gplds{mbvio}{1.5} \garus{mrora}{1.5} \!\!\!\!\!\!\! \garus{mygre}{1.5} \!\!\!\!\!\!\! \garus{mbvio}{1.5}} & $\;\; b_{L}^{\vee} \; \bar{b}_{R}^{\wedge} \;\,$ & $0$ & $0$ & $\pm 1$ & $0$ & $0$ & $\! (  \pm 1 \;$ & $0$ & $\;\;\; 0 \; )$ \\
\noalign{\hrule height 0.2pt}
\raisebox{-.5pt}{\gprus{mrora}{1.5} \!\!\!\!\!\!\! \gprus{mygre}{1.5} \!\!\!\!\!\!\! \gprus{mbvio}{1.5} \galds{mrora}{1.5} \!\!\!\!\!\!\! \galds{mygre}{1.5} \!\!\!\!\!\!\! \galds{mbvio}{1.5}} & $\;\; b_{R}^{\wedge} \; \bar{b}_{L}^{\vee} \;\,$ & $0$ & $\mp 1$ & $0$ & $0$ & $0$ & $\! (  \pm 1 \;$ & $0$ & $\;\;\; 0 \; )$ \\
\noalign{\hrule height 0.2pt}
\raisebox{-.5pt}{\gprds{mrora}{1.5} \!\!\!\!\!\!\! \gprds{mygre}{1.5} \!\!\!\!\!\!\! \gprds{mbvio}{1.5} \galus{mrora}{1.5} \!\!\!\!\!\!\! \galus{mygre}{1.5} \!\!\!\!\!\!\! \galus{mbvio}{1.5}} & $\;\; b_{R}^{\vee} \; \bar{b}_{L}^{\wedge} \;\,$ & $\mp 1$ & $0$ & $0$ & $0$ & $0$ & $\! (  \pm 1 \;$ & $0$ & $\;\;\; 0 \; )$ \\
\noalign{\hrule height 1.0pt}
\end{tabular}
}
\vspace{8pt}
\caption{The $240$ roots of the compact real form, $e_{8(-248)}$, suggestively labeled as elementary particles. \scalebox{.9}{$\om^{\wedge/\vee}_{I,II,III}$} are gravitational spin connection roots related by triality, \scalebox{.9}{$W^\pm$} are weak bosons, \scalebox{.9}{$e^{\wedge/\vee}_{s/T} \ph$} are gravitational frame-Higgs roots, $g$ are gluons, and $X$ are colored X bosons. There is a complete Standard Model generation of fermions with spin, related to two other ``generations'' by triality.}
\label{table:pE8p}
\end{table}

\newpage

\clearpage
\clearpage

\newpage

The particle assignment and canonical triality automorphism, (\ref{tri}), in Table \ref{table:pE8p} and Figure \ref{fig:fE8p}, is consistent with quaternionic $e_{8(-24)}$ or compact $e_{8(-248)}$. Using the split-octonions and/or octonions, corresponding to Clifford algebras $Cl(4,4)$ or $Cl(0,8)$, the Cartan subalgebras are spanned by commuting Clifford basis bivectors, constructed from suitably re-numbered Clifford basis vectors with indices $1,\ldots,8$. We can use the ``$+$'' sign in (\ref{ClD}) for the split-octonion factor and the ``$-$'' sign for octonion factors, giving signatures:
$$
\ba{rcl}
\{ \om_T, \om_s, U, V; p, x, y, z \} &=& \{ \ga_{3'\,4'\,}, \ga_{1'\,2'}, \ga_{5'\,6'} ,\! \ga_{7'\,8'\,}; \ga_{7\,\,8}, \ga_{1\,\,2}, \ga_{3\,\,4}, \ga_{5\,\,6\,} \} \\
 e_{8(-24)}           &:& \{ \ga_{+\,+\,},   \ga_{+\,+\,},    \ga_{-\,-}, \ga_{-\,-};    \ga_{\!--},   \ga_{\!--}, \ga_{\!--}, \ga_{\!--} \} \\
e_{8(-248)}           &:& \{ \ga_{-\,-\,},   \ga_{-\,-\,},    \ga_{-\,-}, \ga_{-\,-};    \ga_{\!--},   \ga_{\!--}, \ga_{\!--}, \ga_{\!--} \} \\
\ea
$$
For $e_{8(8)}$ with compact $so(8)' + so(8)$, one of the three $64$ blocks is compact and two are noncompact, so preservation of the Killing form forbids a real automorphism cycling them. Other embeddings of $e_{8(8)}$ admit compact Cartans compatible with triality.

The only way to accommodate three triality-related linearly-independent fermion generations in $e_8$ is by having Euclidean rather than Lorentzian spacetime and spinors. 

More realistic models, with three triality-related generations, including Lorentzian spacetime and spinors, are achieved by considering the triality eigenspaces, (\ref{teig}), of split-real $e_{8(8)}$ or quaternionic $e_{8(-24)}$, and their triality-related five-gradings, (\ref{grad}). The relevant Cartan subalgebra generators compatible with $e_8$ triality eigenspaces are: 
\beq
\ba{rcl}
\{ \om_t, \om_s, U, V, w, x, y, z \} &=& \{ \ga_{3\,15}, \ga_{1\,\,2\,}, \ga_{5\,\,6\,}, \ga_{7\,\,8\,},   \ga_{4\,16}, \ga_{9\,10}, \ga_{11\,12}, \ga_{13\,14} \} \\
     e_{8(8)}                 &:& \{ \ga_{+\,+},   \ga_{\!++},   \ga_{++},             \ga_{++},        \ga_{\,--},   \ga_{--}, \ga_{\,-\,\,-\,}, \ga_{\,-\,\,-\,} \} \\
     e_{8(-24)}                 &:& \{ \ga_{+\,+},   \ga_{\!++},    \ga_{--},              \ga_{--},          \ga_{\,--},   \ga_{--}, \ga_{\,-\,\,-\,}, \ga_{\,-\,\,-\,} \} \\
\ea
\label{CarC}
\eeq
Within these Cartans, for $e_{8(8)}$ or $e_{8(-24)}$, $w = \ga_{4\,16}$ is the $u(1)$ generator in a triality invariant eigenspace, either $[ so(8,6) + u(1)_w ]$ or $[ so(4,10) + u(1)_w ]$. The corresponding triality automorphism acts as multiplication by $e^{+ \nfr{2 \pi i}{3}}$ on the $[ 14^v_{+1} + 64^{s-}_{- \nha} ]$ and $e^{- \nfr{2 \pi i}{3}}$ on the complex conjugate, $[ 14^v_{-1} + 64^{s+}_{+ \nha} ]$. These same triality automorphisms, $\Th$, act in interesting ways within the five-graded decompositions, (\ref{grad}). The Cartans for the matched five-graded decompositions differ from the Cartans for the triality eigenspaces by a kind of Wick rotation:
\beq
\ba{rcl} 
\{ \om^{\mathbb{R}}_t, \om_s, U, V, w^{\mathbb{R}}, x, y, z \} &=& \{ \ga_{3\,\,4\,}, \ga_{1\,\,2\,}, \ga_{5\,\,6\,}, \ga_{7\,\,8\,}, \ga_{15\,16}, \ga_{9\,10}, \ga_{11\,12}, \ga_{13\,14} \} \\
     e_{8(8)}                 &:& \{ \ga_{+-},   \ga_{\!++},   \ga_{++},             \ga_{++},        \ga_{\,+\,\,-\,},   \ga_{--}, \ga_{\,-\,\,-\,}, \ga_{\,-\,\,-\,} \} \\
     e_{8(-24)}                 &:& \{ \ga_{+-},   \ga_{\!++},    \ga_{--},              \ga_{--},          \ga_{\,+\,\,-\,},   \ga_{--}, \ga_{\,-\,\,-\,}, \ga_{\,-\,\,-\,} \} \\
\ea
\label{CarM}
\eeq
This Wick rotation exchanges the temporal $\ga_{4}$ and a spatial $\ga_{15}$. The resulting $\om^{\mathbb{R}}_t = \ga_{3\,4}$ and $w^{\mathbb{R}}= \ga_{15\,16}$ have real eigenvalues, while $\om_t = \ga_{3\,15}$ and $w= \ga_{4\,16}$ have imaginary eigenvalues.

An explicit description of a continuous Wick rotation within $e_8$ is given by defining a pseudo-similarity transformation acting on $e_8$ Clifford vector, bivector, and spinor generators,\footnotemark[1]
$$
S_\th = e^{i \fr{\th}{2} \ga_{4} \ga_{15}} \s \ga_a \;\mapsto\; \ga^\th_a = S_\th \ga_a S^-_\th \s \ga_{ab} \;\mapsto\; \ga^\th_{ab} = S_\th \ga_{ab} S^-_\th \s \Ps \;\mapsto\; \Ps_\th = S_\th \Ps
$$
with $0 \le \th \le \fr{\pi}{2}$. This corresponds to a Wick rotation of time by $t \mapsto t_\th = e^{i \th} t$. For $\th = \fr{\pi}{2}$ this gives Euclidean time, $t_E = i \, t$, and
$$
S_E = \fr{1}{\sqrt{2}}\lp 1 + i \ga_{4} \ga_{15} \rp \s \ga_4 \;\mapsto\; \ga^E_4 = i \ga_{15} \s \ga_{15} \;\mapsto\; \ga^E_{15} = i \ga_{4} \s \Ps \;\mapsto\; \Ps_E = S_E \Ps
$$
Some Cartan generators change under Wick rotation,
$$
S_E \{ \ga_{3 \, 4}, \ga_{15 \, 16} \} S^-_E = \{ i \ga_{3 \, 15}, i \ga_{4 \, 16} \}
$$
and the spinorial root vectors satisfy $\Ps^E_{\al^E} = S_E \Ps^L_{\al^L}$. For these two Cartan components, $\al^E = -i\al^L$ and $q^E = -q^L$; the other six charges agree. This Wick rotation relates the Euclidean $e_8$ triality eigenspaces, (\ref{teig}),
\beq \!\!\!\!\!\!
\ba{rcl}
e_{8(8)} &\sim& \Big[ 14^v_{-1} + 64^{s+}_{+\nfr{1}{2}} \Big]_{-1} + \Big[ so(8,6) + u(1)_w \Big]_0 + \Big[ 14^v_{+1} + 64^{s-}_{-\nfr{1}{2}} \Big]_{+1} \\[12pt]
e_{8(-24)} &\sim&  \Big[ 14^v_{-1} + 64^{s+}_{+\nfr{1}{2}} \Big]_{-1} + \Big[ so(4,10) + u(1)_w \Big]_0 + \Big[ 14^v_{+1} + 64^{s-}_{-\nfr{1}{2}} \Big]_{+1}
\ea \!\!\!\!\!\!
\label{teigs}
\eeq
to the Lorentzian spaces of the $e_8$ five-gradings, (\ref{grad}). Under $S_E$, the Lorentzian $64^{s+}_{+\nfr{1}{2}}$ maps to the Euclidean $64^{s-}_{-\nfr{1}{2}}$ in the $k=+1$ eigenspace, while its mirrors map to $64^{s+}_{+\nfr{1}{2}}$ in the $k=-1$ eigenspace.

In the spinor representation space of a compact or non-compact spin group, root vectors corresponding to fermions have complex conjugate root vectors called ``conjugate fermions'', which have complex conjugate roots --- the conjugate root coordinates having opposite signs for the compact Cartan generators, such as $\ga_{1\,2}$ spin, and the same signs for non-compact Cartan generators, such as $\ga_{3\,4}$ boost. For spinors of non-compact spin groups there can also be ``mirror fermions'', for which the mirror root coordinates have the same signs for the compact Cartan generators, and opposite signs for non-compact Cartan generators. In the five-graded decompositions, (\ref{grad}),
$$
\ba{rclrcl}
e_{8(8)} &=& 14^v_{-1} + 64^{s+}_{+\nfr{1}{2}} + \Big[ so(7,7) + so(1,1)_{w^\mathbb{R}} \Big] + 64^{s-}_{-\nfr{1}{2}} + 14^v_{+1} &=&  so(8,8) + 128^{s+} \\[12pt]
e_{8(-24)} &=& 14^v_{-1} + 64^{s+}_{+\nfr{1}{2}} + \Big[ so(3,11) + so(1,1)_{w^\mathbb{R}} \Big] + 64^{s-}_{-\nfr{1}{2}}  + 14^v_{+1} &=&  so(4,12) + 128^{s+}
\ea
$$
the $64^{s+}_{+\nfr{1}{2}}$ spinors of $so(7,7)$ or $so(3,11)$ are one generation of Standard Model fermions, and the $64^{s-}_{-\nfr{1}{2}}$ are their mirrors. The triality automorphisms, $\Th$, for which $[ so(8,6) + u(1)_w ]$ and $[ so(4,10) + u(1)_w ]$ are invariant subalgebras, mix the Standard Model fermions, $64^{s+}_{+\nfr{1}{2}}$, and mirror fermions, $64^{s-}_{-\nfr{1}{2}}$, in an interesting way.

For the mixed Cartan, (\ref{CarM}), the Standard Model fermion root vectors, such as $e_L^{\wedge} = \Ps_j$ (for some $j$) and their conjugate antifermions, $\bar{e}_R^{\vee} = \Ps^*_j$, are in $64^{s+}_{+\nfr{1}{2}}$, while their mirrors, $e_L^{\wedge m} = \Ps^m_j$ and $\bar{e}_R^{\vee m} = \Ps^{* m}_j = \Ps^{m *}_j$, are in $64^{s-}_{-\nfr{1}{2}}$, and all are in the $128^{s+}$ of $so(8,8)$ or $so(4,12)$, mixed by $\Th$. Applying the inverse Wick rotation, $\Ps^L = S^-_E \Ps^E$, gives the Lorentzian five-grading, in which triality mixes fermions and their mirrors,
$$
\Th \q : \q \lb \! \ba{c} \Ps_j \\[4pt] \Ps^m_j \ea \! \rb \;\mapsto\;
\lb \! \ba{c} \Ps'_j \\[4pt] \Ps'{}^m_j \ea \! \rb =
\lb \! \ba{cc} -\fr{1}{2} & \fr{\sqrt{3}}{2} i \\[2pt]  \fr{\sqrt{3}}{2} i & -\fr{1}{2}  \ea \! \rb
\lb \! \ba{c} \Ps_j \\[4pt] \Ps^m_j \ea \! \rb 
$$
The $64^{s+}_{+\nfr{1}{2}}$ is inhabited by the real, $\Ps^\mathbb{R}_j$, and imaginary, $\Ps^\mathbb{I}_j$, parts of the complex fermion root vectors, $\Ps_j$. And the $64^{s-}_{-\nfr{1}{2}}$ is inhabited by the real and imaginary parts of the complex mirror fermion root vectors. The triality automorphism mixes these real and imaginary parts of fermions and their mirrors,
$$
\scalebox{.95}{$
\ba{rcl}
\! \Ps^\mathbb{R}_j &=& \ha \lp \Ps_j + \Ps^{*}_j \rp \\[4pt]
\! \Ps^\mathbb{I}_j &=& \fr{1}{2 i} \lp \Ps_j - \Ps^{*}_j \rp \\[4pt]
\! \Ps^{m \mathbb{R}}_j &=& \ha \lp \Ps^m_j + \Ps^{m *}_j \rp \\[4pt]
\! \Ps^{m \mathbb{I}}_j &=& \fr{1}{2 i} \lp \Ps^m_j - \Ps^{m *}_j \rp 
\ea
\s\s
\Th  \q : \q  \lb \! \ba{c} \Ps{}^\mathbb{R}_j \\[4pt] \Ps{}^\mathbb{I}_j \\[4pt] \Ps{}^{m \mathbb{R}}_j \\[4pt] \Ps{}^{m \mathbb{I}}_j \ea \! \rb \;\mapsto\;
\lb \! \ba{c} \Ps'{}^\mathbb{R}_j \\[4pt] \Ps'{}^\mathbb{I}_j \\[4pt] \Ps'{}^{m \mathbb{R}}_j \\[4pt] \Ps'{}^{m \mathbb{I}}_j \ea \! \rb =
\lb \! \ba{cccc} -\fr{1}{2} &  & & -\fr{\sqrt{3}}{2} \\[2pt] & -\fr{1}{2} & \fr{\sqrt{3}}{2} & \\[2pt] & -\fr{\sqrt{3}}{2} & -\fr{1}{2} & \\[2pt] \fr{\sqrt{3}}{2} & & & -\fr{1}{2} \ea  \rb
\lb \! \ba{c} \Ps{}^\mathbb{R}_j \\[4pt] \Ps{}^\mathbb{I}_j \\[4pt] \Ps{}^{m \mathbb{R}}_j \\[4pt] \Ps{}^{m \mathbb{I}}_j \ea \! \rb
$}
$$
a rotation by $120^\circ$ in $64$ independent planes in the $128^{s+}$ of $so(8,8)$ or $so(4,12)$. This triality automorphism acts between three sets of complex state vectors,
$$
\ba{rcl}
\Ps^{I}_j &=& \Ps^\mathbb{R}_j + i \Ps^\mathbb{I}_j = \Ps_j \\[6pt]
\Ps^{II}_j &=& -\fr{1}{2} \Ps^\mathbb{R}_j - \fr{1}{2} i \Ps^{\mathbb{I}}_j  + \fr{\sqrt{3}}{2} i \Ps^{m \mathbb{R}}_j - \fr{\sqrt{3}}{2} \Ps^{m \mathbb{I}}_j  = -\fr{1}{2} \Ps_j + \fr{\sqrt{3}}{2} i \Ps^m_j\\[6pt]
\Ps^{III}_j &=& -\fr{1}{2} \Ps^\mathbb{R}_j - \fr{1}{2} i \Ps^{\mathbb{I}}_j  - \fr{\sqrt{3}}{2} i \Ps^{m \mathbb{R}}_j + \fr{\sqrt{3}}{2} \Ps^{m \mathbb{I}}_j  = -\fr{1}{2} \Ps_j - \fr{\sqrt{3}}{2} i \Ps^m_j
\ea
$$
as a cyclic permutation, $\Th : \Ps^{I}_j \mapsto \Ps^{II}_j \mapsto \Ps^{III}_j \mapsto \Ps^{I}_j$. Using this description, mirror fermion root vectors can be expressed as
\beq
\Ps^m_j = \fr{1}{i \sqrt{3}} \lp \Ps^{II}_j - \Ps^{III}_j \rp
\label{mirrors}
\eeq

The triality automorphism mixes the five-grading spaces $14^v_{\pm 1}$, spanned by $\ga_{x\,15} \mp \ga_{x\,16}$ with $1 \le x \le 14$, with zero-grade bivectors. For $x \!\ne\! 4$,
$$
\Th \q : \q \lb \! \ba{c} \ga_{x\,15} \\[4pt] \ga_{x\,16} \\[4pt] \ga_{x\,4} \ea \! \rb \;\mapsto\;
\lb \! \ba{c} \ga'_{x\,15} \\[4pt] \ga'_{x\,16} \\[4pt] \ga'_{x\,4} \ea \! \rb =
\lb \! \ba{ccc} 1 & 0 & 0 \\[4pt] 0 & -\fr{1}{2} & -\fr{\sqrt{3}}{2} \\[4pt] 0 & \fr{\sqrt{3}}{2} & -\fr{1}{2} \ea \! \rb
\lb \! \ba{c} \ga_{x\,15} \\[4pt] \ga_{x\,16} \\[4pt] \ga_{x\,4} \ea \! \rb
$$
Here $\ga_{x\,4}$ has zero grade. For $x \!\!=\!\! 4$, $\ga_{4\,16}$ is fixed and the same $120^\circ$ rotation acts on $(\ga_{4\,15},\ga_{15\,16})$.

\begin{figure}[p]
    \centering
    \vspace{40pt}
    \includegraphics[height=5.0in]{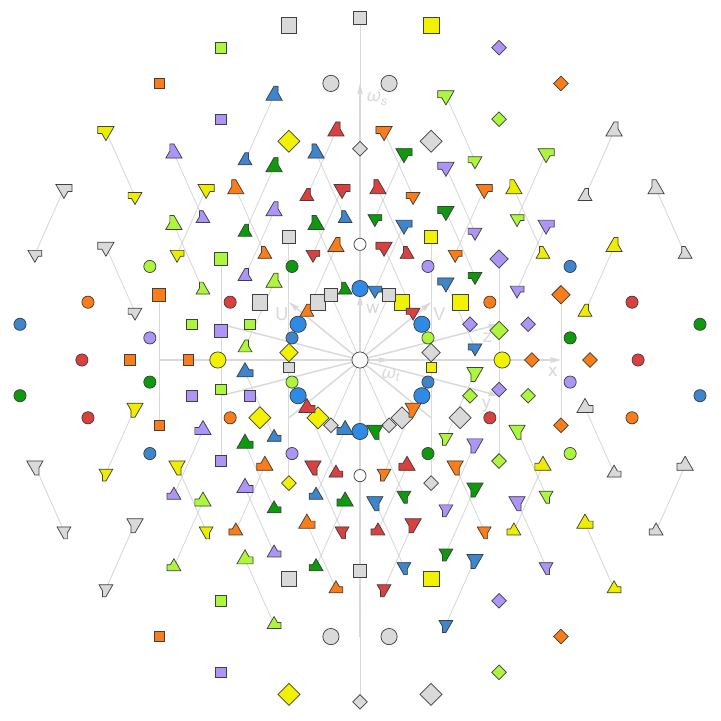}
    \vspace{60pt}
    \caption{The $240$ roots of $e_8$, suggestively labeled as elementary particles, with mirrors related by triality.}
    \label{fig:fE8w}
\end{figure}

Note that we have chosen the simplest possible triality automorphism, $\Th$, compatible with a Wick rotation from a five-grading, (\ref{grad}), to triality eigenspaces. The corresponding elementary particle assignments and the triality relationship to their mirrors are shown in Table \ref{table:pE8w} and Figure \ref{fig:fE8w}. Different choices of $u(1)_w$ within $so(8,8)$ or $so(4,12)$ correspond to different triality automorphisms that can give more complicated mixing between the $64^{s-}_{-\nfr{1}{2}}$ fermions and mirrors, and between the $14^v_{\pm 1}$ vectors. This suggests the three sets of $64$ triality-related generators may correspond to the three generations of fermions in the Standard Model of particle physics.

In the triality decomposition (\ref{e8tri}), the generator $w=\ga_{4\,16}$ and its automorphism are
$$
w = \fr{1}{\sqrt{3}} \lp \ga_1 + Q^-_1 + Q^+_1 \rp
\s\s t = e^{-\nfr{2 \pi}{3} \, \operatorname{ad}_w}
$$
where the subscript $1$ denotes the unit $1\otimes1$ in each of the three blocks.

\newpage

\begin{table}[h!t]
\centering
\scalebox{.52}{
\renewcommand{\arraystretch}{0.95}
\begin{tabular}
{@{\vrule width1.0pt}c@{\vrule width0.2pt}c@{\vrule width1.0pt}c@{\vrule width0.0pt}c@{\vrule width0.0pt}c@{\vrule width0.0pt}c@{\vrule width0.2pt}c@{\vrule width0.0pt}c@
{\vrule width0.0pt}c@{\vrule width0.0pt}c@{\vrule width1.0pt}}
\noalign{\hrule height 1.0pt}

\multicolumn{2}{@{\vrule width1.0pt}c@{\vrule width1.0pt}}{\raisebox{1pt}{$e_8$}} & $\q\; \om^{\mathbb{R}}_t \;\;\;$ {\vrule width0.2pt} & $\q\, \om_s \,\;\;$ {\vrule width0.2pt} & $\q U \;\;\;$ {\vrule width0.2pt} & $\q V \q$ & $\q w^{\mathbb{R}} \;\;\;$ {\vrule width0.2pt} & $\q\; x \;\;\;\;$ {\vrule width0.2pt} & $\q\; y \;\;\;\;$ {\vrule width0.2pt} & $\q\; z \;\q$ \\
\noalign{\hrule height 1pt}

 \raisebox{-1pt}{\gcirs{lgray}{2}}  & $\;\; \om_L^{\wedge/\vee} \;\;$ & $\; \mp 1 \;$ & $\; \pm 1 \;$ & $\; 0 \;$ & $\; 0 \;$ & $0 $ & $0 $ & $0 $ & $0 $ \\
\noalign{\hrule height 0.2pt}
 \raisebox{-1pt}{\gcirs{lgray}{2}}  & $\;\; \om_R^{\wedge/\vee} \;\;$ & $\; \pm 1 \;$ & $\; \pm 1 \;$ & $\; 0 \;$ & $\; 0 \;$  & $0 $ & $0 $ & $0 $ & $0 $ \\
\noalign{\hrule height 0.2pt}
\raisebox{-2pt}{\gcirs{lyell}{2}}  & \raisebox{-.5pt}{$\;\; W^{\pm\phantom{R}}_{\phantom{R}}\!\!\! \;\;$} & $\; 0 \;$ & $\; 0 \;$ & $\; \mp 1 \;$ & $\; \pm 1 \;$  & $0 $ & $0 $ & $0 $ & $0 $ \\
\noalign{\hrule height 0.2pt}
 \raisebox{-1pt}{\gcirs{mywhite}{1.5}}  & \raisebox{-.5pt}{$\;\; W'{}^{\pm\phantom{R}}_{\phantom{R}}\!\!\! \;\;$} & $\; 0 \;$ & $\; 0 \;$ & $\; \pm 1 \;$ & $\; \pm 1 \;$  & $0 $ & $0 $ & $0 $ & $0 $ \\
\noalign{\hrule height 0.2pt}
\raisebox{-1.5pt}{\gcirs{lblue}{2}} & \raisebox{1pt}{$\;\; g^{\phantom{R}}_{\phantom{R}}\!\!\! \;\;$} & $0 $ & $0 $ & $0 $ & $0 $ & $0 $ & $\! (+1 \;$ & $-1 \;$ & $\;\;\; 0 \; ) $  \\
\noalign{\hrule height 0.2pt}

$\;\!$ \raisebox{-1pt}{\gcirs{mred}{1.5} $\!$ \gcirs{mgree}{1.5} $\!$ \gcirs{mblue}{1.5}} $\;\!$ & $\;\, X_{\pm \nfr{2}{3}}^{PS}  \,\;$ & $0 $ & $0 $ & $0 $ & $0 $ & $ 0 $ & $\! ( \mp 1 \;$ & $ \mp 1 \;$ & $\;\;\; 0 \; ) $ \\
\noalign{\hrule height 0.2pt}
$\;\!$  \raisebox{-1pt}{\gcirs{mrora}{1.5} $\!$ \gcirs{mygre}{1.5} $\!$ \gcirs{mbvio}{1.5}}  $\;\!$ & $ Y_{\pm \nfr{1}{3}}^{GG} $ & $0 $ & $0 $ & $ \pm 1 $ & $0 $ & $ 0 $ & $\! ( \mp 1 \;$ & $ 0 $ & $\;\;\; 0 \; )$ \\
\noalign{\hrule height 0.2pt}
$\;\!$ \raisebox{-1pt}{\gcirs{mred}{1.5}  $\!$  \gcirs{mgree}{1.5}  $\!$  \gcirs{mblue}{1.5}}  $\;\!$ $\!\!$ & $\; X_{\mp \nfr{4}{3}}^{GG} \;$ & $0 $ & $0 $ & $0 $ & $ \mp 1 $ & $ 0 $ & $\! ( \pm 1 \;$ & $ 0 $ & $\;\;\; 0 \; )$ \\
\noalign{\hrule height 0.2pt}
$\;\!$  \raisebox{-1pt}{\gcirs{mrora}{1.5} $\!$ \gcirs{mygre}{1.5} $\!$ \gcirs{mbvio}{1.5}}  $\;\!$ & $ Y'_{\pm \nfr{1}{3}} $ & $0 $ & $0 $ & $ \mp 1 $ & $0 $ & $ 0 $ & $\! ( \mp 1 \;$ & $ 0 $ & $\;\;\; 0 \; )$ \\
\noalign{\hrule height .2pt}
$\;\!$ \raisebox{-1pt}{\gcirs{mred}{1.5}  $\!$  \gcirs{mgree}{1.5}  $\!$  \gcirs{mblue}{1.5}}  $\;\!$ $\!\!$ & $\; X'_{\pm \nfr{2}{3}} \;$ & $0 $ & $0 $ & $0 $ & $\pm 1 $ & $ 0 $ & $\! ( \pm 1 \;$ & $ 0 $ & $\;\;\; 0 \; )$ \\
\noalign{\hrule height .2pt}

 \raisebox{-1pt}{\gsqus{lgray}{2} } & $\;\; e_t^{\wedge/\vee} \ph_0  \;\;$ & $\; \pm 1 \;$ & $\; 0 \;$ & $\; + 1 \;$ & $\; 0 \;$  & $0 $ & $0 $ & $0 $ & $0 $ \\
\noalign{\hrule height 0.2pt}
 \raisebox{-1pt}{\gsqus{lgray}{2} } & $\;\; e_s^{\wedge/\vee} \ph_0  \;\;$ & $\; 0 \;$ & $\; \pm 1 \;$ & $\; + 1 \;$ & $\; 0 \;$  & $0 $ & $0 $ & $0 $ & $0 $ \\
\noalign{\hrule height 0.2pt}
 \raisebox{-1pt}{\gdias{lgray}{2} } & $\;\; e_t^{\wedge/\vee} \ph^*_0  \;\;$ & $\; \pm 1 \;$ & $\; 0 \;$ & $\; - 1 \;$ & $\; 0 \;$  & $0 $ & $0 $ & $0 $ & $0 $ \\
\noalign{\hrule height 0.2pt}
 \raisebox{-1pt}{\gdias{lgray}{2} } & $\;\; e_s^{\wedge/\vee} \ph^*_0  \;\;$ & $\; 0 \;$ & $\; \pm 1 \;$ & $\; - 1 \;$ & $\; 0 \;$  & $0 $ & $0 $ & $0 $ & $0 $ \\
\noalign{\hrule height 0.2pt}
 \raisebox{-1pt}{\gsqus{lyell}{2} } & $\;\; e_t^{\wedge/\vee} \ph_+  \;\;$ & $\; \pm 1 \;$ & $\; 0 \;$ & $\; 0 \;$ & $\; +1 \;$  & $0 $ & $0 $ & $0 $ & $0 $ \\
\noalign{\hrule height 0.2pt}
 \raisebox{-1pt}{\gsqus{lyell}{2} } & $\;\; e_s^{\wedge/\vee} \ph_+  \;\;$ & $\; 0 \;$ & $\; \pm 1 \;$ & $\; 0 \;$ & $\; +1 \;$  & $0 $ & $0 $ & $0 $ & $0 $ \\
\noalign{\hrule height 0.2pt}
 \raisebox{-1pt}{\gdias{lyell}{2} } & $\;\; e_t^{\wedge/\vee} \ph_-  \;\;$ & $\; \pm 1 \;$ & $\; 0 \;$ & $\; 0 \;$ & $\; -1 \;$  & $0 $ & $0 $ & $0 $ & $0 $ \\
\noalign{\hrule height 0.2pt}
 \raisebox{-1pt}{\gdias{lyell}{2} } & $\;\; e_s^{\wedge/\vee} \ph_-  \;\;$ & $\; 0 \;$ & $\; \pm 1 \;$ & $\; 0 \;$ & $\; -1 \;$  & $0 $ & $0 $ & $0 $ & $0 $ \\
\noalign{\hrule height 0.2pt}

$\;\!$  \raisebox{0pt}{\gsqus{mrora}{1.5} $\!$ \gsqus{mygre}{1.5} $\!$ \gsqus{mbvio}{1.5}}  $\;\!$ & $\;\; e_t^{\wedge/\vee} \ph_{+ \nfr{1}{3}}  \;\;$ & $ \pm 1 $ & $0 $ & $ 0 $ & $0 $ & $ 0 $ & $\! ( - 1 \;$ & $ 0 $ & $\;\;\; 0 \; )$ \\
\noalign{\hrule height 0.2pt}
$\;\!$  \raisebox{0pt}{\gsqus{mrora}{1.5} $\!$ \gsqus{mygre}{1.5} $\!$ \gsqus{mbvio}{1.5}}  $\;\!$ & $\;\; e_s^{\wedge/\vee} \ph_{+ \nfr{1}{3}}  \;\;$ & $ 0 $ & $ \pm 1 $ & $ 0 $ & $0 $ & $ 0 $ & $\! ( - 1 \;$ & $ 0 $ & $\;\;\; 0 \; )$ \\
\noalign{\hrule height 0.2pt}
$\;\!$  \raisebox{0pt}{\gdias{mrora}{1.5} $\!\!$ \gdias{mygre}{1.5} $\!\!$ \gdias{mbvio}{1.5}}  $\;\!$ & $\;\; e_t^{\wedge/\vee} \ph_{- \nfr{1}{3}}  \;\;$ & $ \pm 1 $ & $0 $ & $ 0 $ & $0 $ & $ 0 $ & $\! ( + 1 \;$ & $ 0 $ & $\;\;\; 0 \; )$ \\
\noalign{\hrule height 0.2pt}
$\;\!$  \raisebox{0pt}{\gdias{mrora}{1.5} $\!\!$ \gdias{mygre}{1.5} $\!\!$ \gdias{mbvio}{1.5}}  $\;\!$ & $\;\; e_s^{\wedge/\vee} \ph_{- \nfr{1}{3}}  \;\;$ & $ 0 $ & $ \pm 1 $ & $ 0 $ & $0 $ & $ 0 $ & $\! ( + 1 \;$ & $ 0 $ & $\;\;\; 0 \; )$ \\
\noalign{\hrule height 0.8pt}

 \raisebox{0pt}{\gsqus{lgray}{1.7} } & $\;\; e_t^{\wedge/\vee} w  \;\;$ & $\; \pm 1 \;$ & $\; 0 \;$ & $\; 0 \;$ & $\; 0 \;$  & $ + 1 $ & $0 $ & $0 $ & $0 $ \\
\noalign{\hrule height 0.2pt}
 \raisebox{0pt}{\gsqus{lgray}{1.7} } & $\;\; e_s^{\wedge/\vee} w  \;\;$ & $\; 0 \;$ & $\; \pm 1 \;$ & $\; 0 \;$ & $\; 0 \;$  & $ + 1 $ & $0 $ & $0 $ & $0 $ \\
\noalign{\hrule height 0.2pt}
\raisebox{-1pt}{\gsqus{lgray}{1.7} \gdias{lgray}{1.6}} & $w \ph_0 \; w \ph^*_0$ & $\; 0 \;$ & $\; 0 \;$ & $\; \pm 1 \;$ & $\; 0 \;$  & $ + 1 $ & $0 $ & $0 $ & $0 $ \\
\noalign{\hrule height 0.2pt}
\raisebox{-1pt}{\gsqus{lyell}{1.7} \gdias{lyell}{1.6}} & $w \ph_{\pm}$ & $\; 0 \;$ & $\; 0 \;$ & $\; 0 \;$ & $\; \pm 1 \;$  & $ + 1 $ & $0 $ & $0 $ & $0 $ \\
\noalign{\hrule height 0.2pt}
\raisebox{-1pt}{\gsqus{mrora}{1.7} \!\!\!\!\!\! \gsqus{mygre}{1.7} \!\!\!\!\!\! \gsqus{mbvio}{1.7} \gdias{mrora}{1.6} \!\!\!\!\!\!\! \gdias{mygre}{1.6} \!\!\!\!\!\!\! \gdias{mbvio}{1.6}} & $\; w \ph_{\pm \nfr{1}{3}} \;$ & $ 0 $ & $ 0 $ & $ 0 $ & $0 $ & $ +1 $ & $\! ( \mp 1 \;$ & $ 0 $ & $\;\;\; 0 \; )$ \\
\noalign{\hrule height .2pt}

\raisebox{-.5pt}{\gplus{lgray}{2} \gplds{lgray}{2}} & $\nu_{eL}^{\wedge/\vee}$ & $\mp \nha \;$ & $\pm \nha \;$ & $- \nha \;$ & $+ \nha \;$ & $ +\nha \;$ & $\!\! \, + \nha \;$ & $\!\! +\nha \,\;$ & $\!\!\! +\nha \,\,$ \\
\noalign{\hrule height 0.2pt}
\raisebox{-.5pt}{\gprus{lgray}{2} \gprds{lgray}{2}}  & $\nu_{eR}^{\wedge/\vee}$ & $\pm \nha \;$ & $\pm \nha \;$ & $+ \nha \;$ & $+ \nha \;$ & $ +\nha \;$ & $\!\! \, + \nha \;$ & $\!\! +\nha \,\;$ & $\!\!\! +\nha \,\,$ \\
\noalign{\hrule height 0.2pt}
\raisebox{-1pt}{\galus{lgray}{2} \galds{lgray}{2}} & $\bar{\nu}_{eL}^{\wedge/\vee}$ & $\mp \nha \;$ & $\pm \nha \;$ & $- \nha \;$ & $- \nha \;$ & $ +\nha \;$ & $\!\! \, - \nha \;$ & $\!\! -\nha \,\;$ & $\!\!\! -\nha \,\,$ \\
\noalign{\hrule height 0.2pt}
\raisebox{-1pt}{\garus{lgray}{2} \gards{lgray}{2}} & $\bar{\nu}_{eR}^{\wedge/\vee}$ & $\pm \nha \;$ & $\pm \nha \;$ & $+ \nha \;$ & $- \nha \;$ & $ +\nha \;$ & $\!\! \, - \nha \;$ & $\!\! -\nha \,\;$ & $\!\!\! -\nha \,\,$ \\
\noalign{\hrule height 0.2pt}
\raisebox{-.5pt}{\gplus{myell}{2} \gplds{myell}{2}}  & $e_L^{\wedge/\vee}$ & $\mp \nha \;$ & $\pm \nha \;$ & $+ \nha \;$ & $- \nha \;$ & $ +\nha \;$ & $\!\! \, + \nha \;$ & $\!\! +\nha \,\;$ & $\!\!\! +\nha \,\,$ \\
\noalign{\hrule height 0.2pt}
\raisebox{-.5pt}{\gprus{myell}{2} \gprds{myell}{2}}  & $e_R^{\wedge/\vee}$ & $\pm \nha \;$ & $\pm \nha \;$ & $- \nha \;$ & $- \nha \;$ & $ +\nha \;$ & $\!\! \, + \nha \;$ & $\!\! +\nha \,\;$ & $\!\!\! +\nha \,\,$ \\
\noalign{\hrule height 0.2pt}
\raisebox{-1pt}{\galus{myell}{2} \galds{myell}{2}} & $\bar{e}_{L}^{\wedge/\vee}$ & $\mp \nha \;$ & $\pm \nha \;$ & $+ \nha \;$ & $+ \nha \;$ & $ +\nha \;$ & $\!\! \, - \nha \;$ & $\!\! -\nha \,\;$ & $\!\!\! -\nha \,\,$ \\
\noalign{\hrule height 0.2pt}
\raisebox{-1pt}{\garus{myell}{2} \gards{myell}{2}} & $\bar{e}_{R}^{\wedge/\vee}$ & $\pm \nha \;$ & $\pm \nha \;$ & $- \nha \;$ & $+ \nha \;$ & $ +\nha \;$ & $\!\! \, - \nha \;$ & $\!\! -\nha \,\;$ & $\!\!\! -\nha \,\,$ \\
\noalign{\hrule height 0.2pt}
$\;$ \raisebox{-.5pt}{\gplus{mred}{2} \!\!\!\!\!\!\! \gplus{mgree}{2} \!\!\!\!\!\!\! \gplus{mblue}{2} \gplds{mred}{2} \!\!\!\!\!\!\! \gplds{mgree}{2} \!\!\!\!\!\!\! \gplds{mblue}{2}} $\;$& $u_{L}^{\wedge/\vee}$ & $\mp \nha \;$ & $\pm \nha \;$ & $- \nha \;$ & $+ \nha \;$ & $ +\nha \;$ & $\!\!\! ( + \nha \;$ & $\!\! -\nha \,\;$ & $\!\! -\nha \, )$ \\
\noalign{\hrule height 0.2pt}
\raisebox{-.5pt}{\gprus{mred}{2} \!\!\!\!\!\!\! \gprus{mgree}{2} \!\!\!\!\!\!\! \gprus{mblue}{2} \gprds{mred}{2} \!\!\!\!\!\!\! \gprds{mgree}{2} \!\!\!\!\!\!\! \gprds{mblue}{2}} & $u_{R}^{\wedge/\vee}$ & $\pm \nha \;$ & $\pm \nha \;$ & $+ \nha \;$ & $+ \nha \;$ & $ +\nha \;$ & $\!\!\! ( + \nha \;$ & $\!\! -\nha \,\;$ & $\!\! -\nha \, )$ \\
\noalign{\hrule height 0.2pt}
\raisebox{-1pt}{\galus{mred}{2} \!\!\!\!\!\!\! \galus{mgree}{2} \!\!\!\!\!\!\! \galus{mblue}{2} \galds{mred}{2} \!\!\!\!\!\!\! \galds{mgree}{2} \!\!\!\!\!\!\! \galds{mblue}{2}} & $\bar{u}_{L}^{\wedge/\vee}$ & $\mp \nha \;$ & $\pm \nha \;$ & $- \nha \;$ & $- \nha \;$ & $ +\nha \;$ & $\!\!\! ( - \nha \;$ & $\!\! +\nha \,\;$ & $\!\! +\nha \, )$ \\
\noalign{\hrule height 0.2pt}
\raisebox{-1pt}{\garus{mred}{2} \!\!\!\!\!\!\! \garus{mgree}{2} \!\!\!\!\!\!\! \garus{mblue}{2} \gards{mred}{2} \!\!\!\!\!\!\! \gards{mgree}{2} \!\!\!\!\!\!\! \gards{mblue}{2}} & $\bar{u}_{R}^{\wedge/\vee}$ & $\pm \nha \;$ & $\pm \nha \;$ & $+ \nha \;$ & $- \nha \;$ & $ +\nha \;$ & $\!\!\! ( - \nha \;$ & $\!\! +\nha \,\;$ & $\!\! +\nha \, )$ \\
\noalign{\hrule height 0.2pt}
\raisebox{-.5pt}{\gplus{mrora}{2} \!\!\!\!\!\!\! \gplus{mygre}{2} \!\!\!\!\!\!\! \gplus{mbvio}{2} \gplds{mrora}{2} \!\!\!\!\!\!\! \gplds{mygre}{2} \!\!\!\!\!\!\! \gplds{mbvio}{2}} & $d_{L}^{\wedge/\vee}$ & $\mp \nha \;$ & $\pm \nha \;$ & $+ \nha \;$ & $- \nha \;$ & $ +\nha \;$ & $\!\!\! ( + \nha \;$ & $\!\! -\nha \,\;$ & $\!\! -\nha \, )$ \\
\noalign{\hrule height 0.2pt}
\raisebox{-.5pt}{\gprus{mrora}{2} \!\!\!\!\!\!\! \gprus{mygre}{2} \!\!\!\!\!\!\! \gprus{mbvio}{2} \gprds{mrora}{2} \!\!\!\!\!\!\! \gprds{mygre}{2} \!\!\!\!\!\!\! \gprds{mbvio}{2}} & $d_{R}^{\wedge/\vee}$ & $\pm \nha \;$ & $\pm \nha \;$ & $- \nha \;$ & $- \nha \;$ & $ +\nha \;$ & $\!\!\! ( + \nha \;$ & $\!\! -\nha \,\;$ & $\!\! -\nha \, )$ \\
\noalign{\hrule height 0.2pt}
\raisebox{-1pt}{\galus{mrora}{2} \!\!\!\!\!\!\! \galus{mygre}{2} \!\!\!\!\!\!\! \galus{mbvio}{2} \galds{mrora}{2} \!\!\!\!\!\!\! \galds{mygre}{2} \!\!\!\!\!\!\! \galds{mbvio}{2}} & $\bar{d}_{L}^{\wedge/\vee}$ & $\mp \nha \;$ & $\pm \nha \;$ & $+ \nha \;$ & $+ \nha \;$ & $ +\nha \;$ & $\!\!\! ( - \nha \;$ & $\!\! +\nha \,\;$ & $\!\! +\nha \, )$ \\
\noalign{\hrule height 0.2pt}
\raisebox{-1pt}{\garus{mrora}{2} \!\!\!\!\!\!\! \garus{mygre}{2} \!\!\!\!\!\!\! \garus{mbvio}{2} \gards{mrora}{2} \!\!\!\!\!\!\! \gards{mygre}{2} \!\!\!\!\!\!\! \gards{mbvio}{2}} & $\bar{d}_{R}^{\wedge/\vee}$ & $\pm \nha \;$ & $\pm \nha \;$ & $- \nha \;$ & $ +\nha \;$ & $ +\nha \;$ & $\!\!\! ( - \nha \;$ & $\!\! +\nha \,\;$ & $\!\! +\nha \, )$ \\
\noalign{\hrule height .8pt}

 \raisebox{0pt}{\gsqus{lgray}{1.5} } & $\;\; e_t^{\wedge/\vee} w^m  \;\;$ & $\; \pm 1 \;$ & $\; 0 \;$ & $\; 0 \;$ & $\; 0 \;$  & $ -1 $ & $0 $ & $0 $ & $0 $ \\
\noalign{\hrule height 0.2pt}
 \raisebox{0pt}{\gsqus{lgray}{1.5} } & $\;\; e_s^{\wedge/\vee} w^m  \;\;$ & $\; 0 \;$ & $\; \pm 1 \;$ & $\; 0 \;$ & $\; 0 \;$  & $ -1 $ & $0 $ & $0 $ & $0 $ \\
\noalign{\hrule height 0.2pt}
\raisebox{-.5pt}{\gsqus{lgray}{1.5} \gdias{lgray}{1.4}} & $\;\; w^m \ph_0 \; w^m \ph^*_0 \;\;$ & $\; 0 \;$ & $\; 0 \;$ & $\; \pm 1 \;$ & $\; 0 \;$  & $ -1 $ & $0 $ & $0 $ & $0 $ \\
\noalign{\hrule height 0.2pt}
\raisebox{-.5pt}{\gsqus{lyell}{1.5} \gdias{lyell}{1.4}} & $w^m \ph_{\pm}$ & $\; 0 \;$ & $\; 0 \;$ & $\; 0 \;$ & $\; \pm 1 \;$  & $ -1 $ & $0 $ & $0 $ & $0 $ \\
\noalign{\hrule height 0.2pt}
\raisebox{-.5pt}{\gsqus{mrora}{1.5} \!\!\!\!\! \gsqus{mygre}{1.5} \!\!\!\!\! \gsqus{mbvio}{1.5} \gdias{mrora}{1.4} \!\!\!\!\!\! \gdias{mygre}{1.4} \!\!\!\!\!\! \gdias{mbvio}{1.4}} & $\; w^m \ph_{\pm \nfr{1}{3}} \;$ & $ 0 $ & $ 0 $ & $ 0 $ & $0 $ & $ -1 $ & $\! ( \mp 1 \;$ & $ 0 $ & $\;\;\; 0 \; )$ \\
\noalign{\hrule height 0.2pt}

\raisebox{-.5pt}{\gplus{lgray}{1.7} \gplds{lgray}{1.7}} & $\nu_{eL}^{\wedge/\vee m}$ & $\pm \nha \;$ & $\pm \nha \;$ & $- \nha \;$ & $+ \nha \;$ & $ -\nha \;$ & $\!\! \, + \nha \;$ & $\!\! +\nha \,\;$ & $\!\!\! +\nha \,\,$ \\
\noalign{\hrule height 0.2pt}
\raisebox{-.5pt}{\gprus{lgray}{1.7} \gprds{lgray}{1.7}}  & $\nu_{eR}^{\wedge/\vee m}$ & $\mp \nha \;$ & $\pm \nha \;$ & $+ \nha \;$ & $+ \nha \;$ & $ -\nha \;$ & $\!\! \, + \nha \;$ & $\!\! +\nha \,\;$ & $\!\!\! +\nha \,\,$ \\
\noalign{\hrule height 0.2pt}
\raisebox{-1pt}{\galus{lgray}{1.7} \galds{lgray}{1.7}} & $\bar{\nu}_{eL}^{\wedge/\vee m}$ & $\pm \nha \;$ & $\pm \nha \;$ & $- \nha \;$ & $- \nha \;$ & $ -\nha \;$ & $\!\! \, - \nha \;$ & $\!\! -\nha \,\;$ & $\!\!\! -\nha \,\,$ \\
\noalign{\hrule height 0.2pt}
\raisebox{-1pt}{\garus{lgray}{1.7} \gards{lgray}{1.7}} & $\bar{\nu}_{eR}^{\wedge/\vee m}$ & $\mp \nha \;$ & $\pm \nha \;$ & $+ \nha \;$ & $- \nha \;$ & $ -\nha \;$ & $\!\! \, - \nha \;$ & $\!\! -\nha \,\;$ & $\!\!\! -\nha \,\,$ \\
\noalign{\hrule height 0.2pt}
\raisebox{-.5pt}{\gplus{myell}{1.7} \gplds{myell}{1.7}}  & $e_L^{\wedge/\vee m}$ & $\pm \nha \;$ & $\pm \nha \;$ & $+ \nha \;$ & $- \nha \;$ & $ -\nha \;$ & $\!\! \, + \nha \;$ & $\!\! +\nha \,\;$ & $\!\!\! +\nha \,\,$ \\
\noalign{\hrule height 0.2pt}
\raisebox{-.5pt}{\gprus{myell}{1.7} \gprds{myell}{1.7}}  & $e_R^{\wedge/\vee m}$ & $\mp \nha \;$ & $\pm \nha \;$ & $- \nha \;$ & $- \nha \;$ & $ -\nha \;$ & $\!\! \, + \nha \;$ & $\!\! +\nha \,\;$ & $\!\!\! +\nha \,\,$ \\
\noalign{\hrule height 0.2pt}
\raisebox{-1pt}{\galus{myell}{1.7} \galds{myell}{1.7}} & $\bar{e}_{L}^{\wedge/\vee m}$ & $\pm \nha \;$ & $\pm \nha \;$ & $+ \nha \;$ & $+ \nha \;$ & $ -\nha \;$ & $\!\! \, - \nha \;$ & $\!\! -\nha \,\;$ & $\!\!\! -\nha \,\,$ \\
\noalign{\hrule height 0.2pt}
\raisebox{-1pt}{\garus{myell}{1.7} \gards{myell}{1.7}} & $\bar{e}_{R}^{\wedge/\vee m}$ & $\mp \nha \;$ & $\pm \nha \;$ & $- \nha \;$ & $+ \nha \;$ & $ -\nha \;$ & $\!\! \, - \nha \;$ & $\!\! -\nha \,\;$ & $\!\!\! -\nha \,\,$ \\
\noalign{\hrule height 0.2pt}
$\;$ \raisebox{-.5pt}{\gplus{mred}{1.7} \!\!\!\!\!\!\! \gplus{mgree}{1.7} \!\!\!\!\!\!\! \gplus{mblue}{1.7} \gplds{mred}{1.7} \!\!\!\!\!\!\! \gplds{mgree}{1.7} \!\!\!\!\!\!\! \gplds{mblue}{1.7}} $\;$& $u_{L}^{\wedge/\vee m}$ & $\pm \nha \;$ & $\pm \nha \;$ & $- \nha \;$ & $+ \nha \;$ & $ -\nha \;$ & $\!\!\! ( + \nha \;$ & $\!\! -\nha \,\;$ & $\!\! -\nha \, )$ \\
\noalign{\hrule height 0.2pt}
\raisebox{-.5pt}{\gprus{mred}{1.7} \!\!\!\!\!\!\! \gprus{mgree}{1.7} \!\!\!\!\!\!\! \gprus{mblue}{1.7} \gprds{mred}{1.7} \!\!\!\!\!\!\! \gprds{mgree}{1.7} \!\!\!\!\!\!\! \gprds{mblue}{1.7}} & $u_{R}^{\wedge/\vee m}$ & $\mp \nha \;$ & $\pm \nha \;$ & $+ \nha \;$ & $+ \nha \;$ & $ -\nha \;$ & $\!\!\! ( + \nha \;$ & $\!\! -\nha \,\;$ & $\!\! -\nha \, )$ \\
\noalign{\hrule height 0.2pt}
\raisebox{-1pt}{\galus{mred}{1.7} \!\!\!\!\!\!\! \galus{mgree}{1.7} \!\!\!\!\!\!\! \galus{mblue}{1.7} \galds{mred}{1.7} \!\!\!\!\!\!\! \galds{mgree}{1.7} \!\!\!\!\!\!\! \galds{mblue}{1.7}} & $\bar{u}_{L}^{\wedge/\vee m}$ & $\pm \nha \;$ & $\pm \nha \;$ & $- \nha \;$ & $- \nha \;$ & $ -\nha \;$ & $\!\!\! ( - \nha \;$ & $\!\! +\nha \,\;$ & $\!\! +\nha \, )$ \\
\noalign{\hrule height 0.2pt}
\raisebox{-1pt}{\garus{mred}{1.7} \!\!\!\!\!\!\! \garus{mgree}{1.7} \!\!\!\!\!\!\! \garus{mblue}{1.7} \gards{mred}{1.7} \!\!\!\!\!\!\! \gards{mgree}{1.7} \!\!\!\!\!\!\! \gards{mblue}{1.7}} & $\bar{u}_{R}^{\wedge/\vee m}$ & $\mp \nha \;$ & $\pm \nha \;$ & $+ \nha \;$ & $- \nha \;$ & $ -\nha \;$ & $\!\!\! ( - \nha \;$ & $\!\! +\nha \,\;$ & $\!\! +\nha \, )$ \\
\noalign{\hrule height 0.2pt}
\raisebox{-.5pt}{\gplus{mrora}{1.7} \!\!\!\!\!\!\! \gplus{mygre}{1.7} \!\!\!\!\!\!\! \gplus{mbvio}{1.7} \gplds{mrora}{1.7} \!\!\!\!\!\!\! \gplds{mygre}{1.7} \!\!\!\!\!\!\! \gplds{mbvio}{1.7}} & $d_{L}^{\wedge/\vee m}$ & $\pm \nha \;$ & $\pm \nha \;$ & $+ \nha \;$ & $- \nha \;$ & $ -\nha \;$ & $\!\!\! ( + \nha \;$ & $\!\! -\nha \,\;$ & $\!\! -\nha \, )$ \\
\noalign{\hrule height 0.2pt}
\raisebox{-.5pt}{\gprus{mrora}{1.7} \!\!\!\!\!\!\! \gprus{mygre}{1.7} \!\!\!\!\!\!\! \gprus{mbvio}{1.7} \gprds{mrora}{1.7} \!\!\!\!\!\!\! \gprds{mygre}{1.7} \!\!\!\!\!\!\! \gprds{mbvio}{1.7}} & $d_{R}^{\wedge/\vee m}$ & $\mp \nha \;$ & $\pm \nha \;$ & $- \nha \;$ & $- \nha \;$ & $ -\nha \;$ & $\!\!\! ( + \nha \;$ & $\!\! -\nha \,\;$ & $\!\! -\nha \, )$ \\
\noalign{\hrule height 0.2pt}
\raisebox{-1pt}{\galus{mrora}{1.7} \!\!\!\!\!\!\! \galus{mygre}{1.7} \!\!\!\!\!\!\! \galus{mbvio}{1.7} \galds{mrora}{1.7} \!\!\!\!\!\!\! \galds{mygre}{1.7} \!\!\!\!\!\!\! \galds{mbvio}{1.7}} & $\bar{d}_{L}^{\wedge/\vee m}$ & $\pm \nha \;$ & $\pm \nha \;$ & $+ \nha \;$ & $+ \nha \;$ & $ -\nha \;$ & $\!\!\! ( - \nha \;$ & $\!\! +\nha \,\;$ & $\!\! +\nha \, )$ \\
\noalign{\hrule height 0.2pt}
\raisebox{-1pt}{\garus{mrora}{1.7} \!\!\!\!\!\!\! \garus{mygre}{1.7} \!\!\!\!\!\!\! \garus{mbvio}{1.7} \gards{mrora}{1.7} \!\!\!\!\!\!\! \gards{mygre}{1.7} \!\!\!\!\!\!\! \gards{mbvio}{1.7}} & $\bar{d}_{R}^{\wedge/\vee m}$ & $\mp \nha \;$ & $\pm \nha \;$ & $- \nha \;$ & $ +\nha \;$ & $ -\nha \;$ & $\!\!\! ( - \nha \;$ & $\!\! +\nha \,\;$ & $\!\! +\nha \, )$ \\

\noalign{\hrule height 1.0pt}
\end{tabular}
}
\vspace{8pt}
\caption{The $240$ roots of quaternionic and split real $e_8$, labeled as particles with mixed-Cartan charges (\ref{CarM}). Their Wick images lie in the triality eigenspaces for (\ref{CarC}): $k=0$ contains the spin connection, weak and weaker bosons, gluons, $X$ and $Y$ bosons, and frame-Higgs; $k=+1$ contains a full Standard Model generation with spin and the mirror frame-w roots, \scalebox{.9}{$e^{\wedge/\vee}_{s/t}w^m$} and \scalebox{.9}{$w^m\ph$}; $k=-1$ contains their mirrors.}
\label{table:pE8w}
\end{table}

\newpage

\clearpage
\clearpage

\newpage

\section{Exceptional Unification}

The suggested assignments of elementary particle labels to roots of $e_6$, $e_7$, and $e_8$ correspond to deeper underlying theories of Exceptional Unification. Each deserves a long description; we only discuss them briefly here. One thing all have in common is a unification of bosons (gauge field $1$-forms) and fermions (spinor Grassmann number fields) in a superconnection valued in an exceptional Lie algebra, usually corresponding to a two-grading. This is sensible within the context of Lie Group Cosmology \cite{Lis15}, with Grassmann number fields understood as $1$-forms along spinorial directions within the Lie group. The supercurvature gives the usual curvature $2$-form of the connection $1$-form plus an exterior-Dirac derivative of the fermions, which can be used with a generalized-Hodge star to build a generalized Yang-Mills action that includes fermions. The exceptional Lie brackets produce the correct interactions between gauge fields and spinorial fermions, which is encouraging. Beyond these commonalities, Exceptional Unification models include the gauge fields of the $SO(10)$ Grand Unified Theory: the electroweak $W$ and $Z$ and $\ga$, and gluons, $g$, as well as the ``weaker'' $W'$ and $Z'$ gauge bosons, Pati-Salam $so(6)$ gauge bosons, $X^{PS}$, Georgi-Glashow $su(5)$ gauge bosons, $X^{GG}$ and $Y^{GG}$, and other GUT gauge bosons, $X'$ and $Y'$. The most minimal model is based on $e_6$, a subalgebra of $e_7$ and of $e_8$.

The particle assignment within $e_6$, shown in Table \ref{table:pE6}, largely ignores triality and corresponds to the $SO(10)$ Grand Unified Theory with one generation of fermions and antifermions. The five-dimensional Cartan subalgebra of $so(10)$ corresponds to charges $(U,V,x,y,z)$, with $U$ and $V$ combining to produce $su(2)$ weak charge, $W$, and $su(2)$ weaker charge, $W'$, while $(x,y,z)$ combine to give strong $su(3)$ color charges and $u(1)$ baryon minus lepton number charge, $B$, (\ref{charges}). These further combine to give hypercharge, $Y$, and electric charge, $Q$. The sixth fundamental Cartan subalgebra element within $e_6$ is scaled helicity, $H$, corresponding to a $u(1)$ gauge field. One generation of fermions, without spin, is embedded in $e_6$, with left-chiral fermions and antifermions in a $16_{s-}$ spinor of $so(10)$, and right-chiral fermions and antifermions in a $16_{s+}$.

The particle assignment within $e_7$, shown in Table \ref{table:pE7}, relates to the work of Dixon, Furey, and Hughes, with $\mathbb{C}\otimes\mathbb{H}\otimes\mathbb{O}$ fermions \cite{Dix94, Fur22, Fur24}. While it is possible to fit three linearly-independent generations of Standard Model fermions into $e_7$, related by triality, one cannot do it with both the left and right-chiral partners and the antiparticle partners -- one must choose either, and is thus forced to use a complex form of $e_7$ to fit three generations, which is problematic for the gauge fields. Alternatively, as shown in Table \ref{table:pE7}, one can include one generation of left-chiral fermions and antifermions, with spin, all in the same triality eigenspace, $[2 \otimes 16^{s-}_{- \nha}]$. There are then two choices. One could consider the right-chiral fermions and antifermions to be in $[2 \otimes 16^{s+}_{+\nha}]$, and call it a day, with a one-generation $e_7$ model. Or, using the triality-rotation of the left-chiral generation within $[2 \otimes 32_{s+}]$, one could consider this a three-generation model, but again only with left-chiral fermions and antifermions. In either case the gravitational frame is missing from $e_7$, and it is not clear what to make of the $\Phi \in 10^v_{\pm 1}$ --- though it includes a needed Higgs field. The $\Phi$ appears to be a scalar or possibly Grassmann field, useful for $SO(10)$ and Standard Model symmetry breaking, but understanding its nature and a consistent E7 Theory awaits future work.

The particle assignment within $e_8$, shown in Table \ref{table:pE8p} and Figure \ref{fig:fE8p}, matches the particle assignment in ``An Exceptionally Simple Theory of Everything'' \cite{Lis07}. Each generation of particles matches to an $\mathbb{O}\otimes\mathbb{O}$ or to an $\mathbb{O}'\otimes\mathbb{O}$, depending on whether the $(\om_T,\om_s,U,V)$ corresponds to an $so(8)$ or $so(4,4)$, and each linearly-independent generation is related to the others by triality. The $(p,x,y,z)$ $so(8)$ charges include a particle or antiparticle charge, $p$, which mixes with $B$ under triality. The main downside of this embedding of the three generation Standard Model in $e_8$ is that it is necessarily Euclidean, and must be part of some larger theory to describe our Lorentzian reality. Also, the natural two-grading of $e_8$ does not match this assignment of fermions. Several variations of this embedding of three linearly-independent generations of fermions within different real forms of $e_8$ are possible, but all have these drawbacks.

A different particle assignment within quaternionic or split real $e_8$, matching its natural two-grading and its triality eigenspaces, is shown in Table \ref{table:pE8w} and Figure \ref{fig:fE8w}. The Wick-rotated root vectors are triality eigenvectors for the compact Cartan, with $\om_t$ and $w$ charges opposite to the tabulated $\om^\mathbb{R}_t$ and $w^\mathbb{R}$ charges. In the mixed-Cartan basis, the triality eigenspaces are inhabited by mixtures of particles and their mirrors. This produces a nice result: triality automorphisms of non-compact $e_8$ can rotate three full generations of fermions between standard fermions and their mirrors. Early criticism of E8 Theory held that these mirror fermions (also called an ``anti-generation'') imply the theory cannot work \cite{Dis09}, but this is not our view. In our model, the mirror fermions are replaced by the second and third generations, related to the first generation by triality. Of course, this implies the three generations of fermions are not linearly-independent, and we do expect there to be generational mixing. Triality also mixes the frame-Higgs root vectors that have a temporal component, such as $\{e_t \ph_0, e_t \ph^*_0, e_t w, e_t w^m\}$. This may introduce multi-fingered time, with these fields interacting differently with the three generations of fermions. This new description of E8 Theory, using triality eigenspaces, has many promising features. All three generations of fermions live in the odd part of $e_8$'s natural two-grading. The gravitational spin connection, frame-Higgs, and $SO(10)$ GUT gauge fields largely live in the triality invariant eigenspace. And the spin and gauge charges of the three generations of fermions, from $\{\om_s,U,V,x,y,z\}$, are the same, producing identical Standard Model quantum numbers for all three generations of fermions. Further development of this model seems warranted. 

\newpage

\section{Division Algebra Automorphisms and \texorpdfstring{$g_2$}{g2}}

The division algebras, and their split algebras, are each invariant under transformation by elements of their \emph{automorphism group}, $\Ph \in G_{\mathbb D}$ \cite{Dix94}. These group transformations leave multiplication invariant, $\Ph(a) \Ph(b) = \Ph(a b)$. The complex numbers are invariant under complex conjugation. The quaternions are invariant under $SO(3)$ rotations of their imaginary elements. The octonions are not invariant under $SO(7)$ rotations of their imaginary elements, but under a subgroup, $G_2$, that preserves octonionic non-associativity. The automorphism groups of the split algebras are similar,
$$
\ba{rclcrcl}
G_{\mathbb C} &=& {\mathbb Z}_2 & \s \s \s \s & G_{\mathbb C'} &=& {\mathbb Z}_2 \\[3pt]
G_{\mathbb H} &=& SO(3) & \s \s \s \s & G_{\mathbb H'} &=& SO(1,2) \\[3pt]
G_{\mathbb O} &=& G_{2(-14)} & \s \s \s \s & G_{\mathbb O'} &=& G_{2(2)}
\ea
$$
Since automorphism group elements leave division algebra multiplication invariant, these groups are subgroups of the corresponding triality group.

For the quaternions, there is an \emph{inner triality automorphism}, $\text{Ad}_t$, corresponding to a $120$-degree rotation around the axis formed by averaging the three unit imaginary quaternions. This rotation cycles the three imaginary unit quaternions,
$$
t = -\ha(e_0 + e_1 + e_2 + e_3)
\s\;\;
t^3 = e_0
\s\;\;
t \, e_1 \, t^- = e_2
\s\;\;
t \, e_2 \, t^- = e_3
\s\;\;
t \, e_3 \, t^- = e_1
$$
If we instead interpret $t$ as an octonion, we see that $\text{Ad}_t$ also preserves octonionic multiplication. Using the multiplication tables (\ref{M}), the rows of the following matrices give the images of the ordered basis $(e_0,\ldots,e_7)$:
$$
\text{Ad}_t \s \sim \s
\scalebox{.8}{$ \lb \ba{cccccccc}
1 & & & & & & & \\[-3pt]
 & & +\nha & +\nha & -\nha & & & -\nha \\
 & +\nha & & +\nha & +\nha & -\nha & & \\
 & +\nha & +\nha & & & +\nha & & +\nha \\
 & +\nha & -\nha & & -\nha & & +\nha & \\
 & & +\nha & -\nha & & -\nha & +\nha & \\
 & & & & -\nha & -\nha & -\nha & +\nha \\
 & +\nha & & -\nha & & & -\nha & -\nha
\ea \rb$}
$$
If we use the same $t$ in the split-octonions, we get the automorphism:
$$
\text{Ad}'_t \s \sim \s
\scalebox{.8}{$ \lb \ba{cccccccc}
1 & & & & & & & \\[-3pt]
 & & 1 & & & & & \\[-3pt]
 & & & 1 & & & & \\[-3pt]
 & 1 & & & & & & \\[-3pt]
 & & & & -\nha & -\nha & -\nha  & -\nha \\
 & & & & +\nha & -\nha & -\nha  & +\nha \\
 & & & & +\nha & +\nha & -\nha  & -\nha \\
 & & & & +\nha & -\nha & +\nha  & -\nha
\ea \rb$}
$$
This same $t$ does not give an automorphism in the split-quaternions. However, if we have $e'_2 e'_2 = -e'_0$, we can have an inner triality automorphism from $t' = -\ha(e'_0 - \sqrt{3} e'_2)$, which rotates $120^\circ$ in the $e'_1-e'_3$ plane.

\newpage

\section{Discussion}

In this work we have explored the deep relationship between division algebras, Clifford algebras, generalized reflections, triality automorphisms, triality Lie algebras, the magic square of Lie algebras, exceptional Lie algebras, root systems, triality eigenspaces, Vinberg $\Th$-groups, and connections with particle physics. From any of these specific subjects, the others can be better understood, so a reader may make their choice of whichever mathematical starting point is more familiar. Although it is expected that current researchers will merely pull useful computational tools from this paper, it is hoped that they will also use this paper to better understand and appreciate the other ways of working with this material for model building.

The true heart of this subject is triality --- a real, cyclic, trilinear function of three elements of a vector space --- which can be used to define the division algebras and their related split composition algebras. These division algebras lend themselves to the explicit matrix representation of certain Clifford algebras, which have a direct geometric interpretation. With this geometric point of view, we can describe reflections through vectors and, using triality, generalized reflections through spinors. These generalized reflections can be combined to define triality automorphisms, cycling vectors and spinors or three division algebra elements. These vectors and spinors combine with the triality algebras of division algebras to produce the triality Lie algebras, understood as generalizations of $su(3)$. Within these triality Lie algebras, the relationships between vectors, chiral spinors, generalized reflections, and triality automorphisms are described using division algebra products or more explicitly using representative Clifford algebra matrices. We encounter the surprising fact that real Lie algebra automorphisms can include explicit $i$'s, provided the automorphisms commute with the complex conjugation used to define the real form of the Lie algebra. The Cartan-Weyl description of these Lie algebras, and their root systems, can be used to visually appreciate their structure and automorphisms. Although useful, and often visually appealing, the root system description of Lie algebras and their automorphisms elides the signs of structure constants and maps between root vectors. These signs can often be guessed or obtained algorithmically, but are obtained directly by the methods presented in this work. The triality Lie algebras combine in pairs to produce the magic square Lie algebras, which are also invariant under generalized reflections and triality automorphisms. Ultimately, building from division algebras, all exceptional Lie algebras and their automorphisms can be understood and described explicitly using these methods. With an explicit description of triality automorphisms in hand, triality eigenspaces are described, and used to define Vinberg $\Th$-groups and $3$-symmetric spaces. A canonical triality automorphism for any magic square Lie algebra in a triality decomposition is generated by the Lie algebra element
$$
T = \fr{1}{\lambda\sqrt{3}} \lp \ga_1 + Q^-_1 + Q^+_1 \rp
$$
where $[\ga_1,Q^-_1]=\lambda Q^+_1$. This is the normalized sum of the three unit elements in the vector, negative chiral spinor, and positive chiral spinor blocks. As an inner automorphism, the canonical triality automorphism is the exponentiation of the adjoint action of this generator, $t = e^{-\nfr{2 \pi}{3} \, \operatorname{ad}_T}$. These tools can be used for model building, and possibly to describe the three generations of fermions in the Standard Model.

The relationship to particle physics largely motivates this work, with triality at its center. The strongest objection to E8 Theory has been the mathematical fact that, under the embedding assumptions of \cite{Dis09}, real forms of $e_8$ necessarily include mirror fermions. While this condemnation is widely accepted by the particle physics community, it aggressively ignores the possibility that mirror fermion degrees of freedom may be inhabited by second and third generation fermions via triality --- a possibly wrong idea proposed and explored in our work here. Although the complete details of describing three fermion generations using triality remain unclear, if triality is an interesting area of exploration for physics, which it almost certainly is, then the explicit tools and descriptions provided in this paper are the keys to the castle. It is hoped that other researchers will use this work to further their own explorations within this rich area of mathematical physics.

\section{Disclosures and Declarations}

\subsection{Data Availability Statement}
This article contains only theoretical analysis, equations, and derivations. No datasets were generated or analyzed in this work.

\subsection{Funding}
The author received no financial support for the research, authorship, or publication of this article.

\subsection{Competing Interests}
The author declares that there are no conflicts of interest regarding the publication of this work.

\bibliographystyle{JHEPg}
\bibliography{DATaEM.6.5}

\end{document}